\documentclass[12pt]{article}
\usepackage[T1]{fontenc}
\usepackage[margin=1in]{geometry}

\usepackage{amsmath}
\usepackage{accents, adjustbox, afterpage, amssymb, amsfonts, amsthm, array, booktabs, calc, caption, color, comment, datetime, dcolumn, enumitem, eurosym, float, geometry, graphics, graphicx, longtable, mathtools, mathbbol, multirow, natbib, pdflscape, pdfpages, ragged2e, rotating, sectsty, setspace, subcaption, subfiles, textpos, threeparttable, tikz, soul, xcolor, xparse
}

\usepackage[hang, flushmargin, bottom]{footmisc}

\usepackage{hyperref}
\setlist[enumerate]{itemsep=0pt,topsep=2pt}
\AtBeginDocument{
  \setlength\abovedisplayskip{4pt}
  \setlength\belowdisplayskip{4pt}
}
\usepackage[sf,sl,outermarks]{titlesec}
\newcommand{\addperiod}[1]{#1.}

\titleformat{\section}[block]
{\normalfont\Large\bfseries}{\thesection.}{.5em}{\Large\bfseries}
\titlespacing*{\section}{0pt}{*1.3}{*0.2}

\titleformat{\subsection}[block]
{\normalfont\large\bfseries}{\thesubsection.}{.5em}{\large\bfseries}
\titlespacing*{\subsection}{0pt}{*1}{*0}

\titleformat{\subsubsection}[block]
{\normalfont\large\bfseries}{\thesubsubsection.}{.5em}{\large\bfseries}
\titlespacing*{\subsubsection}{0pt}{*1}{*0}

\titleformat{\paragraph}[runin]
{\normalfont\bfseries}{}{0em}{\normalsize\bfseries\addperiod}
\titlespacing*{\paragraph}{0pt}{*.15}{*1}

\NewDocumentCommand{\hyref}{m O{}O{}}{\hyperref[#1]{#2 \ref{#1}#3}}
\setcitestyle{authoryear,aysep={},open={(},close={)}}
\bibpunct{(}{)}{;}{a}{,}{,}

\usepackage{etoolbox}

\makeatletter
\DeclareRobustCommand\citepos													
  {\begingroup\def\NAT@nmfmt##1{{\NAT@up##1's}}%
   \NAT@swafalse\let\NAT@ctype\z@\NAT@partrue
   \@ifstar{\NAT@fulltrue\NAT@citetp}{\NAT@fullfalse\NAT@citetp}}

\pretocmd{\NAT@citex}{%
  \let\NAT@hyper@\NAT@hyper@citex
  \def\NAT@postnote{#2}%
  \setcounter{NAT@total@cites}{0}%
  \setcounter{NAT@count@cites}{0}%
  \forcsvlist{\stepcounter{NAT@total@cites}\@gobble}{#3}}{}{}
\newcounter{NAT@total@cites}
\newcounter{NAT@count@cites}
\def\NAT@postnote{}

\def\NAT@hyper@citex#1{
  \stepcounter{NAT@count@cites}%
  \hyper@natlinkstart{\@citeb\@extra@b@citeb}#1%
  \ifnumequal{\value{NAT@count@cites}}{\value{NAT@total@cites}}
    {\if*\NAT@postnote*\else\NAT@cmt\NAT@postnote\global\def\NAT@postnote{}\fi}{}%
  \ifNAT@swa\else\if\relax\NAT@date\relax
  \else\NAT@@close\global\let\NAT@nm\@empty\fi\fi								
  \hyper@natlinkend}
\renewcommand\hyper@natlinkbreak[2]{#1}

\patchcmd{\NAT@citex}															
  {\ifNAT@swa\else\if*#2*\else\NAT@cmt#2\fi
   \if\relax\NAT@date\relax\else\NAT@@close\fi\fi}{}{}{}
\patchcmd{\NAT@citex}
  {\if\relax\NAT@date\relax\NAT@def@citea\else\NAT@def@citea@close\fi}
  {\if\relax\NAT@date\relax\NAT@def@citea\else\NAT@def@citea@space\fi}{}{}
\patchcmd{\NAT@cite}{\if*#3*}{\if*\NAT@postnote*}{}{}
\makeatother

\usepackage{dsfont,fouriernc,charter,libertine}

\newtheoremstyle{newtheorem}{5pt}{2pt}{\itshape}{0pt}{\bfseries}{.}{.4em}{\thmname{#1}\thmnumber{ #2}\textnormal{\thmnote{ (#3)}}}
\theoremstyle{newtheorem}

\newtheorem{definition}{Definition}
\newtheorem{hypothesis}{Hypothesis}

\newtheorem{proposition}{Proposition}

\definecolor{cblue}{rgb}{155, 221, 255}

\hypersetup{
    colorlinks=true,
    linkcolor={red!50!black},
    citecolor={blue!50!black},
    urlcolor={blue!50!black},
    pdffitwindow=false,
}
\newdateformat{DDMonthYYYY}{\THEDAY\, \monthname[\THEMONTH] \THEYEAR}
\newcommand\blfootnote[1]{\begingroup\renewcommand\thefootnote{}\footnote{#1}\addtocounter{footnote}{-1}\endgroup}

\definecolor{mygreen}{RGB}{0, 150, 0}
\definecolor{myred}{RGB}{207,8,8}

\hypersetup{
    pdffitwindow=false,colorlinks=true,linkcolor={red!50!black},citecolor={blue!50!black},urlcolor=black,linkbordercolor={white},
}

\usetikzlibrary{arrows.meta}

\begin{document}
\thispagestyle{empty}
\setcounter{page}{0}
\setcounter{footnote}{0}
\renewcommand{\thefootnote}{\fnsymbol{footnote}}
~\vspace*{-2cm}\\
\begin{center}
    {\noindent
    \Large
    \bfseries  Retractions:\\[.5em] 
    Updating from Complex Information
    }
\end{center}

\makebox[\textwidth][c]{
    \begin{minipage}{1.2\linewidth}
        \Large\centering
        \begin{tabular}{clr}
            Duarte Gonçalves\footnotemark & 
            \quad Jonathan Libgober\footnotemark & 
            \quad Jack Willis\footnotemark 
        \end{tabular}
    \end{minipage}
}
\setcounter{footnote}{1}\footnotetext{
    \setstretch{1} Department of Economics, University College London; \hyperlink{mailto:duarte.goncalves@ucl.ac.uk}{\color{black}duarte.goncalves@ucl.ac.uk}.
}
\setcounter{footnote}{2}\footnotetext{
    \setstretch{1} Department of Economics, University of Southern California; \hyperlink{mailto:libgober@usc.edu}{\color{black}libgober@usc.edu}.
}
\setcounter{footnote}{3}\footnotetext{
    \setstretch{1} Department of Economics, Columbia University; \hyperlink{mailto:jack.willis@columbia.edu}{\color{black}jack.willis@columbia.edu}.
}
\blfootnote{
    An earlier version of this paper was circulated in June 2021 and then as an NBER working paper in November 2021 under the title ``Learning versus Unlearning: An Experiment on Retractions''. 
    We benefited from helpful comments from many individuals, but we especially thank 
    Dan Benjamin, 
    Giorgio Coricelli, 
    Thomas Chaney,  
    Cary Frydman, 
    Terri Kneeland, 
    Rani Spiegler, 
    Michael Thaler, 
    S\'{e}verine Toussaert, 
    Sevgi Yuksel,
    and seminar audiences at Bonn, Nottingham, NYU, Caltech, Ohio State, USC, UCL, 
    and the SWEET, and ESA conferences. 
    Jeremy Ward, Malavika Mani, and Julen Zarate-Pina provided excellent research assistance. 
    We would also like to thank the editor and anonymous referees for their valuable feedback and insightful comments, which significantly improved the quality of our manuscript.
    Funding from IEPR is gratefully acknowledged.\\
	\emph{First posted draft}: 19 June 2021. 
    \emph{This draft}: 
    27 February 2025. 
}
    \vspace*{2em}

\setcounter{footnote}{0} \renewcommand{\thefootnote}{\arabic{footnote}}

\begin{center} \textbf{\large Abstract} \end{center}
\vspace*{1em}
\noindent\makebox[\textwidth][c]{
    \begin{minipage}{.85\textwidth}
        \noindent
        We modify a canonical experimental design to identify the effectiveness of retractions. 
        Comparing beliefs after retractions to beliefs (a) without the retracted information and (b) after equivalent new information, we find that retractions result in diminished belief updating in both cases. 
        We propose this reflects updating from retractions being more complex, and our analysis supports this: we find longer response times, lower accuracy, and higher variability. 
        The results---robust across diverse participant groups and design variations---enhance our understanding of belief updating and offer insights into addressing misinformation. 
        ~
        \\\\
        \textbf{Keywords:} Belief Updating; Retractions; Information; Complexity.\\
        \textbf{JEL Classifications:} D83, D91, C91.
    \end{minipage}
}
\newpage

\setcounter{page}{1}

\section{Introduction}
\label{section:introduction}

Retracted information often influences beliefs even once widely discredited. 
A notorious example is the enduring belief in a link between vaccines and autism, fuelled by a subsequently retracted study in \textit{The Lancet}. 
The article's impact persists as the belief in such an association remains widespread, significantly harming public health \citep[see][]{gabis2022myth,motta2021quantifying,pluviano2017misinformation,Pullan2021}.
While this case is illustrative, retractions are pervasive, and retracted information is rarely erased entirely.\footnote{
    Focusing on tracking retractions of academic papers, the Retraction Watch Database lists over 45,000 articles, with error and failure to replicate constituting a significant fraction of the retraction notices, in addition to misconduct \citep{Science2018}. 
    Of the ten most cited retracted articles as of October 2023 in the Retraction Watch Database, seven had over a hundred citations since retraction, and two that had fewer had only been retracted in 2023 \citep{retractionwatch}. 
    We highlight that many papers that fail to be replicated are not retracted \citep{Serra-GarciaGneezy2021SciAdv}. 
}

Variations of this phenomenon arise in a wide range of situations, from groundless rumours to erroneous earnings reports and from fraudulent research to false political claims. 
Naturally, each case is different, and retraction effectiveness in particular cases can always be attributed to unique intervening factors---e.g., special media coverage, ulterior financial or political motives, or source reliability.
However, while idiosyncratic factors may play important roles, issues in updating from retractions are documented too consistently and in too wide a range of settings for case-by-case explanations to be the whole story. 
This observation suggests moving beyond idiosyncratic factors to identify causes common to retractions generally.

In this paper, we investigate if and why there is a fundamental friction in updating beliefs from retractions. 
To this end, we modify a canonical experimental design to identify and quantify diminished updating from retractions relative to direct evidence, absent a variety of idiosyncratic confounds. 
Our analysis reveals beliefs update significantly less from retractions than from direct evidence, a finding that challenges explanations unrelated to intrinsic characteristics of retractions. 
We propose a simple explanation: retractions convey more complex information than direct evidence. 
To support this hypothesis, we present evidence based on common empirical measures of complexity---specifically, accuracy, response time, and response variability.
We document these basic patterns across numerous variations, show that they are robust to certain alternative implementation details, and argue against several natural competing explanations unrelated to the processing of retractions.

Identifying the diminished effectiveness of retractions requires a clear benchmark against which updating from retractions can be measured. 
The aforementioned confounding factors in particular settings complicate assessing how individuals should interpret any given retraction.
Whether the retraction was prompted by negligence or malfeasance, casts doubt on other evidence, is politically motivated, or is disputed all influence a retraction's correct interpretation but may not be precisely quantifiable.
At the same time, as individuals may err when interpreting \emph{any} information, the mere presence of an error does not imply differential treatment of retractions compared to other pieces of new evidence. 
Indeed, previously documented updating biases may appear capable of explaining the diminished effectiveness of retractions. 
Perhaps most notable among these is \emph{confirmation bias}---updating more when information confirms one's prior beliefs than when it does not \citep{RabinSchrag1999}---as it suggests individuals resist disregarding information supporting their beliefs, such as a discredited study.

We develop a variation on the classic balls-and-urns experiment to identify and quantify updating from retractions absent a variety of idiosyncratic confounds.
This canonical experimental design is widely used to study limitations in information processing, for example, in belief updating \citep{Benjamin2019,AugenblickEtAl,BaBohrenImas2022}, social learning \citep{AndersonHolt1997AER,Weizsacker2010AER,Angrisanietal2020}, and asset pricing \citep{HalimRiyantoRoy2019JF}. 
Our version allows us to repeatedly provide retractions that are informationally equivalent to new observations to participants facing identical problems. 
At the same time, we, as analysts, have access to quantifiable information about the objective truth. 
These properties are essential to distinguish belief updating issues specific to retractions.

We briefly describe how we modify the canonical balls-and-urns design to accommodate retractions. 
As is standard, participants are presented with draws of balls from a box (with replacement), which are either blue or yellow. 
In our version, balls can be ``noise balls,'' which are blue and yellow in equal proportion, or a ``truth ball,'' which is either blue or yellow. 
Instead of asking if the box has a majority of blue or yellow balls, we elicit beliefs about whether the truth ball is blue or yellow, an equivalent event. 
After a number of draws, in which participants are shown the colour but not the truth/noise status of the ball drawn, we then either present another such draw or inform participants whether a randomly chosen earlier ball draw was the truth ball or a noise ball. 
We refer to the disclosure that an earlier draw was noise as a \emph{retraction}. 
In our formulation, retractions \emph{only} provide information that a given signal was noise, analogous to a researcher having fabricated data or a news article relying on made-up claims.\footnote{
    \label{footnote:BroockmanKalla2016}
    Retractions of scientific articles are often due to problems with experimental conduct, suggesting uninformative findings but leaving open the possibility that the tested hypotheses are true. 
    One example illustrating this possibility is the retracted study on the impact of contact on opinion formation; despite the fabrication of evidence from an early study on this topic, \citet{broockman2016durably} subsequently conducted an experiment that did indeed provide evidence for one of its key hypotheses.
} 
In some practical instances, a retraction may also be coupled with additional information contradicting the initial subsequently retracted statement. 
Our design decouples these, as these are decoupled in several applications; however, our design also allows us to study updating from retractions with additional information.

We test for retraction effectiveness by comparing beliefs over the truth ball's colour after updating from retractions to (a) beliefs without having observed the retracted observation in the first place and to (b) beliefs updated from new draws with identical Bayes updates (i.e., a new draw of colour opposite the retracted observation). 
These comparisons identify whether participants update less from retractions than from either (a) the retracted observation or (b) a new informationally-equivalent observation. 
We find participants update less from retractions in both comparisons. 
The magnitude of this diminished updating is significant: beliefs update on average about 50\% less from retractions than new draws (see \hyref{section:ineffectiveness}[Section]).

Why are retractions less effective? 
The minimality of our design strongly suggests that any explanation should be intrinsic to how retractions are processed. 
Consistent with this intuition, we consider a general class of \emph{quasi-Bayesian} belief updating models that nests---but also accommodates usual deviations from---Bayesian updating. 
We show that results cannot be reconciled with any explanation that does not treat retractions as inherently different despite identical informational content (see \hyref{proposition:onlyproposition}[Proposition] in \hyref{section:framework:setup}[Section]).
Notably, widely documented deviations, including confirmation bias, cannot rationalise our findings.

Our explanation is that retractions are more complex than direct information. 
Borrowing from \citet{PearlCausality}, we formally articulate a distinctive feature of retractions: Unlike the evidence they typically refer to, which is \emph{directly informative} about the state---in our setting, the colour of a draw---retractions are only \emph{indirectly informative} about the state, that is, they are only informative in light of the retracted evidence.\footnote{
    We say that $x$ is directly informative about $y$ if $x$ and $y$ are neither independent nor independent conditionally on some third variable $z$.
    In our setting, the colour of a draw is directly informative about the state, but learning about its noise status is indirectly informative: only by conditioning on its colour can it be informative, and it is otherwise independent.
} 
Consequently, retractions always require additional contingent reasoning relative to direct evidence.
Indeed, recent literature has shown not only that considering more contingencies renders problems more complex and leads to inference errors in various domains  \citep{AliEtAl2021,EspondaVespa2014,EspondaVespa2021,MartinezMarquinaetal} but also that complexity considerations can explain several well-documented behavioural biases \citep{Oprea2020,Oprea2022Simplicity, BaBohrenImas2022, EnkeGraeberOprea2022}.
This background motivates our hypothesis that the greater complexity inherent to retractions explains diminished updating.

To test this hypothesis, we analyze three empirical complexity measures: (1) accuracy of belief reports, (2) speed of decision, and (3) variability of belief reports. 
All three of these data types are borrowed from past work in which they were used as measures of complexity and cognitive noise---see, for example, 
\citet{CaplinCsabaLeahyNov2020REStud} and \citet{EnkeShubatt2023} for the first; 
\citet{WrightAyton1988MC}, \citet{KrajbichLuCamererRangel2012Frontiers}, and \citet{FrydmanJin2022QJE} for the second; and 
\citet{KhawLiWoodford2021} and \citet{EnkeGraeber2021} for the third.
All proxies are larger when updating from retractions compared to equivalent new information, as well as compared to when the retracted signal had never been seen. 
These patterns suggest higher complexity for retractions, as proposed.

We leverage natural variation provided by our design, which suggests variation in the relative complexity of retractions, and verify that these covary with updating strength. 
First, we compare updating from retractions of more or less recent observations.
If the most recent observation is retracted, participants can simply ``go back'' to a past belief, making updating easier. 
This point suggests that retractions of more recent evidence are less complex, corroborated by our empirical complexity measures.
In line with our mechanism, participants also update more when the most recent observation is retracted than when retractions refer to an earlier observation.
Second, we examine updating from new observations after retractions. 
Beliefs respond less to new observations after retractions, and our empirical measures indicate inference is more complex.\footnote{
    This finding is relevant for situations where (i) some evidence is found inaccurate and (ii) further contradictory evidence is subsequently revealed. 
    The diminished updating from retractions under (i) and the diminished updating \emph{following} retractions in (ii) indicate that both elements contribute to a diminished updating from retractions.
}

We further examine how updating patterns vary across histories. 
We use standard \citet{Grether1980QJE} log-odds regressions to compare biases from retractions to those typically documented in updating from new observations. 
While updating from new evidence exhibits confirmation bias, retractions entail both underinference and anticonfirmation bias.
In line with this, confirmatory retractions are least effective (relative to equivalent new evidence) at histories inducing more extreme beliefs.
These findings offer valuable insights into the unique influence of retractions on belief-updating behaviour.

We conducted a wide range of robustness checks to ensure the validity and generalisability of our results. 
First, we assessed whether our results simply reflect limited participant understanding and inattention despite our screening measures and attention checks.
We consider removing participants who are ``noisy'' or prone to mistakes, as well as those who did not correctly answer unincentivised comprehension questions on the first try.
We can also rule out misinterpreting that the draws are with replacement. 
A theme that emerges is that our results are maintained, if not strengthened, when restricting to participants who appear to have understood the task better.\footnote{
    This finding is perhaps unsurprising since documenting any effect requires that participants act differently for retractions; if participants answered randomly or always answered 50-50, we would not document any difference.
    But it is worth emphasising that most of our sample did very well on unincentivised comprehension questions, confirming our assertion that our design achieved its desired simplicity despite the richness it contains. 
}

Second, we explored variations in participant characteristics.
We find that our results on the diminished updating from retractions and its greater complexity are robust to whether participants 
perform better or worse in quantitative tasks,
are more or less confident about their belief updating, 
are more or less experienced with the task, 
or more or less Bayesian in updating from observations. 
Although we are not powered for a fully-fledged within-participant analysis, inspection of individual heterogeneity in our results indicates that the diminished effectiveness of retractions compared with new observations is a general phenomenon in our sample.

Third, we examined the impact of design variations, such as having shorter histories, omitting the history of past draws, garbling information so that the state is never perfectly learned, and different wording for retractions. 
These variations allowed us to assess whether our main findings were driven by specific features of the information process or details of the experimental. 
We found that our results remained robust across all these different experimental setups. 
Notably, our results are robust even when beliefs are only elicited at the end of each round---dispelling concerns that our findings are driven by information being hard to disregard once it has been ``acted upon,'' as would be suggested by a cognitive dissonance explanation. 
Overall, our comprehensive analysis underscores the robustness and reliability of our findings across various conditions and contexts.

These observations support the claim that our work provides some of the first evidence that diminished retraction effectiveness could have origins (at least partially) in fundamental information processing properties. 
An advantage of showing this in a setting where beliefs can be elicited directly is that it suggests a unified and systematic approach to analyzing patterns in belief updating from retractions. 
Of course, retractions in practice will differ from those we present to participants in our experiment. 
Indeed, we expect many elements deliberately precluded by design, such as memory frictions, salience, or motivated reasoning, to play a significant role in many settings where retractions appear less effective.

Our results are both of practical value and theoretical interest.
We designed the experiment to connect the diminished effectiveness of retractions to information processing errors.\footnote{
    In this sense, our paper is part of a sizable literature that, while motivated by anecdotal or domain-specific evidence of biases, uses fundamental belief updating tasks to highlight a relevant theoretical mechanism; see, for example, \citet{OpreaYuksel2020}, \citet{EOY2022}, \citet{HHI2021}, and \citet{AgranovEtAl2022}.
} 
From a theoretical standpoint, our findings motivate the development of theoretical models of costly information processing that treat indirect information differently from direct information---even when their informational content is the same. 
From a practical standpoint, our analysis provides guidelines regarding how individuals respond to retractions, potentially relevant to campaigns targeting misinformation.
The fact that retraction failures arise due to information processing errors suggests limits to the ``this time is different'' logic policy-makers may adopt---it is generally unreasonable to expect a retraction to be entirely successful in correcting beliefs. 
In many real-world cases, appreciating the inability to correct beliefs with retractions ex post may very well have changed the calculus regarding decisions to disseminate information ex ante.\footnote{
    We do not speak to issues of how these biases interplay with information \emph{preferences}, although this might influence some of these decisions in practice; see \citet{MasatliogluEtAl2021}, \citet{GulNatenzonPesendorfer2021Ecta}, \citet{AmbuehlLi2018}, or \citet{Charnessetal2020} for papers studying this element.
}

\subsection{Past Work on Causes and Consequences of Continued Influence} 
\label{section:introduction:CIE}

The closest precedent for the diminished effect of retractions comes from the literature on the \emph{continued influence effect} in psychology. 
Reviewing this literature, \citet{EckerEtAl2022} define this effect as the finding that ``misinformation can often continue to influence people's thinking even after they receive a correction and accept it as true.'' 
\citet{JohnsonSeifert94} provided an early articulation of such a result, asking participants to recount the cause of the start of a fire and finding that they would still rely upon discredited information.\footnote{
    A more extreme reaction is \emph{backfiring}, in which participants believe more strongly in the retracted information.
    \citet{NyhanReifler2010} documented this pattern when providing participants with information about the presence of weapons of mass destruction in Iraq during the early 2000s (and subsequently providing corrections). 
    But unlike continued influence, backfiring has not been replicated for the most part. 
    See \citet{NyhanSurvey2021} for an authoritative discussion.
} 
\citet{ChanEtAl2017} and \citet{WalterTukachinsky2020} provide meta-analyses of the literature---across experiments that range from stories to advertising, scientific retractions, and beyond---and find that corrections fail to fully correct beliefs.
These and similar patterns have been extensively documented in many settings; \ref{appendix:related-literature} discusses specific applications.

\citet{EckerEtAl2022} and \citet{LewandowskyEtAl2012} discuss a number of channels for continued influence to emerge. 
These include biases related to memory storage (e.g., in terms of ``mental models'' individuals used) and retrieval,\footnote{
    In particular, the mere passage of time may affect the perception of evidence \citep{JacobyEtAl1989}.
} as well as explanations based on the perceived credibility of a retraction and the extent to which it clashes with an individual's worldview.\footnote{
    As illustrated by \citet{SussmannWegener2022}, a possible reason underlying this belief-updating pattern is that it reflects an implied cognitive dissonance \citep{HarmonJones}, owing to the psychological discomfort following from holding two contradicting ideas that retractions induce. 
} 
A confounding factor, however, is that in many existing papers, the ``continued influence effect'' and related `biases' could actually be consistent with Bayesian updating, depending on the implementation of retractions \citep[see][]{RandPaperHowToThink,RandPaperGuide}.

Our implementation of retractions within a balls-and-urns design differs from the existing literature in that we, as analysts, know a retraction's objective informational content. 
This advantage facilitates the identification of differences in information processing \emph{due to information being a retraction}, and our proposed mechanism is intrinsically tied to how retractions generate information. 
Further, while each explanation above is undoubtedly important in some circumstances and less relevant in others, our design allows us to differentiate our proposed mechanism from these setting-specific explanations---issues discussed in more detail in \hyref{section:robustness}[Section].

\subsection{Other Work on Belief Updating Biases}

Our paper builds on the experimental literature studying belief updating. 
\citet{Benjamin2019} provides a comprehensive survey; of independent interest, we replicate many of its key findings.\footnote{
    For recent papers studying these patterns in belief updating, see, for instance, \citet{AmbuehlLi2018}, \citet{Coutts2019b}, and \citet{AugenblickEtAl}.
}

We aim to identify and distinguish the updating from retractions and other well-known biases. 
For instance, we document \emph{base-rate neglect} (whereby agents underweight the prior when updating; see, e.g., \citealt{Espondaetal2020}), as well as \emph{confirmation bias}, discussed above \citep[see also][]{RabinSchrag1999}.\footnote{
    To avoid confounding factors, our design features exogenous information; \citet{Charnessetal2020} study how biases may influence participants' \emph{choice} of sources of information.
} 
We show in our theoretical framework that the diminished effectiveness of retractions is \emph{distinct from these biases} and cannot be explained by models that do not treat retractions inherently differently.

Our analysis suggests that ``indirect information'' is more complex to process than ``direct information.'' 
Though our focus on retracting information is new, the idea that contingent reasoning entails higher cognitive effort has been illustrated in different settings. 
One of the first documented difficulties of contingent reasoning was \citet{CharnessLevin2005} for the winner's curse.\footnote{
    See \citet{EspondaVespa2014} and  \citet{MartinezMarquinaetal} for more on difficulties in contingent reasoning in particular games.
} 
Closer to our study is \citet{Enke2020}, which documents in a pure prediction setting that many participants consistently fail to account for the informational content from the absence of observations, suggesting a failure of contingent reasoning.
One microfoundation driving greater complexity for ``indirect information'' than ``direct information'' is that participants face \emph{higher cognitive imprecision} in their understanding of the informativeness of a retraction than of an observation---see \citet{Woodford2020ARE} for a survey, and \citet{EnkeGraeber2020} and \citet{AugenblickEtAl} for recent applications to belief updating.

\section{Framework and Design}
\label{section:framework}

\subsection{Information Arrival: Draws and Retractions} 
\label{section:framework:setup} 
Our experiments consider a simple belief updating problem. 
Participants form beliefs over a state $\theta$, which takes one of two values with equal probability, say $\theta \in \{\text{\emph{yellow}, \emph{blue}}\}$. 
We write $\hat p_{t}$ to denote a participant's belief that $\theta =\text{\emph{yellow}}$, given all the information observed by period $t$. 
We use the term ``signal'' as a generic term for information throughout. 
Our interest is in two kinds of information participants may have access to: \emph{draws} and \emph{retractions}.

\paragraph{Draws} 
In a given period $t$, participant $i$ may observe $s_{t}\in \{\text{\emph{yellow}, \emph{blue}}\}$, a signal informative about $\theta$ and drawn independently conditional on $\theta$. 
We refer to this kind of information as an ``observation'' or ``draw''.
In our baseline experiment, each observation $s_{t}$ can correspond either to the \emph{truth}, in which case $s_{t}=\theta$, or to \emph{noise}, in which case it is given by an independent $\epsilon_{t} \in \{\text{\emph{yellow}, \emph{blue}}\}$.
Denoting the former event by $\{n_{t} = T\}$ and the latter by $\{n_{t} = N\}$, we focus on cases where these events are independent of $\theta$. 
To summarise, we have $s_{t}=\theta$ if $n_{t} = T$, and $s_{t}=\epsilon_{t}$ if $n_{t} = N$, where $n_{t}\in \{T,N\}$ and $n_{t}$, $\epsilon_{t}$ and $\theta$ are independent. 
For simplicity, we write $S_{t}=\{s_{1},\ldots, s_t\}$. 
In this setup, if $n_{t}=T$, then $s_{t}$ reveals the state. 
In one variant, we additionally allow $s_{t}$ to be \emph{im}perfectly informative even when $n_{t}=T$, but we defer our discussion of this possibility.

\paragraph{Retractions}
The second kind of signal a participant may receive in period $t$ is a \emph{retraction}. Formally: 
\begin{definition} \label{def:retraction}
    A \emph{retraction} of the $\rho$-th observation informs that it was noise, i.e., $n_{\rho}=N$.
\end{definition}
\noindent Retractions provide information about \emph{past signals}. 
The process by which retractions are determined---for example, how observation $\rho$ is chosen to be retracted---matters for how they should be interpreted, a theme we return to later. 
Important for identification in our experimental paradigm, we focus on the following type of retraction:
\begin{definition} 
    A \emph{verifying retraction} of the $\rho$-th observation is a retraction in which $\rho$ (the period that the retraction refers to) is chosen independently of that or other observations' truth value. 
\end{definition} 
\noindent Our experiment implements verifying retractions by selecting $\rho$ uniformly at random from all past observations and subsequently revealing $n_{\rho}$, that is, whether this observation was noise.\footnote{ 
    This implementation implies one learns that past information was not noise when $n_{\rho}=T$, which, in the current setting, perfectly reveals $\theta$ in turn. 
} 
We refer to the signal that informs the participant of $n_{\rho}$ as a \emph{verification}, noting that a verification is a retraction when $n_{\rho}=N$. 
The indicator variable $r$ denotes the occurrence of a retraction, whereby $r_t=1$ if a retraction occurs in period $t$ and $r_t=0$ otherwise.

\subsection{Experimental Design} 
\label{section:framework:design} 
We turn to how we operationalised this information arrival process in our experiments. 
Here we focus on our baseline setup and subsequently discuss how we modified it in our variants.

In each \emph{round} of the experiment, we provided information about a state across up to four \emph{periods}:
\begin{enumerate}[noitemsep]
    \item At the start of the round, a \emph{truth ball} (corresponding to the state $\theta$) is chosen at random to be either yellow or blue, with equal probability. 
    The truth ball is then placed into a box with four \emph{noise balls}, two yellow and two blue (corresponding to $P(n_{t}=N)=4/5$ and $P(\epsilon_{t}=\text{\emph{yellow}})=1/2$).
    \item In periods one and two, participants obtain a \emph{new observation}: a draw from the box with replacement. 
    They see the ball's colour ($s_{t}$) but not whether it is the truth ball or a noise ball ($n_{t}$). 
    \item In periods three and four, and independently across periods, participants either obtain a new observation (as above) or observe a verification of an earlier observation ($\rho$) from the same round, with equal probability. 
    Under a verification, one of the previous draws is chosen uniformly at random, and it is revealed whether that draw was a noise ball ($n_{\rho}=N$)---a \emph{retraction}---or the truth ball ($n_{\rho}=T$). 
    If the draw turns out to have been the truth ball, the round ends, as at that point, the state (the colour of the truth ball) is fully revealed.
\end{enumerate}

\begin{figure}[th!]\setstretch{1.1}
\centering\small
        \includegraphics[width=.45\linewidth]{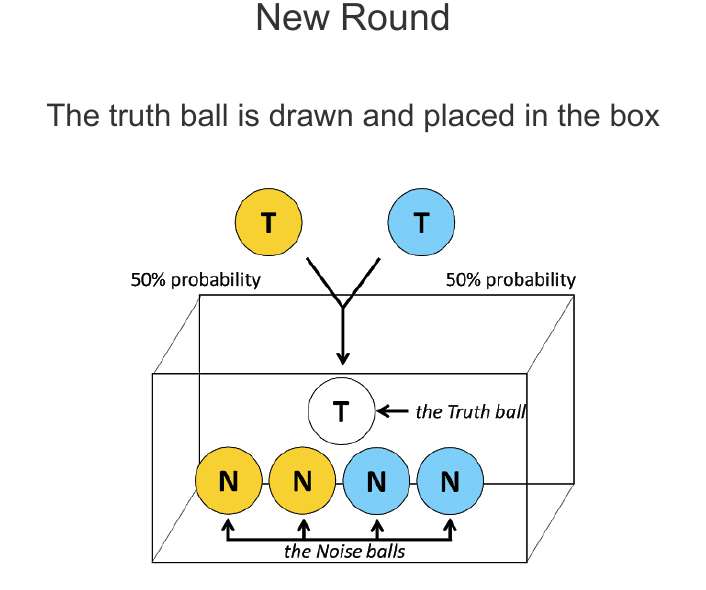}
            \includegraphics[width=.45\linewidth]{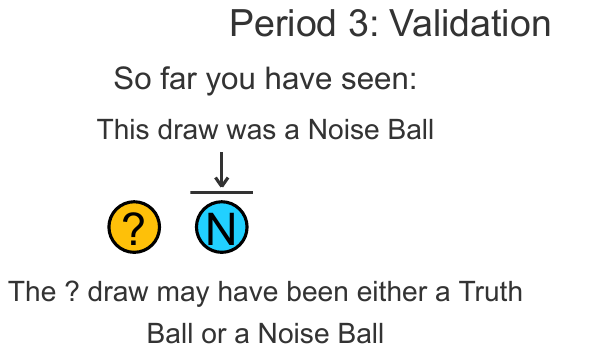}
    \begin{minipage}{1\linewidth}
        \small
        \vspace*{.5em}
        \caption{Screenshots of the Experimental Implementation
        }
         \label{figure:information-structure-screenshots}
        \vspace*{-1.5em}
        \small
        \singlespacing \emph{Notes}:
        This figure provides screenshots of the visuals provided to the participants corresponding to the operationalisation of the information structure. 
     \end{minipage}
\end{figure}

\noindent Participants report their belief regarding the probability that the truth ball is blue or yellow ($\hat p_{t}$) at the end of each period, that is, after each new signal (observation or retraction). 
These reports are incentivised, as detailed in \hyref{section:framework:implementation}[Section].
Each participant plays a total of 32 rounds, and no feedback on performance is provided until performance-based payouts are made at the end of the experiment.

\paragraph{Variants} \hyref{section:ineffectiveness}[Sections], \ref{section:complexity}, and \ref{section:different-biases} present results using the described implementation. 
However, in total we ran four experiments with six main, across-participant treatments (including the baseline). 
\hyref{table:treatmentssummary}[Table] summarises these treatments and details where the paper discusses them. 
These variants aimed to demonstrate the robustness of our findings and to investigate the underlying mechanisms. 
\hyref{table-sample-characteristics}[Table] presents sample characteristics for each treatment. 
We defer detailed descriptions of each variant until \hyref{section:robustness}[Section].  

\begin{table}[th!]
    \begin{tabular}{clccccc}
        \hline\hline
        Experiment & Treatment & Venue & \# Participants & Duration & Payment & Sections  \\  
        \hline
        A & Baseline & MTurk & 211 & 31 min & \$11.96 & Throughout \\
        A & Elicit at End & MTurk & 204 & 24 min & ~~\$8.14 & \ref{section:robustness:design} \\
        B & Garbled Information & MTurk & 164 & 40 min & \$11.03 & \ref{section:robustness:design} \\
        C & Baseline & Prolific & 155 & 49 min & \$11.64 &  Throughout \\
        C & Retraction Information & Prolific & 164 & 52 min & \$11.76 & \ref{section:robustness:understanding} \\
        C & No History & Prolific & 164 & 51 min & \$11.80 & \ref{section:robustness:design} \\
        D & Short Histories & Prolific & 150 & 26 min  & \$12.02 & \ref{section:robustness:design} \\
        \hline\hline
    \end{tabular}
    \begin{minipage}{1\linewidth}
        \small
        \vspace*{.5em}
        \caption{Summary of Treatments}
        \label{table:treatmentssummary}
        \vspace*{-1.5em}
        \singlespacing \emph{Notes}: 
        This table summarises the four experiments, their respective treatments, and the sections of the paper where they are discussed. 
        ``Duration'' and ``Payment'' refer to the average time spent in the experiment in minutes and to the average payment in USD, respectively.
    \end{minipage}
\end{table}

\subsection{Implementation Details} 
\label{section:framework:implementation}

This section discusses implementation details for all experiments discussed in the paper. 

\paragraph{Experimental Interface} 
\hyref{figure:comparisons}[Figure] summarises the explanatory visuals shown to participants in our baseline treatment, and \hyref{online-appendix:instructions} contains the experiment's instructions. 
Participants reported beliefs using a slider, which displayed the stated probability assigned to the truth ball being yellow and the complementary probability assigned to it being blue. 
After the instructions, participants were given two rounds of unincentivised ``practice'' to familiarise themselves with the interface.

\paragraph{Participant Pool} 
We ran four experiments (labelled A--D) comprising different treatments as described in \hyref{table:treatmentssummary}[Table]. 
The first (A and B) were on Amazon Mechanical Turk (MTurk), and the remaining two (C and D) were on Prolific, with different experiments corresponding to different requests for participants.\footnote{
    On each platform, we excluded participants who participated in the earlier experiment on that platform.
} 
The latter two experiments were run to address the mechanisms underlying our results, in response to reviewer feedback. 
To ensure that the choice of venue did not influence our main findings, we ran our baseline treatment on both platforms. 
We recruited 1,212 participants in total; \hyref{appendix:sample-characteristics} presents sample characteristics for all experiments.
The assignment of participants to treatments was randomised within each experiment.
We took several steps to ensure that our participant pool was of high quality; \hyref{section:robustness:understanding}[Section] describes these steps in greater detail.

\paragraph{Payments} 
We incentivised participants to report their beliefs truthfully using a binarised scoring rule \citep{HassanOkui2013,NiederleEtAl2013}. 
By reporting $\hat p_{t} \in [0, 1]$, a participant would receive \$\emph{High} with probability $(1-(\mathbf 1\{\theta=\text{\emph{yellow}}\}-\hat p_{t})^2)$ and \$\emph{Low} with complementary probability, where \$\emph{High} and \$\emph{Low} correspond to \$12.00 and \$6.00 for experiments A and B (ran in 2020 and 2021, respectively) and to \$13.00 and \$7.00 for experiments C and D (ran in 2024). 
To determine payments, we used a report from a single randomly selected period of a randomly selected round.\footnote{
    \citet{AzrieliEtAl2018} show that random selection is essentially the unique problem-selection mechanism inducing incentive compatibility when preferences satisfy state-wise monotonicity, namely that participants prefer higher payments given any realisation of uncertainty (selected problem/underlying states). 
}

In the instructions---but not in the main interface---we provided information on the elicitation procedure, phrased as eliciting the probability that the truth ball was either yellow or blue. 
The instructions explained that the procedure was meant to ensure they were incentivised to answer truthfully. 
As the elicitation scheme we used may appear complicated, we sought to limit the extent to which participants were required to focus on it while maintaining transparency. 
\citet{DanzEtAl2023} show that the binarised scoring rule can introduce noise and ``pull beliefs toward the center'', although the magnitude appears to vary across participant pools and might be lower for online platforms \citep[see][]{HealyKagelWP}.
As our primary focus is on how updating from retractions compares to direct information, any difference is still meaningful. 
Moreover, any potential underreaction would make it \emph{harder} to detect an effect of retractions, not easier. 

We also asked additional questions on mathematical ability, which were incentivised via a \$0.50 reward if a randomly chosen question was answered correctly.
The average duration and compensation were 31 minutes and \$9.98 (\$24.36/hour) for experiments A and B and 45 minutes and \$11.81 (\$20.52/hour) for C and D.
For comparison, this rate is higher than the MTurk experiment of \citet{EnkeGraeber2020} and four times the MTurk average of \$5.00.

\paragraph{Preregistration} 
Experiment A was registered using the AEA RCT Registry under RCT ID AEARCTR-0003820, while Experiment B was registered using the AEA RCT Registry under RCT ID AEARCTR-0006106. 
The experimental design and recruitment targets for these experiments were pre-registered. 
Of our four main hypotheses presented below, Experiment A's preregistration formulated \hyref{hypothesis:retractions-beliefs}[Hypotheses], \ref{hypothesis:harder-retractions} and \ref{hypothesis:after-retractions}, in addition to the analysis in Section \ref{section:different-biases}. 
While our registration discusses difficulty of updating from retractions as a mechanism, our formal hypothesis on complexity, \hyref{hypothesis:retractions-time-variance}[Hypothesis], was introduced subsequently, as feedback we received convinced us they provided evidence for our proposed mechanism. 
Experiments C and D were run to test hypotheses suggested by reviewers at this journal and not preregistered.

\section{Diminished Updating from Retractions}
\label{section:ineffectiveness}

\subsection{Theoretical Predictions}
\label{section:ineffectiveness:theory}
We start by clarifying how our design enables us to identify if and how updating differs between retractions and direct information.
The core of our identification strategy comes from our result that, in our setting, any difference in updating would be inconsistent with any explanation that does not treat retractions differently from direct information---including the general class of frameworks used to explain many known deviations from Bayesian updating. 
In the process, we clarify why seemingly similar paradigms fail to do so and the extent to which continued influence could be consistent with rational belief updating. 

Let $P(\cdot )$ denote \emph{objective} probabilities associated with the data generating process, and $\hat P(\cdot)$ denote $i$'s \emph{subjective} beliefs. 
For a Bayesian decision-maker, subjective beliefs about $\theta$, $\hat p_{t}:= \hat P(\theta \mid \mathcal H_{t})$ coincide with the objective probability that $p_t:= P(\theta \mid \mathcal H_{t})$, where $\mathcal H_{t}$ represents the entire history at period $t$, that is, the set of all the draws observed as well as any retractions, fixing the order. 
Past work has routinely rejected this hypothesis. 
A common alternative is to assume there is a strictly increasing $f_i$ such that $\hat p_{t}=f_i(p_t)$.
It follows that, upon observing some event $E$ at $t$, updating of beliefs $\hat p_{t-1}$ is given by the following identity: 
\begin{equation} 
    \mathcal L(f_i^{-1}(\hat p_{t}))=\mathcal L(f_i^{-1}(\hat p_{t-1}))+K(E), 
    \label{equation:logodds2}
\end{equation}
\noindent where 
$\mathcal L(p):=\ln \left( \frac{p}{1-p} \right)$ denotes the log-odds of $p\in (0,1)$, and $K(E):=\ln\left(\frac{P(E|\theta=\text{\emph{yellow}})}{P(E|\theta=\text{\emph{blue}})}\right)$ the log-likelihood of $E$, with the understanding that $\mathcal H_{t}=\mathcal H_{t-1} \cup E$. 
As long as $\hat p_{t} = f_i(p_{t})$, this relationship holds for all histories $\mathcal H_{t}$; this point will be useful in our analysis. 

Inspired by \citepos{Cripps2019} axiomatic work, we call a decision-maker who updates according to (\ref{equation:logodds2}) ``quasi-Bayesian'':
\begin{definition} \label{def:asifbayesian}
    We say that a decision-maker is \emph{quasi-Bayesian} if there exists a strictly increasing $f_i$ such that, for any information $\mathcal H_{t-1}$ and event $E$, 
    $\hat p_{t}=\hat P(\theta \mid \mathcal H_{t-1},E)$ 
    can be derived from 
    $\hat p_{t-1}=\hat P(\theta \mid \mathcal H_{t-1})$ 
    according to (\ref{equation:logodds2}).
\end{definition}

\noindent Note that, to accommodate some forms of confirmation bias, it may be necessary to allow the function $f_i$ to depend on the prior belief from which participants update; we strive to be as agnostic as possible and our comparisons will hold across a number of possible assumptions. 
We return to a discussion of possible microfoundations for distortions under quasi-Bayesianism in our discussion of mechanisms in \hyref{section:complexity}[Section]. 

Our main comparisons in the paper relate to the following subjective beliefs: 
\begin{enumerate}[label=(\arabic*),noitemsep]
    \item $\hat P(\theta|S_{t},n_\rho=N)$: the participant's belief after observing the retraction $n_{\rho}=N$ in period $t+1$; 
    \item $\hat P(\theta|S_{t}\setminus s_{\rho})$: the participant's belief had the retracted observation $s_{\rho}$ never been observed; and
    \item $\hat P(\theta|S_{t}\cup s_{t+1})$: the participant's belief following a new observation $s_{t+1}$ instead of the retraction.
\end{enumerate}

\begin{proposition} \label{proposition:onlyproposition}
    Suppose retractions are verifying. For any quasi-Bayesian, 
    \begin{enumerate}[label=(\alph*),noitemsep]
        \item their belief after observing the retraction $n_{\rho}=N$ in period $t+1$ is the same as their belief had the retracted observation $s_{\rho}$ never been observed, i.e., $\hat P(\theta|S_{t},n_\rho=N)=\hat P(\theta|S_{t}\setminus s_{\rho})$;
        \item their belief after observing the retraction $n_{\rho}=N$ in period $t+1$ is the same as their belief following a new draw $s_{t+1}$ instead of the retraction, i.e., $\hat P(\theta|S_{t},n_{\rho}=N)=\hat P(\theta|S_{t}\cup s_{t+1})$, if and only if the log-likelihood of the new draw is negative of the retracted observation, $K(s_{t+1})=-K(s_{\rho})$.
    \end{enumerate} 
\end{proposition}
\noindent 
The proof of this proposition essentially follows from an application of Bayes rule and the observation that quasi-Bayesian updating rules still satisfy this identity under the transformation $f_i^{-1}$. 
An identical argument could be used to introduce additional history dependence into the updating rule; our identification strategy below would remain valid. 
More generally, while our framework allows decision-makers to exhibit a plethora of biases, any differences between (1) and (2) or (3) in our experimental setup will require retractions to be treated as intrinsically different.

We focus on verifying retractions to ensure equivalence to signal histories with only new draws and that updating is equivalent to simply never having observed the retracted evidence, and nothing more. 
In particular, the log-likelihood of retracting an observation exactly offsets the log-likelihood of the retracted observation, i.e., $K(n_\rho=N)=-K(s_\rho)$. 
This property contrasts with setups where participants consider restricted information structures, a factor \citet{MillerSanjurjo2019} argue leads to mistakes in probabilistic reasoning, such as those in the Monty Hall Problem.\footnote{
    In the Monty Hall Problem, a participant selects one of three doors, one of which hides a prize.
    After making a choice, an \emph{un}selected door that does \emph{not} hide the prize is opened. 
    The participant can then switch choices. 
    Since only unselected doors \emph{without a prize} are opened, the other unselected door is more likely to hide a prize, making switching optimal. 
    \citet{Friedman1998} finds participants err with striking consistency, often choosing to keep their choices.
} 
Here, the \emph{selection} of a signal is independent of its and other observations' truth value, making our implementation of retractions \emph{unrestricted}. 
In fact, \hyref{proposition:onlyproposition}[Proposition] no longer generally holds if retractions are not verifying and unrestricted.

A provocative implication of this observation is that sometimes ``continued influence'' or related ``biases'' could simply reflect Bayesian updating \citep{RandPaperGuide}. 
If, for instance, only uninformative signals are selected ($\rho =t$ implies $n_{t}=N$), retracting an observation gives more credence to \emph{nonretracted} evidence, which can lead to updating patterns resembling ``continued influence''  \citep{JohnsonSeifert94}---as well as patterns resembling backfiring \citep[discussed in][]{NyhanSurvey2021}.\footnote{
    Related to this point, \citet{RandPaperHowToThink} mentions that studies often obtain different results depending on whether they vary the extent to which participants are shown exclusively fake news versus a mix.
} 
While in many important settings, disclosure is targeted and retractions are restricted, verifying retractions allow direct comparisons and serve as a natural starting point---thus implying that a (quasi-)Bayesian agent would \emph{not} exhibit continued influence.\footnote{
    In ongoing research, we examine a version of this experiment using targeted (i.e., nonverifying) retractions; the results are largely consistent, although direct comparisons between the two are unwarranted. 
    These results are available from the authors upon request.
}

\subsection{Hypothesis and Identification}
\label{section:ineffectiveness:hypothesis}

This paper aims to study updating from retractions and, in particular, compare it to updating from direct information. 
Our first hypothesis concerns our two basic approaches to doing so: 
\begin{hypothesis}[Retractions are Less Effective]
    \label{hypothesis:retractions-beliefs} 
    Participants (a) fail to internalise retractions fully and (b) treat retractions as less informative than an otherwise equivalent piece of new information. 
\end{hypothesis}
\noindent 
We emphasise that our usage of ``retractions'' reflects the meaning in \hyref{def:retraction}[Definition], with ``otherwise equivalent'' reflecting the last case of \hyref{proposition:onlyproposition}[Proposition]. 
In our experimental setting, the log-likelihood of a \emph{blue} draw exactly offsets that of a \emph{yellow} draw ($K(\text{\emph{blue}})=-K(\text{\emph{yellow}})$), so a retraction of a \emph{blue} draw is informationally equivalent to a new \emph{yellow} draw, and vice versa.

We will refer to retractions having \emph{diminished effectiveness} as the finding that belief updates are diminished when generated by retractions, reflecting either part of this hypothesis.
\hyref{proposition:onlyproposition}[Proposition] shows retractions should be as effective as new direct information unless participants treat these two types of information differently.
Therefore, we identify the diminished effectiveness of retractions as a phenomenon distinct from belief-updating biases that are not intrinsically related to retractions. 

\begin{figure}[th!]\setstretch{1.1}
\centering\small
    \begin{subfigure}{1\textwidth} \centering
        \begin{minipage}{.4\linewidth}
            \includegraphics[width=1\textwidth]{Figures/Box/YBN.pdf}
        \end{minipage}
        \begin{minipage}{.1\linewidth}
            \Large vs.
        \end{minipage}
         \begin{minipage}{.4\linewidth}
            \includegraphics[width=1\textwidth]{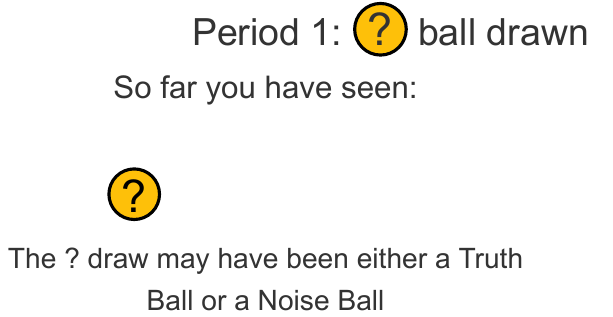}
        \end{minipage}
        \vspace{2mm}
         \caption{\small 
         {Retraction vs. No Retracted Draw} 
         \vspace{3mm}}
      \end{subfigure}
      \begin{subfigure}{1\textwidth} \centering
        \begin{minipage}{.4\linewidth}
            \includegraphics[width=1\textwidth]{Figures/Box/YBN.pdf}
        \end{minipage}
        \begin{minipage}{.1\linewidth}
            \Large vs.
        \end{minipage}
         \begin{minipage}{.4\linewidth}
            \includegraphics[width=1\textwidth]{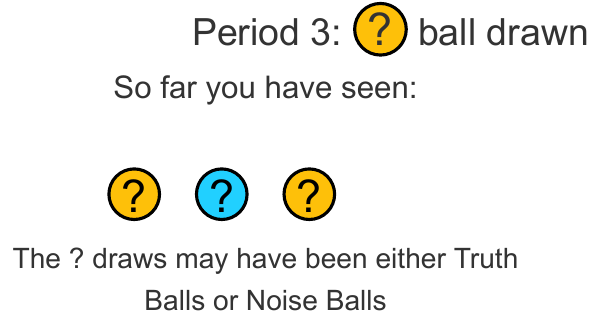}
        \end{minipage}
        \vspace{2mm}        
        \caption{\small 
        {Retraction vs. Equivalent New Draw} 
        } 
     \end{subfigure}
    \begin{minipage}{1\linewidth}
        \small
        \vspace*{.5em}
        \caption{Illustrative Examples to Explain the Empirical Strategy
        }
         \label{figure:comparisons}
        \vspace*{-1.5em}
        \small
        \singlespacing \emph{Notes}:
        This figure provides an illustrative example of the empirical strategy for testing \hyref{hypothesis:retractions-beliefs}[Hypothesis]. 
        According to \hyref{proposition:onlyproposition}[Proposition], for any quasi-Bayesian, beliefs should be identical for each of the displayed histories. 
        Panel (a), comparing beliefs following {(yellow, blue, \emph{retraction of the} blue)} to those following {(yellow)}, tests if beliefs are the same when evidence gets retracted and when such evidence was never observed. 
        Panel (b), comparing beliefs following {(yellow, blue, \emph{retraction of the} blue)} to those following {(yellow, blue, yellow)}, tests whether updating from retractions is the same as from otherwise equivalent direct information.
     \end{minipage}
\end{figure}

In our context, parts (a) and (b) of \hyref{hypothesis:retractions-beliefs}[Hypothesis] correspond to the following comparisons, which we will make repeatedly in the paper, explained visually in \hyref{figure:comparisons}[Figure]: 
\begin{enumerate}[label=(\alph*),noitemsep]
    \item \emph{Comparing beliefs with retractions and without the retracted observation}: Are participants' beliefs after seeing a retraction the same as if the retracted observation had never been observed in the first place? 
    
    \item \emph{Comparing beliefs with retractions and with equivalent new observation}: Are participants' beliefs following a retraction of a \emph{yellow} signal the same as when observing a new \emph{blue} draw?
\end{enumerate}

While (a) and (b) can both be used to assess whether retractions are less effective, and although one conclusion may be \emph{suggestive} of the other, they are ultimately distinct. 
In principle, both new observations and retractions could be treated as equivalent and less informative than an earlier observation, leading to (a) without (b)---diminished updating from retractions could be driven by a feature of belief updating common to both retractions and new information. 
Conversely, new observations and retractions could be treated differently, but with retracted evidence treated as if it had never been seen, and with over-reaction to new observations driven by some other channel---leading to (b) without (a).

\subsection{Estimation Strategy} 
\label{section:ineffective:empirical-strategy}

We start by noting that belief updates in log-odds should be $\pm K(\text{\emph{yellow}})$, no matter the signal (a draw or a retraction) and no matter the prior (moderate or extreme).\footnote{
    In contrast, the change in levels is lower the farther away from 1/2 the prior belief is.
} 
This is because a Bayesian would have constant log-odds updates for any prior. 
Therefore, since using log-odds beliefs allows us to more easily compare and interpret our results, and in keeping with standard practice in the literature on belief updating \citep[as in][]{Benjamin2019}, we will specify all our regressions using log-odds of beliefs, defined as $\hat \ell_{t} := \mathcal L(\hat p_{t})$.
Our conclusions are, however, robust to relying either on log-odds or level beliefs, as shown below.

The key element of our estimation strategy relies on precisely defining fixed effects based on the comparisons described in \hyref{proposition:onlyproposition}[Proposition] (and illustrated in \hyref{figure:comparisons}[Figure]) to identify diminished effectiveness of retractions.
For this, we will pair histories $\mathcal H_{t}$ with and without retractions. 
Recall that $\mathcal H_{t}$ denotes the history up to and including period $t$: the set of all the draws observed and any retractions, fixing the order. 
Except for \hyref{section:complexity}[Section], where we explicitly consider updating \textit{after} retractions, we do not include histories in which there was previously a retraction or where the truth ball was revealed so as to avoid any confounding factors.

For \emph{comparing beliefs with retractions and without the retracted evidence}, test (a), we define the \emph{compressed history}, $C(\mathcal H_{t})$: the history with the retracted observations removed, as if they had never occurred to begin with.
Taking as an example the top panel of \hyref{figure:comparisons}[Figure], the compressed history of \emph{(\emph{yellow}, \emph{blue}, retraction of the \emph{blue})} is simply \emph{(\emph{yellow})}.\footnote{
    Note that compressed histories do not distinguish between the retracted observation having been drawn in period 1 or period 2.
    For example, both \emph{(\emph{yellow}, \emph{blue}, retraction of the \emph{blue})} and \emph{(\emph{blue}, \emph{yellow}, retraction of the \emph{blue})} have the same compressed history,  \emph{(\emph{yellow})}.
} According to \hyref{hypothesis:retractions-beliefs}[Hypothesis][a]---based on \hyref{proposition:onlyproposition}[Proposition][(a)]---histories sharing a common compressed history should also share common beliefs and, therefore, the same log-odds beliefs.

We then test \hyref{hypothesis:retractions-beliefs}[Hypothesis][a] with the following regression, 
\begin{equation} 
    \hat \ell_{i,t} = \beta_0 \cdot r_{i,t} + \beta_1 \cdot r_{i,t} \cdot K(s_{i,\rho_{i,t}}) + \gamma_{C(\mathcal H_{i,t})}+\varepsilon_{i,t},\label{equation:basic-regression}
\end{equation} 
where $i$ denotes the participant, $r_{i,t}$ denotes a dummy variable indicating in period $t$ there is a retraction ($r_{i,t}=1$) or a new observation ($r_{i,t}=0$), $s_{\rho_{i,t}}$ denotes the colour of the retracted observation, $\gamma_{C(\mathcal H_{i,t})}$ are fixed effects for compressed history, and $\varepsilon_{i,t}$ is a noise term. Note that we do not include $K(s_{i,\rho_{i,t}})$ as a term in this regression, since this term is the same for every observation with the same compressed history.

The coefficient of interest is $\beta_1$.
In the context of the illustrative example, our compressed-history fixed effects allow us to take differences in beliefs across histories that induce the same compressed history, \emph{(yellow)}, such as \emph{(yellow)} and \emph{(yellow, blue, retraction of blue)}.
As $K(\text{\emph{blue}})<0$, retracting \emph{blue} should increase the belief that $\theta=$\emph{yellow}, and so $\beta_1$ captures how much less beliefs update from a retraction compared to how much they update from the retracted observation when it was first observed.
\hyref{hypothesis:retractions-beliefs}[Hypothesis][a] corresponds to $\beta_1>0$.\footnote{
    The scaling by $K(s_{\rho_t})$ will prove useful when discussing how much participants infer from observations in the same log-likelihood scale to enable a comparison to Bayesian updating.
    Note that, upon observing $s_t$, Bayesian updating implies that $\ell_t = \ell_{t-1}+K(s_t)$, where $\ell_t := \mathcal L(p_t)$.
}

For \textit{comparing retractions to new evidence}, test (b), we define \emph{sign history}, $S(\mathcal H_t)$, which is the history without distinguishing whether signals were new observations or retractions. 
For example, as illustrated in the bottom panel of \hyref{figure:comparisons}[Figure], \emph{(yellow, blue, retraction of blue)} and  \emph{(yellow, blue, yellow)} both have the same sign history. 
We then run the same regression as before, \hyref{equation:basic-regression}[Equation], except with sign-history fixed effects, $\gamma_{S(\mathcal H_t)}$, instead of compressed-history fixed effects, $\gamma_{S(\mathcal H_t)}$. 
$\beta_1$ again is the coefficient of interest, measuring how much less beliefs update from retractions than from (informationally) equivalent new observations.

\subsection{Updating from New Observations} 
\label{section:ineffectiveness:new-observations}
As a first step in our analysis, and in part as a test of the validity of our experimental setting, we examined participants' belief updating from (nonretracted) new observations using a standard empirical approach in this literature. 
Here, we simply note that our findings are consistent with existing literature---we present the results more in-depth in \hyref{section:different-biases}[Section], where we investigate how retractions affect belief-updating patterns.

In the absence of a retraction, the design is similar to many others surveyed by \citet{Benjamin2019}. 
Participants appear to correctly understand the setting, with reported beliefs tracking Bayesian posteriors closely.\footnote{
    \hyref{online-appendix:beliefs-by-history}[Online Appendix] presents beliefs and Bayesian posteriors disaggregated by history; in \hyref{online-appendix:comparison-to-bayes}, we report the difference and the distance between beliefs and Bayesian posteriors.
}
We consider Grether-style \citep{Grether1980QJE} regressions---a workhorse model of analysis in this literature---enabling a direct comparison to existing experimental results on belief updating. 
Specifically, we replicate common patterns in belief updating, such as base-rate neglect and confirmation bias. 
While participants depart from Bayesian updating, our theoretical framework implies that any additional departure due to retractions cannot be attributed to explanations that are not specific to the nature of the information source.
We first focus on how belief updating from retractions differs from updating from new observations, deferring the detailed reporting and discussion of general departures from Bayesian updating to \hyref{section:different-biases}[Section].

\subsection{Updating from Retractions}
\label{section:ineffectiveness:maincomparison}
We now present our first central finding: empirical support of \hyref{hypothesis:retractions-beliefs}[Hypothesis]. 
We estimate the differences in beliefs specific to retractions using Equation (\ref{equation:basic-regression}) on our baseline treatments.

We find a diminished effectiveness of retractions: participants update beliefs less from retractions than from both the retracted observation (Retraction vs. No Retracted Draw) and an equivalent new observation (Retraction vs. Equivalent New Draw). 
\hyref{table-retractions-beliefs}[Table] presents our estimates for our baseline treatments.
Belief updates are significantly lower for retractions than new information: by 0.586 for retractions compared to belief updates had the retracted evidence never been observed and by 0.603 compared to equivalent new draws.

\begin{table}[t]\setstretch{1.1}
	\centering\small
	\begin{tabular}{l@{\extracolsep{4pt}}cc@{}}
\hline\hline
\multicolumn{1}{r}{Retraction vs.} &\multicolumn{1}{c}{No Retracted Draw}  &  \multicolumn{1}{c}{Equivalent New Draw} \\
\cline{2-2} \cline{3-3}
& (1) & (2) \\
& $\hat \ell_t$ & $\hat \ell_t$ \\
\hline
Retraction ($r_t$) & 0.011 & -0.019 \\
 & (0.018) & (0.025) \\ [.1em]
Retracted Draw ($r_t\cdot K(s_{\rho_t})$) & 0.586$^{***}$ & 0.603$^{***}$ \\
 & (0.067) & (0.087) \\ [.1em]

Compressed History FEs & Yes & No  \\
Sign History FEs       & No  & Yes \\
\hline
R$^2$ & 0.26 & 0.27 \\
N & 39162 & 39162 \\
\hline\hline
\multicolumn{3}{l}{\footnotesize Clustered standard errors at the subject level in parentheses.}\\
\multicolumn{3}{l}{\footnotesize $^{*}$ \(p<0.1\), $^{**}$ \(p<0.05\), $^{***}$ \(p<0.01\)}\\
\end{tabular}
    \begin{minipage}{1\linewidth}
        \small
        \vspace*{.5em}
        \caption{Updating from Retractions (\hyref{hypothesis:retractions-beliefs}[Hypothesis])}
        \vspace*{-1.5em}
        \label{table-retractions-beliefs}
        \singlespacing \emph{Notes}: 
        Column (1) tests \hyref{hypothesis:retractions-beliefs}[Hypothesis][a] by estimating \hyref{equation:basic-regression}[Equation]. 
        Column (2) tests \hyref{hypothesis:retractions-beliefs}[Hypothesis][b] by estimating a variant of \hyref{equation:basic-regression}[Equation], in which compressed-history fixed effects are replaced with sign-history fixed effects. 
        The sample includes all observations of participants in the baseline treatment, excluding periods in which the truth ball is disclosed or in which there was a retraction in an earlier period.
    \end{minipage}
\end{table}

\begin{figure}[t]\setstretch{1.1}
    \centering\small
    \begin{subfigure}{.495\textwidth} 
        \centering
        \includegraphics[width=1\linewidth]{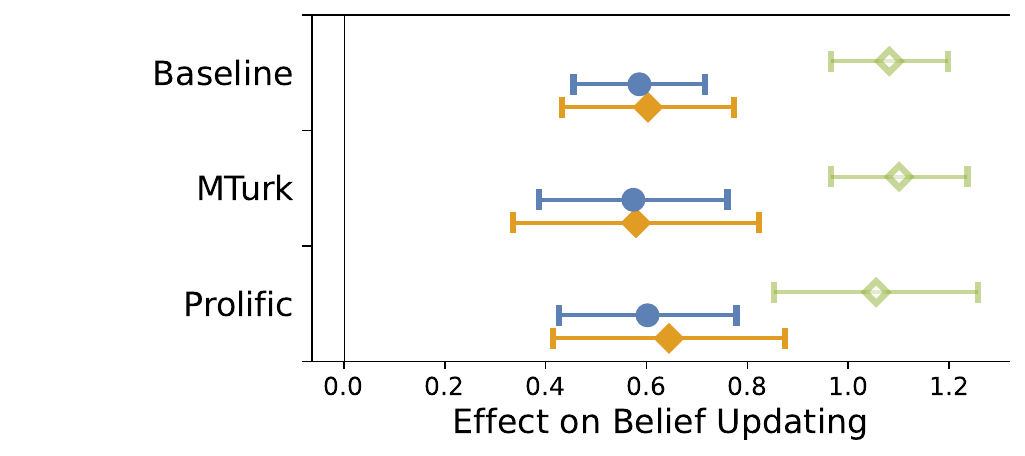}
        \vspace*{-2em}
        \caption{Log-Odds}
    \end{subfigure}
    \begin{subfigure}{.495\textwidth}
        \centering
        \includegraphics[width=1\linewidth]{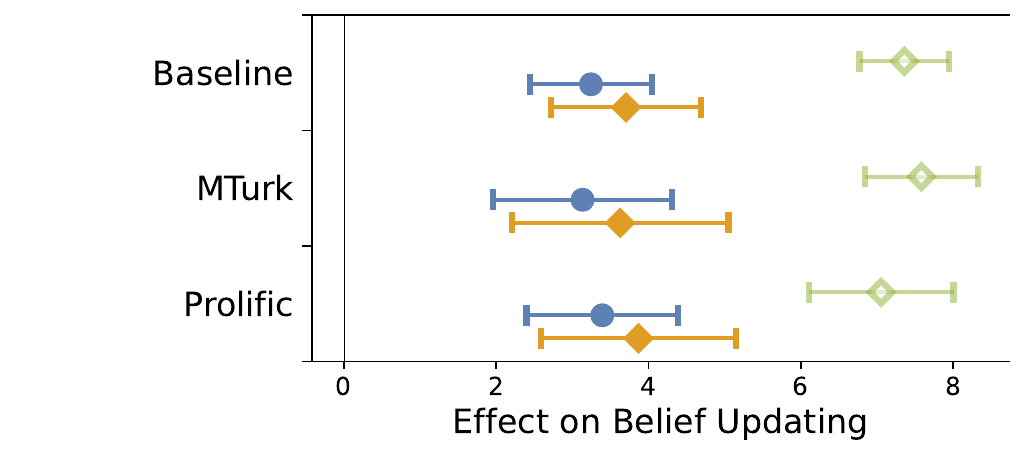}
        \vspace*{-2em}
        \caption{Levels}
    \end{subfigure}
    \begin{subfigure}{1\textwidth} 
        \centering
        \includegraphics[width=.8\linewidth]{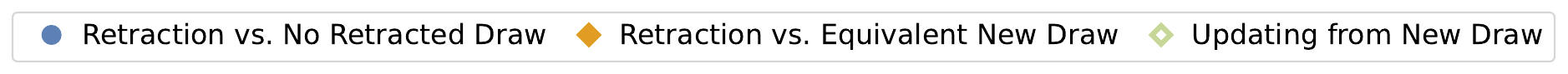}
    \end{subfigure}
    \begin{minipage}{1\linewidth}
        \small
        \caption{Retractions Are Less Effective  (\hyref{hypothesis:retractions-beliefs}[Hypothesis])}
        \label{figure-ate-retractions-beliefs}
        \vspace*{-1.5em}
        \small
        \singlespacing \emph{Notes}: 
        This figure depicts the effects of retractions on belief updating, showing how much less participants update beliefs from retractions than from the retracted evidence (Retractions vs. No Retracted Draw; blue solid circle) and from new direct evidence (Retractions vs. Equivalent New Draw; orange solid diamond).
        The green hollow diamond depicts how much beliefs update on average from new draws, for comparison. 
        Panel (a) shows these estimates for beliefs in log-odds ($\hat \ell_t$) as per \hyref{equation:basic-regression}[Equation], while panel (b) provides the analogous estimates for beliefs in levels ($\hat p_t$), measured in percentage points (0-100\%). 
        The sample includes all observations of participants in the baseline treatment, excluding periods in which the truth ball is disclosed or in which there was a retraction in an earlier period. 
        The figure displays results both pooled (Baseline) and separated by recruitment platform. Plot whiskers represent 95\% confidence intervals.
     \end{minipage}
\end{figure}

In order to contextualise this number, we compare it to the estimate of how much beliefs update following a new observation. 
This estimate is given by the coefficient $\beta_1$ from the regression specification  $\Delta\hat \ell_{i,t} = \beta_0  + \beta_1 \cdot K(s_{i,t}) + \gamma_{S(\mathcal H_{i,t})}+\varepsilon_{i,t}$, where $\Delta \hat \ell_{i,t}=\hat \ell_{i,t}-\hat \ell_{i,t-1}$, restricted to histories $\mathcal H_{i,t}$ consisting only of new draws.\footnote{
    Note that when we restrict to histories without retractions, compressed and sign histories are the same: $C(\mathcal H_t)=S(\mathcal H_t)$; hence, this normalisation is appropriate for both comparisons.
}
Since the left-hand side is $\Delta\hat \ell_{i,t}$, Bayesian updating corresponds to $\beta_{0}=0$ an $\beta_{1}=1$. We find that a new draw moves beliefs by 1.081 times the log-likelihood of a new draw, providing a rough estimate of how much \emph{less} beliefs update from retractions relative to new draws: .603/1.081, approximately 55\%. 
Panel (a) of \hyref{figure-ate-retractions-beliefs}[Figure] provides a visualisation of these estimates. 
Panel (b) provides analogous estimates using levels ($\hat p_t$), instead of log-odds, and exhibits consistent results.
Specifically, we find that following retractions (i) beliefs update insufficiently and remain on average 3.2 percentage points away from the beliefs held absent the retracted evidence, and (ii) participants update beliefs on average 3.7 percentage points less than from new draws---about 50\% of the average belief updates from observations of 7.4 percentage points.

We conclude that participants infer substantially less from retractions than direct evidence. 
Furthermore, this difference does not depend on whether test (a) or test (b) is considered. 

\begin{figure}[t]\setstretch{1.1}
    \centering\small
    \begin{subfigure}{.495\textwidth} 
        \centering
        \includegraphics[width=1\linewidth]{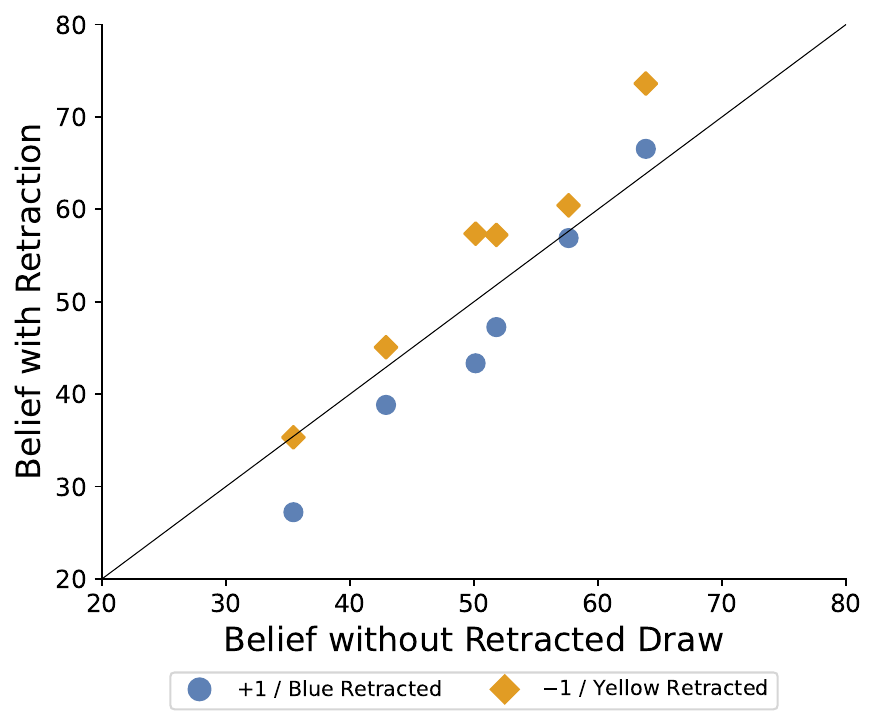}
        \caption{Retractions vs. No Retracted Draw}
    \end{subfigure}
    \begin{subfigure}{.495\textwidth}
        \centering
        \includegraphics[width=1\linewidth]{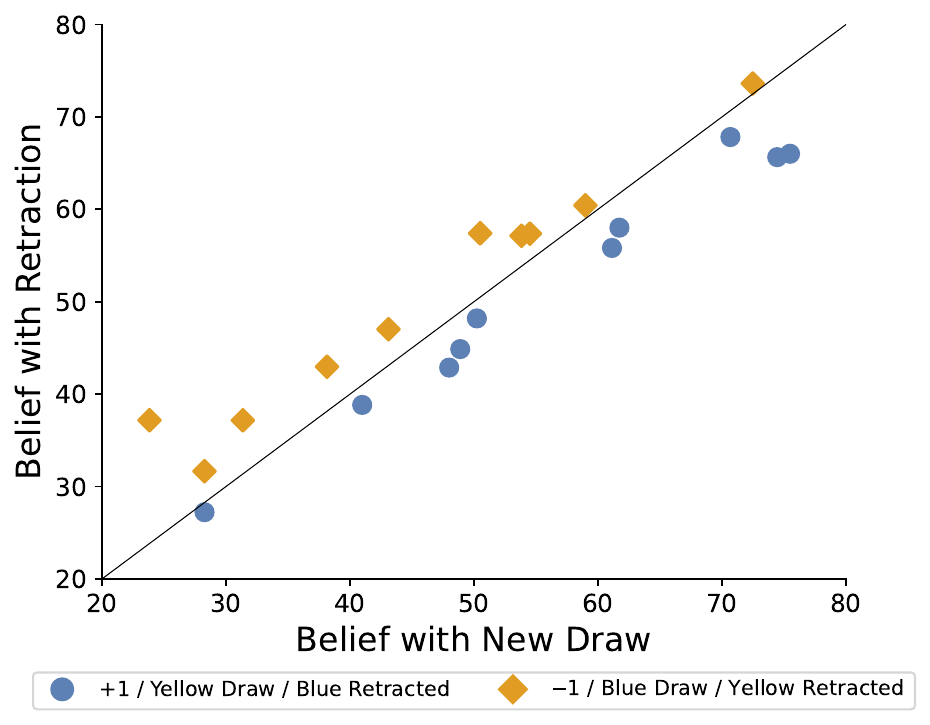}
        \caption{Retractions vs. Equivalent New Draw}
    \end{subfigure}
    \begin{minipage}{1\linewidth}
        \small
        \vspace*{.5em}
        \caption{Retractions Are Less Effective (\hyref{hypothesis:retractions-beliefs}[Hypothesis])}
        \label{figure-beliefs-byhist}
        \vspace*{-1.5em}
        \small
        \singlespacing \emph{Notes}: 
        This figure exhibits the effect of retractions on belief updating across the fixed effects used in our baseline specifications, reported in \hyref{table-retractions-beliefs}[Table]. 
        Each marker in panel (a) represents average beliefs with a retraction ($y$-axis) and without the retracted draw ($x$-axis) for a specific compressed history. 
        Analogously, each marker in panel (b) represents average beliefs with a retraction ($y$-axis) and with an equivalent new draw ($x$-axis) for a specific sign history. 
        Blue dots correspond to cases in which a blue draw is retracted and orange diamonds to those in which the retraction refers to a yellow draw. 
        Retractions being less effective corresponds to blue dots being below the 45-degree line and orange diamonds above. 
        The sample includes all observations of participants in the baseline treatments, excluding periods in which the truth ball is disclosed or in which there was a retraction in an earlier period.
     \end{minipage}
\end{figure}

These findings represent average estimates, and a natural question is the extent to which there is heterogeneity in the effects across histories.  
Throughout, we will discuss different meaningful dimensions of heterogeneity, namely with respect to how recent retracted observations are and the number of draws observed (\hyref{section:complexity:variation}[Section]), as well as if the retraction is confirmatory (reinforces the prior belief) or not (\hyref{section:different-biases}[Section]).
While we lack statistical power at the most disaggregated level, 
\hyref{figure-beliefs-byhist}[Figure] provides indicative evidence that our results are robust across histories, 
and we also report results fully disaggregated by history, with consistent conclusions across histories (see \hyref{online-appendix:table-hte-by-history-beliefs}[Online Appendix]).

We note that we collected data for our baseline design twice, on Amazon Mechanical Turk in 2020 and again on Prolific in 2024. 
We obtained remarkably similar estimates of the effect across both platforms, as seen in \hyref{figure-ate-retractions-beliefs}[Figure].
In \hyref{appendix:mturk-vs-prolific}, we show there are no significant differences between the two recruitment platforms across all our main specifications.
We discuss the robustness of our results further in \hyref{section:robustness}[Section], only mentioning for now that restricting to particular rounds or to participants that appear to perform better does not affect our conclusions.

\section{Informational Complexity and Diminished Updating from Retractions}
\label{section:complexity}

Having documented differences in beliefs in updating from retractions, we now turn to a discussion of mechanisms. 
We divide our analysis of possible mechanisms into two parts. In this section, we propose and analyze the hypothesis that retractions are less effective because they entail greater informational complexity.
We defer to the following sections the discussion of alternative explanations that could plausibly generate our results---and show that they do not.

\subsection{Retractions Provide More Complex Information} 
\label{section:complexity:retractions}
While the informational content (as captured by the log-likelihood) of a retraction is the same as that of a new observation, we argue that properties inherent to retractions render it more complex and lead to the observed diminished belief updates. 

One such property refers to the kind of information retractions provide. 
In contrast to observations that provide direct evidence about the state (e.g., statements, trials, data), retractions provide only indirect information. 
To see this, note that retractions' meaning is obtained by informing about the quality or properties of direct evidence and are hence ``one step removed'' from the state relative to observations. 
Inference from retractions, therefore, necessitates an additional layer of contingent reasoning compared to observations, which renders them more complex.
Indeed, there is abundant evidence that contingent reasoning renders problems more complex and explains deviations from optimality. 
These include failure to incorporate pivotality considerations in voting \citep{EspondaVespa2021}, neglecting correlation in information sources \citep{EnkeZimmermann2019}, or in common value auctions \citep{Eyster2019}. 
Even in very simple environments, an added layer of contingent reasoning entails a significantly greater propensity for suboptimal choices \citep{MartinezMarquinaetal}.

\afterpage{
\begin{figure}[t]
    \centering
    \begin{tikzpicture}
        \node[circle, draw=black] (s) at (0,0) {$s_t$};
        \node[circle, draw=black] (theta) at (-1,1.25) {$\theta$};
        \node[circle, draw=black] (epsilon) at (0,1.5) {$\epsilon_t$};
        \node[circle, draw=black] (n) at (1,1.25) {$n_t$};
        \draw[-{Stealth[length=.5em, width=.65em]}] (theta) -> (s);
        \draw[-{Stealth[length=.5em, width=.65em]}] (epsilon) -> (s);
        \draw[-{Stealth[length=.5em, width=.65em]}] (n) -> (s);
    \end{tikzpicture}

    \begin{minipage}{1\linewidth}
        \small
        \vspace*{1em}
        \caption{Graphical Model Representation of a New Draw}
        \label{figure:dag}
        \vspace*{-1.5em}
        \small
     \end{minipage}
\end{figure}
}

In our setup, this additional layer of contingent reasoning can be precisely seen using a simple causal model as given by a directed acyclical graph \citep{PearlCausality}.
\hyref{figure:dag}[Figure] represents how $\theta, s_t, \epsilon_{t}$, and $n_{t}$ are related, whereby an arrow from variable $x$ to variable $y$ means that $x$ determines (in part) the value of $y$.
We say that an observation $s_t$ provides \emph{direct information} about the state $\theta$, since $s_{t}$ is directly connected to $\theta$, with $\theta$ directly influencing the distribution over the observation's realisation. 
However, information obtained from a retraction---disclosing $n_t$---is only indirectly informative about $\theta$, as $\theta$ and $n_t$ are independent. 
Dependence emerges only through conditioning on $s_{t}$: information that an observation is or is not noise ($n_t$) is only informative about $\theta$ \emph{contingent on} $s_{t}$. 
\citet{PearlCausality} refers to such connections as \emph{indirect}. 
\citet{PearlWhy} argue that this phenomenon---that is, that independent variables can become correlated conditional on another variable---is responsible for several apparent logical paradoxes.\footnote{
    For instance, the Monty Hall problem is central among the paradoxes described by \citet{PearlWhy}, connecting this observation to our discussion of \citet{MillerSanjurjo2019} from \hyref{section:ineffectiveness:theory}[Section]. 
    Other related phenomena are the observed difficulty people have in thinking through problems involving higher-order reasoning, 
    expressed in aversion to compound lotteries \citep{AbdellaouiKlibanoffPlacido2015MnSc,DeanOrtoleva2019PNAS} 
    and in mistaken higher-order beliefs in strategic settings \citep{CrawfordCosta-GomesIriberri2013JEL,Kneeland2015Ecta,AlaouiPenta2016REStud,AlaouiJanezicPenta2020JET}.
}

In our subsequent analyses, we turn to measuring complexity and identifying its prominent association with updating from retractions.

\subsection{Tracing Retraction Complexity}
\label{section:complexity:indicators}
We now turn to our empirical measures of complexity. 
A common microfoundation for deviations from Bayesian updating is the hypothesis that the agent faces cognitive imprecision, as posited by models of cognitive uncertainty, efficient coding, and sequential sampling.\footnote{\label{footnote:seq-sampling}
    While distinct, the literatures are closely related.
    Efficient coding \citep{WeiStocker2015NN} and cognitive uncertainty models have been increasingly popular in economics; e.g., \citet{KhawLiWoodford2021}, \citet{FrydmanJin2022QJE}, \citet{EnkeGraeber2020}, and \citet{AugenblickEtAl}.
    Models of sequential sampling provide a relationship between cognitive uncertainty and time through evidence accumulation \citep{Krajbichetal,BhuiGershman2018PsyRev}.
    See \citet{RatcliffSmithBrownMcKoon2016Trends} for a survey of sequential sampling models in psychology and neuroscience, and \citet{FudenbergStrackStrzalecki2018AER}, \citet{Alos-FerrerFehrNetzer2021JPE}, and \citet{Goncalves2022WP} for recent applications in economics.
}  
Our hypothesis is that this cognitive imprecision is higher for retractions. 
We provide evidence for this using two broad strategies. 
First, we consider different empirical measures of complexity borrowed from the literature and show that these generally are larger for retractions.
Second, we consider treatments of and variation in our baseline design where retraction complexity would appear to increase, showing that this correspondingly strengthens the effect.

Before presenting our evidence for such a mechanism, we briefly sketch a model in the spirit of this literature, which ties complexity to empirical measures that we can infer from the data. 
Suppose decision-maker $i$ faces uncertainty about how to interpret the likelihood of evidence $E$ and update beliefs. 
In particular, for tractability, we assume the decision-maker's prior about $\theta$ is Gaussian, with $K(E)\sim \mathcal N(0,\sigma^2)$, and that they obtain $T$ noisy estimates $K(s_{t})+\sigma_\zeta\cdot\zeta$, where $\zeta \sim \mathcal N(0,1)$ denotes (Gaussian) noise.
Using the Bayesian updating formulas for normal distributions, this yields posterior log-odds updates as
\[
    \hat \ell_{t}
    = 
    \hat \ell_{t-1} 
    + \beta K(E)+\beta\frac{\sigma_\zeta}{\sqrt{T}} \zeta,
\]
with $\beta=(1+\sigma_\zeta^2/(\sigma^2 T))^{-1}$. 
\hyref{section:ineffectiveness}[Section] shows that $\beta$ is lower for retractions. 
The hypothesis that retractions increase complexity is reflected in an increase of $\sigma_{\zeta}$.

We test falsifiable predictions from this setup that could explain our results. 
For that, we use three behavioural markers of complexity: (1) \emph{accuracy}, i.e., how close belief reports are to Bayesian posteriors; (2) \emph{speed}, i.e., decision times; and (3) \emph{variability} in belief reports.

\paragraph{Accuracy} 
Our first indicator measures the distance between belief reports and the Bayes posterior.
This variable captures accuracy since, based on our incentivisation, the optimal report given the provided information coincides with the Bayesian posterior, and the expected payoff is decreasing in the absolute error of beliefs, that is, the distance between the belief reported and the Bayes posterior, $|\hat p_t- p_t|$.

\paragraph{Speed} 
Our second indicator captures how much effort individuals exert.
A standard approach in the literature associates $T$ with decision time, the idea being that the decision-maker obtains one such signal per unit of time spent deliberating (see footnote \ref{footnote:seq-sampling}). 
In line with the general finding that decision-makers take more time and do less well on {simple} tasks when these tasks become less immediately apparent, we will interpret longer decision times, together with lower accuracy, as suggestive evidence that complexity is higher in the updating problem.\footnote{
    Early evidence for this observation can be found in, for instance, \citet{BanksEtAl}, \citet{BuckleyGillman}, or \citet{Ratcliff78}; see \citet{Goncalves2024WP} for a formal treatment.
} 

\paragraph{Variability}
Our third measure is the variability in the belief reports; following \citet{KhawLiWoodford2021} and \citet{EnkeGraeber2021}, we adopt it as an indicator of the underlying complexity. 
The underlying intuition is that greater cognitive imprecision generates less precise choices.
In our model, given the above, an increase in $\sigma_\zeta^2$ increases the variance of log-odds posterior beliefs insofar as the posterior variance about $K(E)$ is at most half of the prior variance about $K(E)$, i.e., $\sigma^2$---see \hyref{online-appendix:variance-proof}[Online Appendix].

\subsection{Retraction Complexity}
\label{section:complexity:results}
The preceding discussion motivates the following hypothesis, which we proceed to analyze: 
\begin{hypothesis}[Retractions Are More Complex]
    \label{hypothesis:retractions-time-variance} 
    Inference from retractions is more difficult than processing new observations, resulting in (a) lower belief accuracy, (b) longer decision time, and (c) higher belief variance. 
\end{hypothesis}

To test this hypothesis, we use an identification strategy similar to the one used to test the effects of retractions on belief updating (\hyref{section:ineffective:empirical-strategy}[Section]).
Specifically, in \hyref{table-ate-complexity}[Table], we estimate versions of the following:
\begin{align}
    \label{equation:retractions-time}
    y_{i,t} &= \beta_1 \cdot r_{i,t} + \gamma_{i,t} + \varepsilon_{i,t},
\end{align}
where $y_{i,t}$ is a dependent variable and $\gamma_{i,t}$ are the relevant fixed effects, as in \hyref{section:ineffective:empirical-strategy}[Section] and under the same sample restrictions. 

Specifically, to test if belief accuracy is lower and decision times longer when participants face a retraction, the dependent variable $y_{i,t}$ corresponds to participant $i$'s absolute error in beliefs ($|\hat p_{i,t}-p_{i,t}|$), and to log decision time ($\ln(T_{i,t})$), respectively.
We perform both comparisons outlined in \hyref{hypothesis:retractions-beliefs}[Hypothesis]: (a) retractions compared to histories where the draw was never observed, using compressed history fixed effects ($\gamma_{i,t}=\gamma_{C(\mathcal H_{i,t})}$), and (b) retractions versus an equivalent new draw, relying on sign history fixed effects ($\gamma_{i,t}=\gamma_{S(\mathcal H_{i,t})}$).

We test if retractions increase belief variance by taking the dependent variable to be the sample variance of beliefs computed at the participant level and conditional on (i) whether a retraction was observed and (ii) either the compressed history ($\text{Var}(\hat \ell_{i,t}\mid C(\mathcal H_{i,t}),r_{i,t})$) or the sign history ($\text{Var}(\hat \ell_{i,t}\mid S(\mathcal H_{i,t}),r_{i,t})$).
Here, due to power considerations, we treat compressed/sign histories that are the same up to permutations as the same, and therefore, estimate within-participant belief variance at a given (permuted) compressed/sign history---for notational simplicity, we maintain the same notation. 

\begin{table}[t]\setstretch{1.1}
	\centering\small
	\begin{tabular}{l@{\extracolsep{4pt}}cccccc@{}}
\hline\hline
\multicolumn{1}{r}{Retraction vs.} &\multicolumn{3}{c}{No Retracted Draw}  &  \multicolumn{3}{c}{Equivalent New Draw} \\
\cline{2-4} \cline{5-7}
& (1) & (2) & (3) & (4) & (5) & (6) \\
& $|\hat p_t-p_t|$ & ln(T$_t$) & Var($\hat \ell_t \mid h_t$) & $|\hat p_t-p_t|$ & ln(T$_t$) & Var($\hat \ell_t \mid h_t$) \\
\hline
Retraction ($r_t$) & 2.765$^{***}$ & 0.064$^{***}$ & 1.240$^{***}$ & 1.111$^{***}$ & 0.084$^{***}$ & 0.580$^{***}$ \\
 & (0.266) & (0.012) & (0.172) & (0.282) & (0.014) & (0.171) \\ [.1em]

Mean Decision Time     &   & 8.830 &   &   & 8.830 &   \\
Compressed History FEs & Yes & Yes & Yes & No & No & No  \\
Sign History FEs & No & No & No & Yes & Yes & Yes  \\
\hline
R$^2$ & 0.07 & 0.01 & 0.03 & 0.08 & 0.01 & 0.03 \\
N & 39162 & 39162 & 5236 & 39162 & 39162 & 5236 \\
\hline\hline
\multicolumn{7}{l}{\footnotesize Clustered standard errors at the subject level in parentheses.}\\
\multicolumn{7}{l}{\footnotesize $^{*}$ \(p<0.1\), $^{**}$ \(p<0.05\), $^{***}$ \(p<0.01\)}\\
\end{tabular}
    \begin{minipage}{1\linewidth}
        \small
        \vspace*{.5em}
        \caption{Effect of Retractions on Complexity Indicators (\hyref{hypothesis:retractions-time-variance}[Hypothesis])}
        \vspace*{-1.5em}
        \label{table-ate-complexity}
        \singlespacing \emph{Notes}: 
        This table provides estimates of the effect of retractions on three indicators of complexity, following  \hyref{equation:retractions-time}[Equation].
        There are two types of comparison: (a) updating from a retraction vs. without the retracted observation (Columns (1)--(3)) and (b) updating from a retraction vs. an equivalent new draw (Columns (4)--(6)). 
        Columns (1) and (4) refer to the accuracy in belief updating as given by the absolute error in beliefs, defined as the absolute difference between beliefs and Bayesian posteriors.
        Columns (2) and (5) refer to the speed of response, defined as log decision time. 
        Columns (3) and (6) refer to the variability of updating, defined as participant-level history-contingent log-odds belief variance. 
        Decision time is measured in seconds. 
        The sample includes all observations of participants in the baseline treatment, excluding periods in which the truth ball is disclosed or in which there was a retraction in an earlier period.
    \end{minipage}
\end{table}

\hyref{table-ate-complexity}[Table] confirms \hyref{hypothesis:retractions-time-variance}[Hypothesis].
Retractions decrease accuracy in that the absolute error in beliefs increases both compared to not having seen the retracted draw (by almost 3 percentage points---Column (1)) and compared to an equivalent new observation (by over 1 point---Column (4)).
Participants also take longer in reporting beliefs---approximately 6\% compared without the retracted observation (Column (2)) and 10\% longer when compared to an equivalent new draw (Column (5))---a conclusion that remains valid when controlling for experience and considering only later rounds.\footnote{
    See \hyref{section:robustness:heterogeneity}[Section].
    While our results show participants take less time in later rounds, the increase in decision time caused by retractions remains consistent in later rounds, when participants have had more experience observing retractions.
    Note that participants are fully informed they may see a retraction prior to any round where they do, and the interface is as similar as possible for new draws and retractions; hence, it appears unsurprising that we do not detect a difference depending on whether participants have seen more retractions in the past.
} 
Columns (3) and (6) provide an analogous comparison for the (log-odds) belief variance estimated at the participant level, where retractions increase significantly---by over one-third in either case. 
In both cases, we see that belief variance increases following a retraction.
\hyref{figure-ate-complex}[Figure] below provides a visualisation of the results in \hyref{table-ate-complexity}[Table].

Our results suggest that retractions are not only treated differently but also involve greater complexity. 
In line with the literature on cognitive imprecision, one interpretation consistent with our results is that such increased complexity is reflected in a noisier perception of a retraction's informativeness relative to direct information about the state of the world.

\subsection{Validating and Varying Complexity} 
\label{section:complexity:variation}

We now show that variation in the strength of belief updating moves together with predictions that would emerge from a complexity-based mechanism. 
In particular, we complement our analysis by assessing whether, in situations that we would expect to be more complex, our proxies for complexity are aligned, and if beliefs are correspondingly less responsive to more complex information.

We first exploit the natural variation in our experimental design to consider cases in which retractions should be less complex.
If, at time $t$, the observation received at $t-1$ is retracted, participants need only to revert to the belief they held at $t-2$, that is, before receiving that observation. 
In contrast, inferring from a retraction of previous evidence involves forming beliefs about a dataset not previously observed, thus involving counterfactual reasoning.
Hence, we expect retractions of more recent observations to be easier to process than retractions of less recent observations and, consequently, more effective in moving beliefs:

\begin{hypothesis}[Retracting Recent Observations Is Easier]
    \label{hypothesis:harder-retractions} 
    Retractions of recent observations are (a) more effective and (b) less complex, compared to those in the overall sample.
\end{hypothesis}
To assess \hyref{hypothesis:harder-retractions}[Hypothesis], we use the same regression specifications and contrast the estimates of the effect of retractions on belief updating (\hyref{table-retractions-beliefs}[Table]), belief accuracy, decision time, and belief variance (\hyref{table-ate-complexity}[Table]) in our baseline treatments to the estimates one obtains when considering only retractions of the more recent observation.

We also examine how retractions affect inference from subsequent new evidence. 
Our posited mechanism suggests that if a retraction is harder to process, then it may be more difficult to update following a retraction. 
To see why, we note that a signal history $S_{t}$ will generally influence how a participant should respond to $s_{t+1}$ via its implications on $\theta$; the added complexity of retractions would then imply spillovers as participants would correspondingly face greater difficulty understanding what this implication should be. 
This idea underlies another expression of our proposed mechanism, which we articulate as a related hypothesis:
\begin{hypothesis}[Updating after Retractions]
    \label{hypothesis:after-retractions} 
    Following a retraction, (a) participants update less from new observations, and (b) inference is more difficult.
\end{hypothesis}
Since participants update differently from a retraction than from an equivalent new draw, a difference in beliefs $\hat \ell_t$ following a retraction in period $t-1$ may just be an expression of the difference in the history at $t-1$.
In order to test if participants update less after a retraction, one needs to now explicitly consider how the \emph{change} in log-odds beliefs at a particular sign history is affected by having observed a retraction in the previous period. For this reason, we use the change in log-odds beliefs, $\Delta \hat \ell_{i,t}:=\hat \ell_{i,t}-\hat \ell_{i,t-1}$, as our dependent variable when testing this hypothesis. Thus, we estimate the following:
$
\Delta \hat \ell_{i,t}=\beta_0+\beta_1 \cdot r_{i,t-1} + \gamma_{S(\mathcal H_{i,t})} + \varepsilon_{i,t},
$ 
where $\gamma_{S(\mathcal H_{i,t})}$ denotes sign-history fixed effects. 
To test \hyref{hypothesis:after-retractions}[Hypothesis][b], we consider an analogous version of Equation (\ref{equation:retractions-time}): 
$y_{i,t} = \beta_1 \cdot r_{i,t-1} + \gamma_{S(\mathcal H_{i,t})} + \varepsilon_{i,t},$ 
where $y_{i,t}$ is a dependent variable.
We exclude periods in which the truth ball was revealed for obvious reasons.

\begin{figure}[t]\setstretch{1.1}
    \centering\small
    \begin{subfigure}{.495\textwidth} 
        \centering
        \includegraphics[width=1\linewidth]{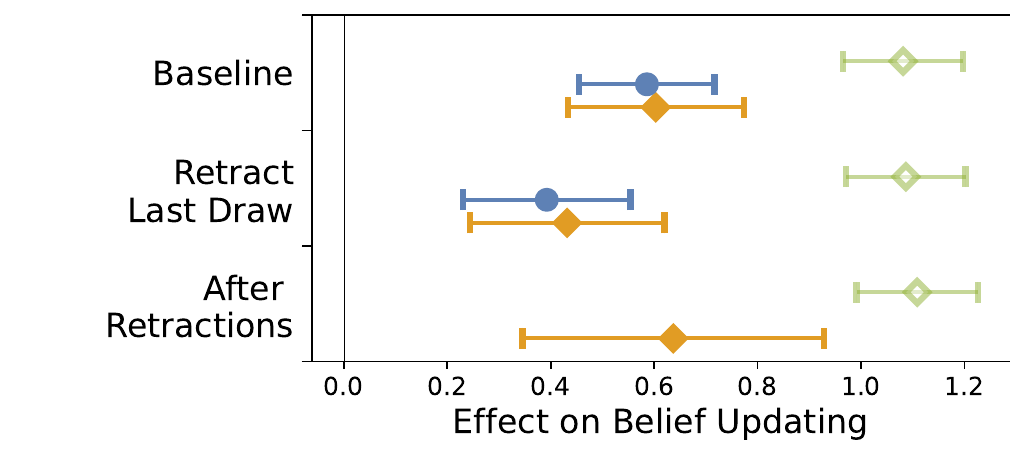}
        \vspace*{-2em}
        \caption{Beliefs}
    \end{subfigure}
    \begin{subfigure}{.495\textwidth}
        \centering
        \includegraphics[width=1\linewidth]{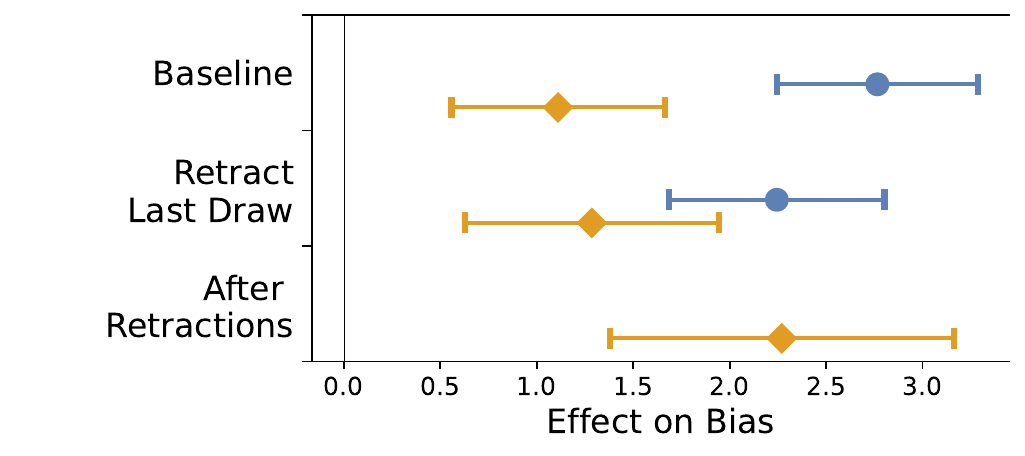}
        \vspace*{-2em}
        \caption{Accuracy}
    \end{subfigure}
    \begin{subfigure}{.495\textwidth} 
        \centering
        \includegraphics[width=1\linewidth]{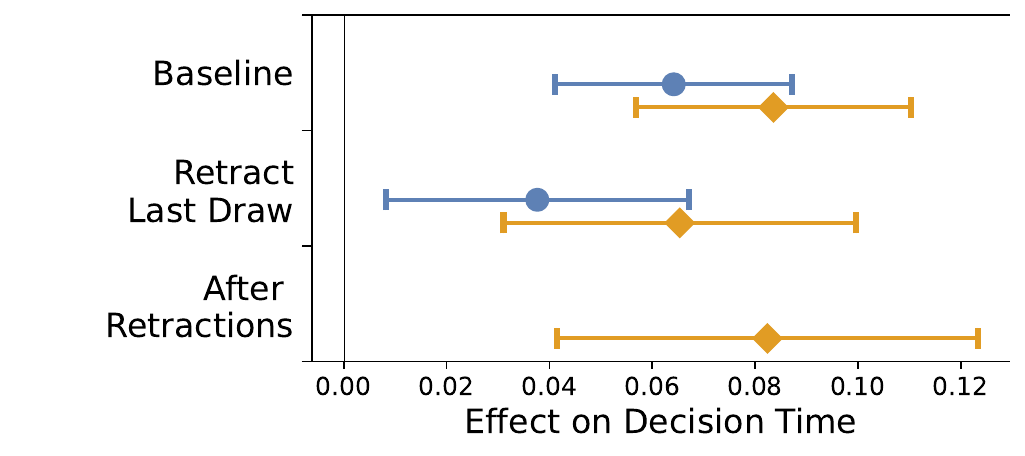}
        \vspace*{-2em}
        \caption{Speed}
    \end{subfigure}
    \begin{subfigure}{.495\textwidth}
        \centering
        \includegraphics[width=1\linewidth]{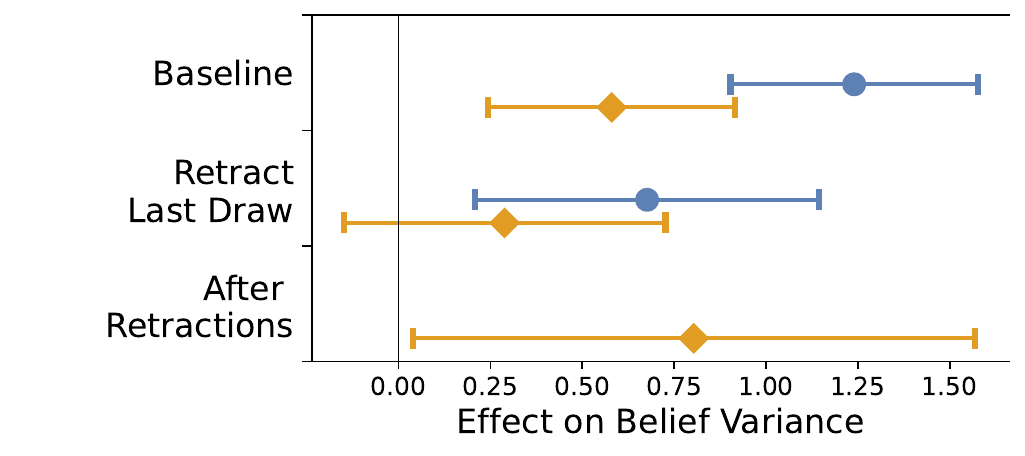}
        \vspace*{-2em}
        \caption{Variability}
    \end{subfigure}
    \begin{subfigure}{1\textwidth} 
        \centering
        \includegraphics[width=.8\linewidth]{Figures/figure-legend.pdf}
    \end{subfigure}
    \begin{minipage}{1\linewidth}
        \small
        \caption{Retracting Recent Evidence and Evidence after Retractions (\hyref{hypothesis:harder-retractions}[Hypotheses] and \ref{hypothesis:after-retractions})}
        \label{figure-ate-complex}
        \vspace*{-1.5em}
        \small
        \singlespacing \emph{Notes}: 
        This figure provides estimates for the effect of retractions on belief updating and on three complexity indicators, across settings in which we expect complexity to change. 
        ``Retract Last Draw'' restricts the sample of retractions to retractions in which the most recent draw is retracted, corresponding to \hyref{hypothesis:harder-retractions}[Hypothesis]. 
        ``After Retractions'' considers updating from new draws contingent on whether or not a retraction occurred in the past, corresponding to \hyref{hypothesis:after-retractions}[Hypothesis]. 
        Panel (a) displays the effect of retractions on belief updating, $\hat \ell_t$, under the same specifications as for \hyref{figure-ate-retractions-beliefs}[Figure]. 
        Panels (b)-(d) display effects on our three complexity indicators---accuracy ($|\hat p_t-p_t|$), speed (ln(T$_t$)), and variability (Var($\hat \ell_t \mid h_t$))---under the same specifications as for \hyref{table-ate-complexity}[Table]. Plot whiskers represent 95\% confidence intervals.
     \end{minipage}
\end{figure}

We find support for both \hyref{hypothesis:harder-retractions}[Hypotheses] and \hyref{hypothesis:after-retractions}. 
As shown in \hyref{figure-ate-complex}[Figure][(a)], retractions of more recent observations are significantly more effective. 
Specifically, participants update about 35--40\% less from retractions of recent draws than from equivalent new draws, in contrast to approximately 50-55\% in our baseline. 
In line with greater effectiveness, we find that belief reporting is starkly faster when the retraction refers not to an earlier but to the last draw (panel (c)) and also that retractions of more recent observations induce lower belief variances (panel (d)). 
We further observe that belief accuracy is attenuated (panel (b)), although not significantly different from our baseline in one case. 
Regarding \hyref{hypothesis:after-retractions}[Hypothesis], we find that participants update less after retractions than after equivalent new draws (a), are less accurate (b), take longer (c), and exhibit higher variability in their reports (d).

To summarise, consistent with our posited mechanism, the data suggest retractions of more recent observations are less cognitively demanding and that inference from new draws is more complex if they follow a retraction. 
In both cases, the intensity of belief updates aligns with our complexity indicators.

\section{Belief Updating Patterns under Retractions}
\label{section:different-biases}

So far, we have provided evidence that complexity considerations can explain the diminished effectiveness of retractions. 
Here, we discuss how retractions entail significantly different belief-updating patterns compared to updating from new direct evidence.

While our results imply that retractions---indirect information---are treated differently from direct information, one possibility is that retractions simply magnify known updating biases. 
To examine this, we rely on \citet{Grether1980QJE} log-odds regressions, the main workhorse in the existing literature \citep[cf.][]{Benjamin2019}. 
Starting from the observation that, with Bayesian updating, the log-odds posterior probability equals the prior log-odds plus the log-likelihood ($\ell_{i,t} = \ell_{i,t-1} + K_{i,t}$), a Grether regression relaxes the weight on the prior log-odds and the log-likelihood, allowing them to be different from one, i.e., $\hat \ell_{i,t} = \beta_0 + \beta_1 \cdot \hat \ell_{i,t-1} + \beta_2 \cdot  K_{i,t}$. 
Following \citet{Benjamin2019}, we estimate variants of the following:
\begin{equation}
    \hat \ell_{i,t} = \beta_0 + \beta_1 \cdot \hat \ell_{i,t-1} + \beta_2 \cdot K_{i,t} + \beta_3 \cdot K_{i,t} \cdot c_{i,t} + \varepsilon_{i,t}, \label{equation:grether-baseline}
\end{equation}
where $\hat \ell_{i,t}$ denotes $i$'s log-odds belief at period $t$, $K_{i,t}$ the log-likelihood of the signal---that is, $K(s_{i,t})$ in the case of a new draw $s_{i,t}$, and $-K(s_{i,\rho_{i,t}})$ for retractions---and $c_{i,t}$ an indicator variable that equals 1 whenever the signal observed confirms the prior belief ($\text{sign}(\hat{\ell}_{i,t-1})=\text{sign}(K_{i,t})$) and 0 if otherwise.
Bayesian updating implies that $\beta_1=1$, $\beta_2=1$, and $\beta_3 = 0$.
Base rate neglect, for instance, corresponds to $\beta_1 < 1$; under- and overinference are expressed by $\beta_2<1$ and $>1$, respectively; and confirmation bias, to updating relatively more from signals when these confirm one's prior belief, that is, $\beta_3>0$.

In examining how patterns in updating from retractions differ from updating from direct evidence, we fully interact the specification given above with the dummy variable $r_{t}$ indicating whether or not the signal corresponds to a retraction or a new draw:
\begin{equation}
    \hat \ell_{i,t} = \beta_0 + \beta_1 \cdot \hat \ell_{i,t-1} + \beta_2 \cdot K_{i,t} + \beta_3 \cdot K_{i,t} \cdot c_{i,t} + r_{i,t}\cdot [\gamma_0 + \gamma_1 \cdot \hat \ell_{i,t-1} + \gamma_2 \cdot K_{i,t} + \gamma_3 \cdot K_{i,t} \cdot c_{i,t}] + \varepsilon_{i,t}. \label{equation:grether-retractions}
\end{equation}
\noindent The interaction terms allow us to examine how previously documented deviations from Bayesian updating vary depending on whether or not the signal is a retraction. 
\hyref{table-grether}[Table] presents these results.

\begin{table}[t]\setstretch{1.1}
	\centering\small
	\begin{tabular}{l@{\extracolsep{1pt}}cc@{}}
\hline\hline
& (1) & (2) \\
& $\hat \ell_t$ & $\hat \ell_t$ \\
\hline
Signal ($K_t$) & 1.102$^{***}$ & 0.907$^{***}$ \\
 & (0.060) & (0.060) \\ [.1em]
Prior ($\hat \ell_{t-1}$) & 0.801$^{***}$ & 0.747$^{***}$ \\
 & (0.032) & (0.032) \\ [.1em]
Confirmatory Signal ($K_t \cdot c_t$) & -- & 0.651$^{***}$ \\
 &  & (0.097) \\ [.1em]
Retraction ($r_t$) x Signal ($K_t$) & -0.768$^{***}$ & -0.516$^{***}$ \\
 & (0.071) & (0.074) \\ [.1em]
Retraction ($r_t$) x Prior ($\hat \ell_{t-1}$) & 0.042 & 0.106$^{***}$ \\
 & (0.037) & (0.039) \\ [.1em]
Retraction ($r_t$) x Confirmatory Signal ($K_t \cdot c_t$) & -- & -0.807$^{***}$ \\
 &  & (0.130) \\ [.1em]

\hline
R$^2$ & 0.42 & 0.42 \\
N & 39162 & 39162 \\
\hline\hline
\multicolumn{3}{l}{\footnotesize Clustered standard errors at the subject level in parentheses.}\\
\multicolumn{3}{l}{\footnotesize $^{*}$ \(p<0.1\), $^{**}$ \(p<0.05\), $^{***}$ \(p<0.01\)}\\
\end{tabular}
    \begin{minipage}{1\linewidth}
        \small
        \vspace*{.5em}
        \caption{Belief Updating Patterns under Retractions: Grether Regressions}
        \label{table-grether}
        \vspace*{-1.5em}
        \singlespacing \emph{Notes}: 
        This table shows that patterns in belief updating from retractions do not simply reflect a strengthening of known updating biases. 
        It reports estimates of \hyref{equation:grether-retractions}[Equation] interacting the independent variables with whether or not the signal was a retraction ($r_{t}$). 
        The sample includes all observations of participants in the baseline treatment, excluding periods in which the truth ball is disclosed or in which there was a retraction in an earlier period. 
    \end{minipage}
\end{table}

As foreshadowed in \hyref{section:ineffectiveness:new-observations}[Section], we replicate known updating patterns.
In line with results by \citet{AugenblickEtAl},\footnote{
    \citet{AugenblickEtAl} provide evidence that participants overinfer (resp. underinfer) from signals in similar symmetric environments whenever $P(s_t=\theta\mid \theta)\geq 1/2$ is below (resp. above) approximately 3/5, coinciding with our parameters in the experimental design.
} 
we find $\hat \beta_2=1.102$, indicating weak overinference from new observations, although not statistically different from 1.
Once we consider whether the signal is confirmatory, we then obtain underinference from new observations, with $\hat \beta_2=0.907$ and not statistically different from 1, while $\hat \beta_3=0.651>0$ indicates confirmation bias, resulting in over-inference from confirmatory information ($\beta_2+\beta_3>1$)---a phenomenon previously documented by, for example, \citet{CharnessDave2017}.
Together, this finding suggests that our participants slightly overreact to new observations. 
However, this conclusion is primarily driven by confirmation bias: participants update more from a signal when it corroborates their prior belief.
We also verify another deviation from Bayesian updating identified in the literature: participants exhibit base-rate neglect.
In other words, they underweight the prior, as evidenced by $\beta_1<1$.

A striking difference emerges: while updating from new draws exhibits slight overinference $(\beta_2 \geq 1)$ driven by confirmation bias $(\beta_3>0)$, updating from retractions leads to marked \emph{under}inference $(0<\beta_2+\gamma_2<1)$ and \emph{anti}confirmation bias $(\beta_3+\gamma_3<0)$.
In sum, belief updating from retractions exhibits biases opposite those that emerge when updating from new draws, a conclusion which is robust across specifications.
This nuance strengthens our finding that retractions are treated differently from new signals, as the behavioural responses to retractions are not simply accentuating pre-existing biases. 
In fact, retractions induce opposite biases in belief-reporting behaviour.

\begin{figure}[t]\setstretch{1.1}
    \centering\small
    \begin{subfigure}{.495\textwidth}
        \centering
        \includegraphics[width=1\linewidth]{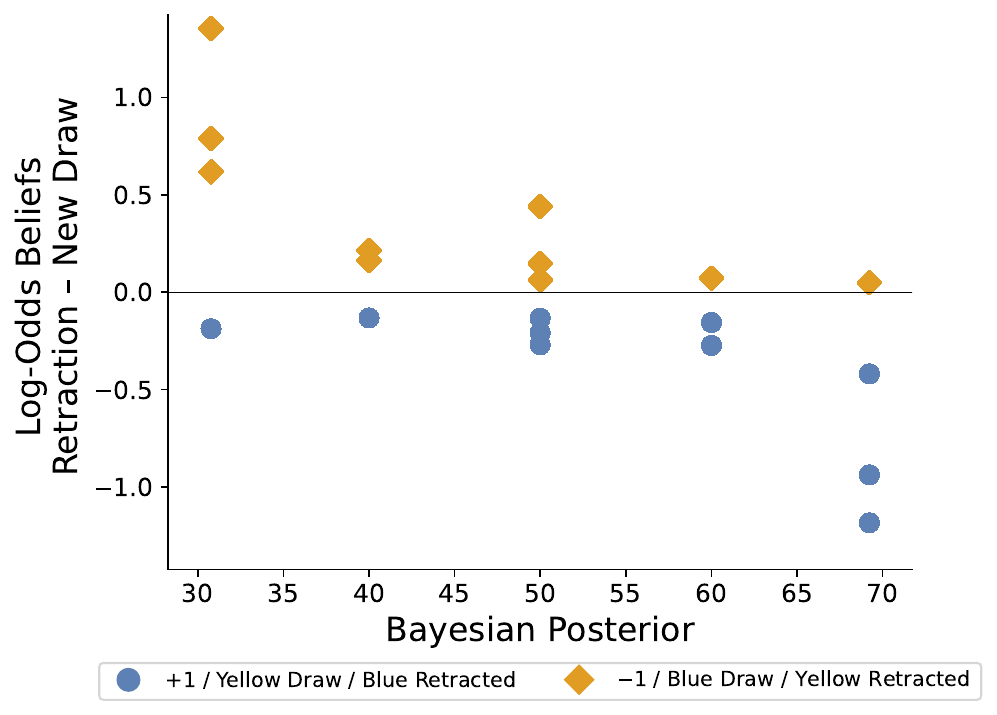}
        \vspace*{-2em}
        \caption{Log-Odds}
    \end{subfigure}
    \begin{subfigure}{.495\textwidth} 
        \centering
        \includegraphics[width=1\linewidth]{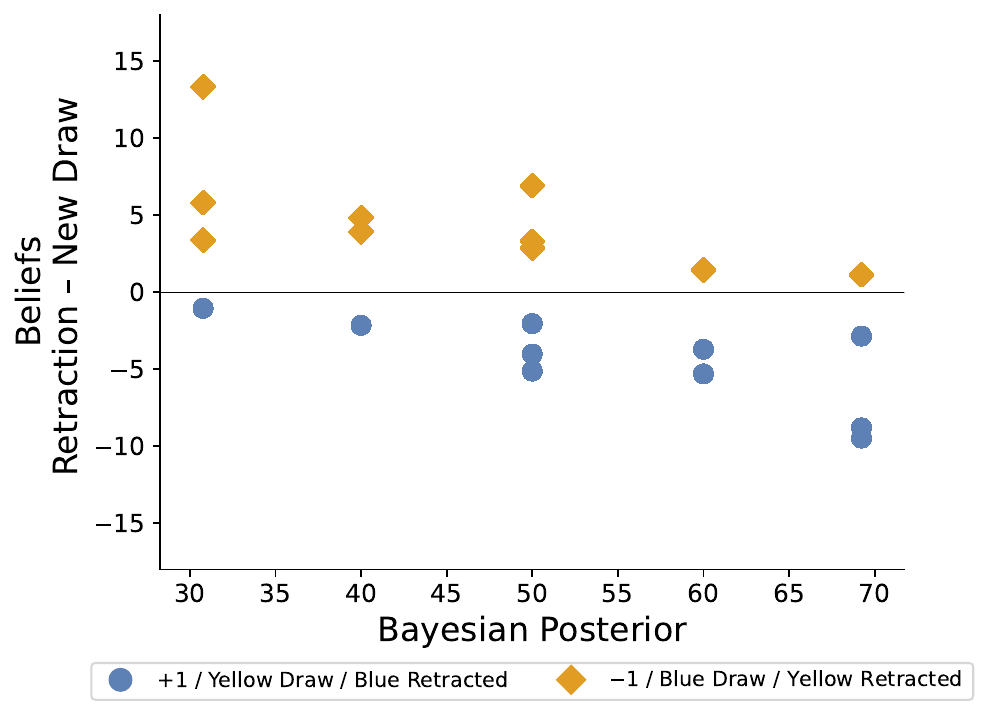}
        \vspace*{-2em}
        \caption{Levels}
    \end{subfigure}
    \begin{minipage}{1\linewidth}
        \small
        \caption{Updating from Retractions: Heterogeneity by History}
        \label{figure-ate-beliefs-sh}
        \vspace*{-1.5em}
        \small
        \singlespacing \emph{Notes}: 
        This figure displays the difference between beliefs following retractions versus equivalent new draws disaggregated by sign history. 
        Blue circles represent sign histories in which the last signal was either the retraction of a blue draw or a new yellow draw.
        Orange diamonds represent sign histories in which the last signal was either the retraction of a yellow draw or a new blue draw.
        Panel (a) presents results in log-odds, while panel (b) presents results in levels. 
        In both cases, the $x$-axis is the Bayesian posterior of the sign history. 
        The sample includes all observations of participants in the baseline treatment, excluding periods in which the truth ball is disclosed or in which there was a retraction in an earlier period.
     \end{minipage}
\end{figure}

These results suggest a specific form of heterogeneity in the diminished effect of retractions across different histories. 
We examine this heterogeneity using our baseline identification strategy (\hyref{section:ineffective:empirical-strategy}[Section]). 
\hyref{figure-ate-beliefs-sh}[Figure] shows the difference between beliefs updated from retractions and equivalent new draws for each sign history. 
In line with the documented expression of anticonfirmation bias in \hyref{table-grether}[Table], participants update less from confirmatory retractions than from confirming new draws at extreme histories, following which they hold more extreme beliefs. 
\hyref{table-grether}[Table] documents (1) a general diminished updating from retractions relative to new draws, (2) confirmatory bias from new draws, and (3) anticonfirmatory bias from retractions. 
\hyref{figure-ate-beliefs-sh}[Figure] illustrates this finding: it is exactly at more extreme histories, entailing more extreme beliefs, when observing a confirmatory signal induces participants to update less from retractions relative to new draws, as (2) and (3) there enhance (1). 
In contrast, (2) and (3) counter (1) for disconfirmatory signals, explaining why the difference between beliefs following retractions and new draws is small in this case, even if with the anticipated sign.
We emphasise that this result does not speak to which beliefs are more difficult to update from\footnote{
    Indeed, evidence for our complexity indicators is mixed, suggesting one should not infer that the heterogeneity across histories is motivated by varying degree complexity in updating. 
    While we do find that decision time patterns by sign history are strongly related to those in \hyref{figure-ate-beliefs-sh}[Figure], the difference in the absolute error in beliefs when updating from retractions and new draws, however, is greater for histories inducing more moderate posteriors, and a similar phenomenon seems to occur with belief variability---see \hyref{online-appendix:different-biases:figures}[Online Appendix].
}---rather, it speaks to the differential impact of retractions.

\section{Robustness of the Findings}
\label{section:robustness}

We performed extensive robustness checks to assess the validity of our results.
In this section, we examine the extent to which our results (1) are driven by participant understanding, (2) reflect a general feature of behaviour or rather depend on specific individual characteristics, and (3) are affected by design choices.

\subsection{Robustness 1: Participant Screening and Understanding}
\label{section:robustness:understanding}

\paragraph{Participant Screening}
We strove to ensure that our results were not driven by inattentive participants. 
While the behaviour of participants on Amazon Mechanical Turk and Prolific has been shown to approximate well representative population samples, it can sometimes be ``noisy'' relative to traditional laboratory participants \citep{SnowbergYariv2021AER,GuptaRigottiWilson2021WP}. 
To ensure our data was of high quality, we restricted participation to US residents with high approval rates (over 95\%) and held our study during business hours (Eastern Standard Time), added captchas throughout the experiment, employed an incentivisation scheme involving a high baseline and reward pay (see \hyref{section:framework:implementation}[Section]), and precluded the possibility of repeating the experiment. 
Additionally, we included comprehension questions in the instructions, which participants had to answer correctly to proceed.
These quality checks were important for us to be able to meaningfully test our hypotheses.
If participants were simply answering randomly, they would be biased relative to Bayesian updating but would exhibit no difference between updating from retractions relative to direct evidence.

\afterpage{
\begin{figure}[th!]\setstretch{1.1}
    \centering\small
    \begin{subfigure}{.495\textwidth} 
        \centering
        \includegraphics[width=1\linewidth]{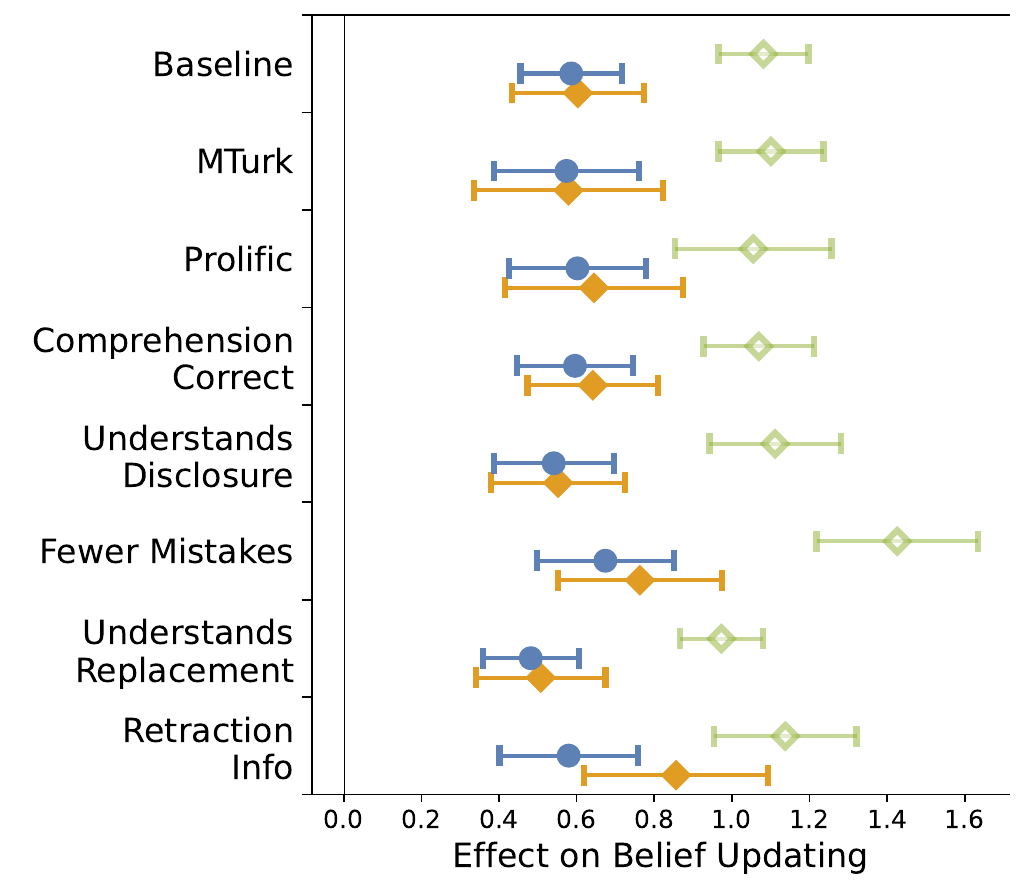}
        \vspace*{-2em}
        \caption{Beliefs}
    \end{subfigure}
    \begin{subfigure}{.495\textwidth}
        \centering
        \includegraphics[width=1\linewidth]{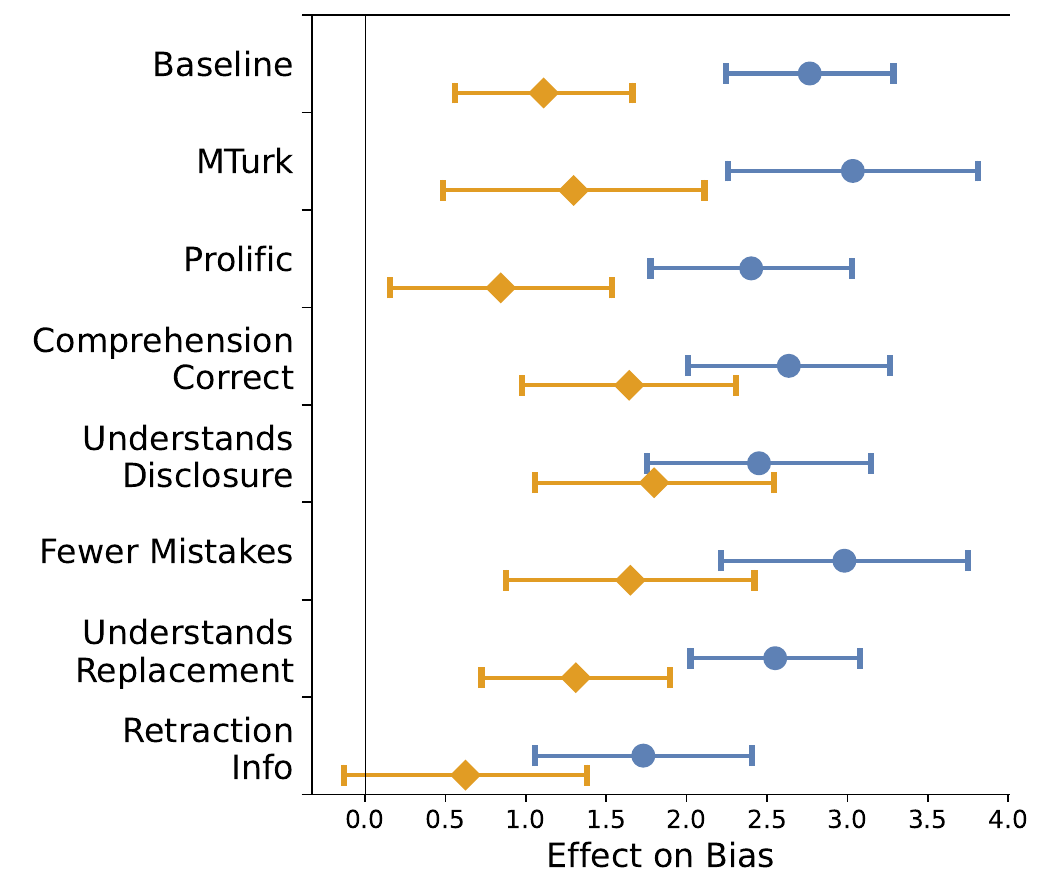}
        \vspace*{-2em}
        \caption{Accuracy}
    \end{subfigure}
    \begin{subfigure}{.495\textwidth} 
        \centering
        \includegraphics[width=1\linewidth]{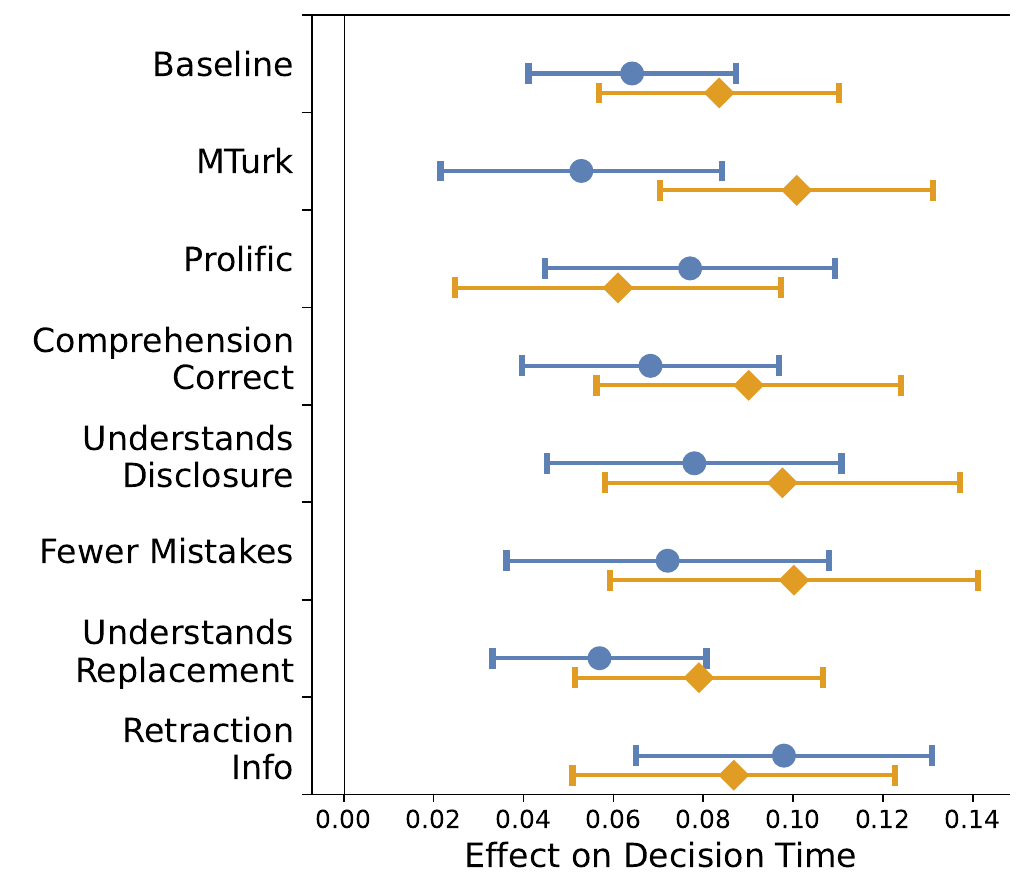}
        \vspace*{-2em}
        \caption{Speed}
    \end{subfigure}
    \begin{subfigure}{.495\textwidth}
        \centering
        \includegraphics[width=1\linewidth]{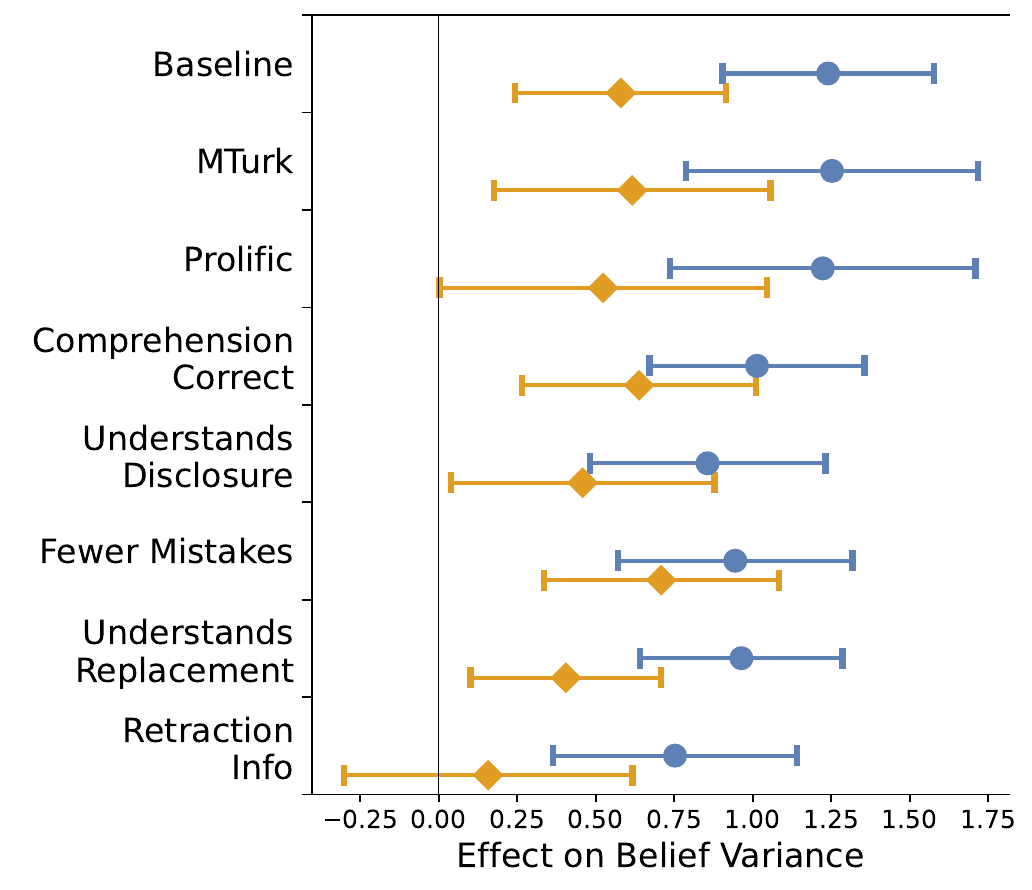}
        \vspace*{-2em}
        \caption{Variability}
    \end{subfigure}
    \begin{subfigure}{1\textwidth} 
        \centering
        \includegraphics[width=.8\linewidth]{Figures/figure-legend.pdf}
    \end{subfigure}
    \begin{minipage}{1\linewidth}
        \small
        \caption{Robustness 1: Participant Screening and Understanding}
        \label{figure-ate-robustness-understanding}
        \vspace*{-1.5em}
        \small
        \singlespacing \emph{Notes}: 
       This figure provides estimates of the effect of retractions on belief updating and on our three complexity measures across sample restrictions and experimental variants designed to test robustness to participant understanding. 
       ``Baseline'' is the pooled sample of our baseline treatments; ``MTurk'' and ``Prolific'' split the sample by those platforms. 
       We restrict the baseline sample to participants who appear attentive in four ways: 
       ``Comprehension Correct'' restricts to participants who answered all experimental comprehension questions correctly on their first try; 
       ``Understand Disclosure'' restricts to participants who, when the state is revealed, correctly report that they know the state; 
       ``Fewer Mistakes'' removes participants who update in the opposite direction to the signal more than 10\% of the time; 
       ``Understands Replacement'' excludes participants who could be mistaking sampling with and without replacement. 
       ``Retraction Info'' reports results from experiment C, where participants are told retracted observations should be ignored.
       Panel (a) displays the effect of retractions on belief updating, $\hat \ell_t$, under the same specification as for \hyref{figure-ate-retractions-beliefs}[Figure]. 
       Panels (b)--(d) display effects on our three complexity indicators---accuracy ($|\hat p_t-p_t|$), 
       (ln(T$_t$)), and variability (Var($\hat \ell_t \mid h_t$))---under the same specifications as for \hyref{table-ate-complexity}[Table]. Plot whiskers represent 95\% confidence intervals.
     \end{minipage} 
\end{figure}
\clearpage
}

\paragraph{Participant Understanding}
We further examined the robustness of our results to excluding participants based on different measures of inattentiveness.
The results are robust, and if anything slightly stronger, when restricting the sample to those participants who appear attentive, as defined in four different ways.
First, using the comprehension questionnaire, we restrict our sample to participants who answered all questions correctly on their first try (``Comprehension Correct''). 
While unincentivised, the majority of the participants demonstrated clear understanding: approximately 60\% and 90\% answered all questions correctly on the first and second try, respectively; when answering randomly, the probability of answering all correctly on the first try would be 0.2\% (see \hyref{appendix:robustness:understanding:comp}).
Second, we further restrict the sample to participants who, when the state is revealed, correctly report that they know the state  (``Understands Disclosure'').
Third, we remove participants whose belief reports are excessively noisy, which we define as updating in the opposite direction to the signal more than 10\% of the time  (``Fewer Mistakes'').\footnote{
    We considered various degrees of mistake-propensity: 1\%, 5\%, 10\%, 20\%; our conclusions remain the same.
    We also note that these checks are correlated. 
    For example, the first two samples contain a substantially smaller fraction of participants with excessively noisy reports.
}
Fourth, we exclude participants who could be mistaking sampling with and without replacement (``Understands Replacement'').\footnote{
    If sampling were without replacement, observing three draws of the same colour would reveal the colour of the truth ball.
    Less than 10\% of all participants hold extreme beliefs (close to 1 or 0) in these cases.
    Removing these participants from the sample leaves results virtually unchanged.
}

In \hyref{figure-ate-robustness-understanding}[Figure], we exhibit the estimates of the coefficient of interest corresponding to our baseline tables (\hyref{table-retractions-beliefs} and \hyref{table-ate-complexity}); the supporting regression tables can be found in \hyref{online-appendix:robustness:understanding:regression-tables}[Online Appendix].
The robustness of the results is consistent with noisy participants if anything attenuating the effect, and shows that inattention is not driving our results.

\paragraph{Participant Confidence}
We examine the possibility that retractions are associated with lower confidence, which would express greater cognitive uncertainty. 
For this, we included a question regarding participant confidence in all treatments in experiment C: similar to \citet{EnkeGraeber2020}, following the input of a belief report of $\hat p\in [0,100]$, we ask participants 
``Out of 100, how certain are you that the optimal estimate of the Truth Ball being yellow lies between $\hat p-1$ and $\hat p+1$?''
Participants then report a value between 0, labelled ``completely uncertain'', and 100, ``completely certain''.

In line with \citet{EnkeGraeber2020}, higher confidence is associated with participants inferring more from new draws. 
However, this effect seems to be driven by greater confidence being associated with greater reliance on \emph{confirmatory} signals.\footnote{
    Specifically, we find that participants that are, on average, more confident than the median infer slightly more, especially from confirmatory new draws.
    Perhaps more interesting is that when \emph{within}-participant confidence is higher---that is, using measures of confidence normalised for each participant---participants do infer significantly more from signals, even though, again, this effect seems to be driven by inference from confirmatory signals.
    However, we find no significant correlation between confidence and absolute error in beliefs---if anything, there is a weak positive correlation.
} 
Furthermore, confidence increases from approximately 60 out of 100, on average, to about 90 out of 100 when the truth ball is disclosed---a figure that is even closer to 100 for any of the sample restrictions discussed above.

While we do not find a significant difference in updating from retractions and direct evidence (see \hyref{table-hte-chigh}[Table]) depending on whether participants are more or less confident, we do observe an effect of updating from retractions (relative to new draws) on participant confidence, albeit a small one: about 2 ``confidence points'' on average, and about 0.1 standard deviations in confidence, normalised within-participant (see \hyref{online-appendix:robustness:understanding:confidence}[Online Appendix]). 
This suggests that participants are aware of, but ultimately underestimate, the greater complexity associated with updating from retractions, indicating---in the terminology of \citet{EnkeGraeberOprea2023AER} and \citet{EnkeShubatt2023}---that objective complexity (as revealed by behaviour) is more severe than participants' subjective perception, as given by reported confidence or cognitive certainty. 
Still, this finding provides reassurances that our results are not driven by participants being uncomfortable with retractions or considering their interpretation insufficiently clear. 

\paragraph{Additional Retraction Information}
We examine if providing additional information about retractions improves outcomes significantly.
In experiment C, we included a treatment (``Retraction Info'') in which, when presented with a retraction, participants are not only informed that a particular earlier draw was a noise ball but also told that 
``A noise ball is not informative about the colour of the Truth Ball and you should ignore that you have seen it''.
The treatment is identical to our baseline but for this extra information.
While some outcomes seem to improve (e.g., accuracy increases, as does belief variance, and confidence in updating from retractions increases), the differences with respect to our baseline are not statistically significant---see \hyref{figure-ate-robustness-understanding}[Figure] and \hyref{online-appendix:robustness:understanding:info}[Online Appendix].
This suggests participants err in interpreting the indirect evidence provided by retractions.\footnote{
    This pattern is also reminiscent of findings documented in experimental tests of the Monty Hall problem, where individuals often fail to recognise the error even when told the correct way to reason through it \citep[e.g.][]{Friedman1998}. 
    \citet{PearlWhy} discuss famous anecdotal instances of sophisticated individuals unwilling to admit errors in paradoxes involving reasoning with colliders.
}

\subsection{Robustness 2: Consistency across Heterogeneity}
\label{section:robustness:heterogeneity}

\afterpage{
\begin{figure}[th!]\setstretch{1.1}
    \centering\small
    \begin{subfigure}{.495\textwidth} 
        \centering
        \includegraphics[width=1\linewidth]{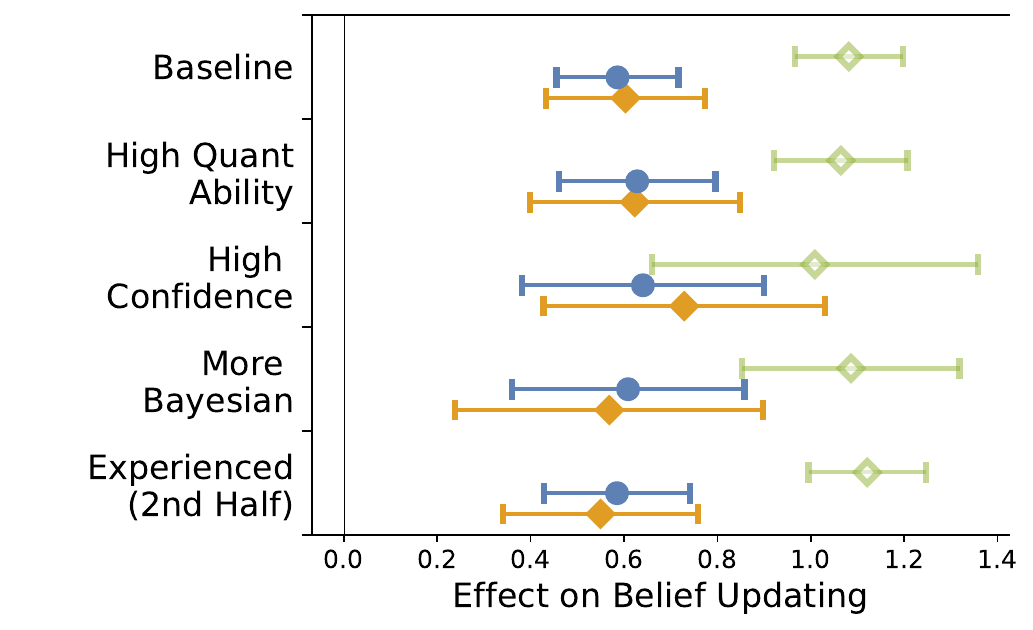}
        \vspace*{-2em}
        \caption{Beliefs}
    \end{subfigure}
    \begin{subfigure}{.495\textwidth}
        \centering
        \includegraphics[width=1\linewidth]{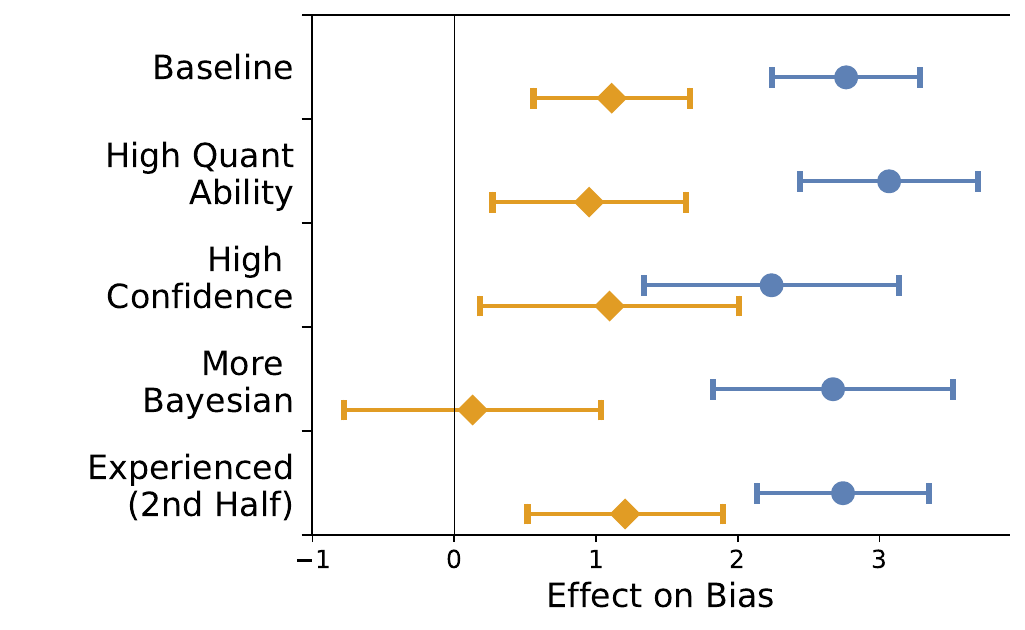}
        \vspace*{-2em}
        \caption{Accuracy}
    \end{subfigure}
    \begin{subfigure}{.495\textwidth} 
        \centering
        \includegraphics[width=1\linewidth]{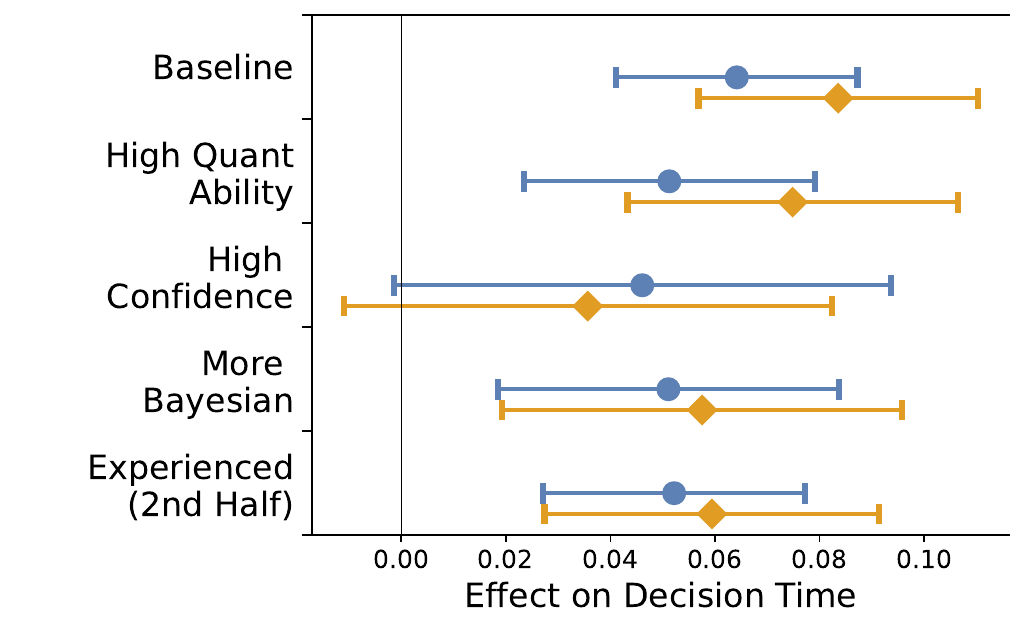}
        \vspace*{-2em}
        \caption{Speed}
    \end{subfigure}
    \begin{subfigure}{.495\textwidth}
        \centering
        \includegraphics[width=1\linewidth]{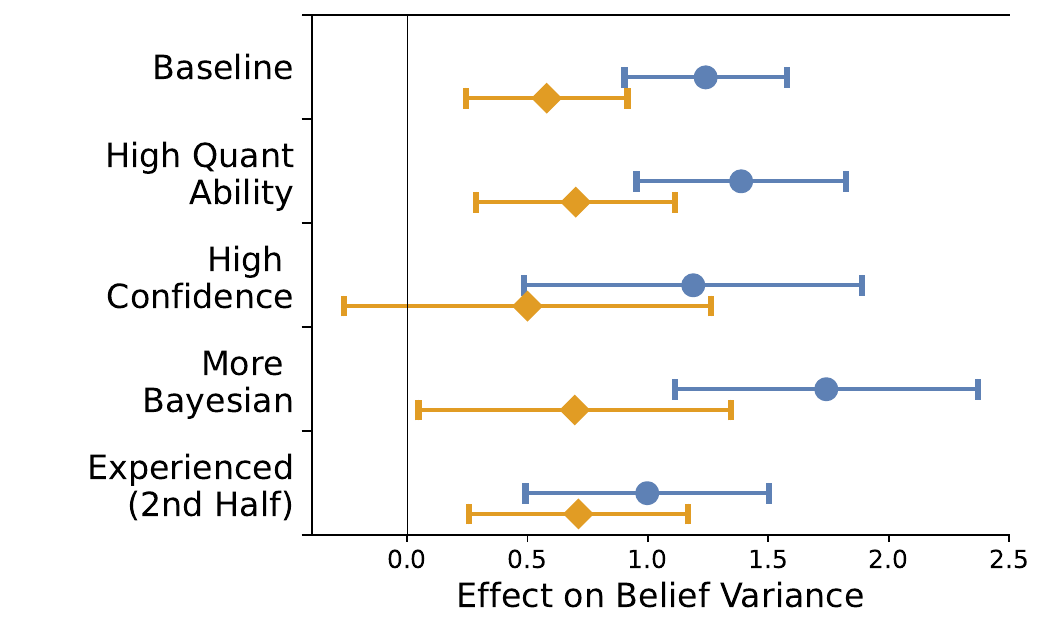}
        \vspace*{-2em}
        \caption{Variability}
    \end{subfigure}
    \begin{subfigure}{1\textwidth} 
        \centering
        \includegraphics[width=.8\linewidth]{Figures/figure-legend.pdf}
    \end{subfigure}
    \begin{minipage}{1\linewidth}
        \small
        \caption{Robustness 2: Consistency across Heterogeneity}
        \label{figure-ate-robustness-heterogeneity}
        \vspace*{-1.5em}
        \small
        \singlespacing \emph{Notes}: 
          This figure provides estimates of the effect of retractions on belief updating and on our three complexity measures across sample restrictions designed to test for heterogeneity. 
          ``Baseline'' is the pooled sample of our baseline treatment. 
          We restrict the baseline sample to test for heterogeneity in four ways: 
          ``High Quant Ability'' restricts to participants with above median score on a quantitative test in the experiment; 
          ``High Confidence'' restricts to participants with above median confidence in their beliefs; 
          ``More Bayesian'' restricts to those who are more Bayesian than the median participant when updating from new draws; 
          and ``Experienced'' restricts to the second half of rounds for each participant. 
          Panel (a) displays the effect of retractions on belief updating, $\hat \ell_t$, under the same specification as for \hyref{figure-ate-retractions-beliefs}[Figure]. 
          Panels (b)--(d) display effects on our three complexity indicators---accuracy ($|\hat p_t-p_t|$), speed (ln(T$_t$)), and variability (Var($\hat \ell_t \mid h_t$))---under the same specifications as for \hyref{table-ate-complexity}[Table]. Plot whiskers represent 95\% confidence intervals.
     \end{minipage}
\end{figure}
}

\paragraph{Heterogeneous Treatment Effects}

We explore heterogeneity in updating from retractions across multiple dimensions.
We consider heterogeneity by whether participants (i) have higher quantitative ability, as proxied for by their scores on incentivised quantitative multiple-choice questions which were asked at the end of the experiment (``High Quant Ability''); (ii) are more confident on average than the median participant (``High Confidence''); and (iii) are on average closer to the Bayesian posterior when updating from new draws than the median participant (``More Bayesian'').
We reestimate our main specifications on these groups (\hyref{figure-ate-robustness-heterogeneity}[Figure]) and expand our main specifications with interaction terms to account for heterogeneity (\hyref{online-appendix:robustness:heterogeneity:regression-tables}[Online Appendix]), failing to find any relevant deviations from our baseline.

We also examine whether experience with the task affects our results.
For this, we perform a similar heterogeneity analysis considering the second half of the experiment (rounds 17--32), at which point almost all participants will have encountered a retraction.
Again, we find no significant difference.

Finally, attesting to the robustness of our findings, we highlight that we replicated results using our baseline treatment in two different recruitment platforms, Amazon Mechanical Turk and Prolific, two years apart (see \hyref{appendix:mturk-vs-prolific}).

\paragraph{Individual Heterogeneity}
Underinference from retractions appears to be a robust feature within our sample, reflecting the overwhelming majority of participants' behaviour rather than a small minority. 
To show this, we estimate the specifications in \hyref{table-retractions-beliefs}[Table] at the \emph{participant} level.
We report summary statistics on the participant-level estimates of the coefficient of interest in \hyref{online-appendix:robustness:heterogeneity:individual}[Online Appendix].
It is difficult to fully decompose the heterogeneity in these estimates into underlying population heterogeneity versus sampling noise, given the small number of belief reports per participant. 

That said, the following observations are notable: 
First, the estimates are strictly negative for most participants (approx. 70\%). 
Second, the mean estimate is higher than the median; thus, while most participants infer less from retractions (with the median participant's absolute error still substantial), the distribution is skewed. 
Bootstrapped standard errors for both mean and median coefficients of interest show that these estimates are several standard deviations above 0, implying that these estimates are sufficiently precise to conclude that the diminished effectiveness of retractions is the rule, not the exception, among our participant pool. 
Finally, the individual-level estimates are single-peaked around the mean, pointing to a continuous spectrum of intensity of diminished inference from retractions rather than clearly distinguishable heterogeneous types.

\subsection{Robustness 3: Variations on the Design}
\label{section:robustness:design}

We now discuss our experimental design.
We begin by revisiting how it contributes to our identification of mechanisms and subsequently examining the robustness of our results with respect to various design features.

\subsubsection{Alternative Explanations Ruled Out by Design}
We first take stock of alternative explanations for retraction failure that we rule out based on the design itself.

First, our use of a balls-and-urns design was motivated by our desire to tie the limited effectiveness of retractions to belief updating itself, minimising the role of explanations related to particular domains (e.g., scientific understanding or political preferences). 
The fact that motivated reasoning is often at play in political domains might suggest it plays a crucial role in the limited effectiveness of retractions.
While it could magnify it, we find this effect even without motivated reasoning. 
Additionally, even if we recognise memory is bound to play an important role in many settings, our baseline design also precludes memory-based explanations for retraction's limited effectiveness, as all information remained on the screen making the recollection of past signals simple.\footnote{
    \citepos{Ratcliff78} seminal paper already provided evidence that recall is imperfect even when referring to very short periods of time---and the more so, the greater the elapsed time.
} 
Furthermore, issues of whether retractions lead to questioning the source's reliability, while interesting in their own right, are also precluded in our setting: a Bayesian decision-maker should be able to update beliefs from retractions without any ambiguity.\footnote{
    This lack of ambiguity distinguishes our experiment from \citet{Liang2020}, \citet{OrtolevaShishkin2021}, and \citet{EpsteinHalevy2020}.
}

Second, as \hyref{proposition:onlyproposition}[Proposition] demonstrates, only explanations specific to retractions can rationalise retraction's diminished effectiveness. 
Indeed, we designed the experiment to compare retractions to informationally equivalent direct evidence.  
The paradigm we build on allows us to quantify objectively correct beliefs, which is difficult or impossible in domains where beliefs are subjective or, perhaps more problematically, not concretely defined. 
We can thus distinguish retraction failures from any explanation that applies to all forms of information processing and belief updating, such as confirmation bias.
Our results studying such biases further show that they are also \emph{qualitatively} different for retractions as compared to new observations, as shown above in \hyref{section:different-biases}[Section]: biases in updating from retractions are not simply accentuated versions of known biases.

\afterpage{
\begin{figure}[th]\setstretch{1.1}
    \centering\small
    \begin{subfigure}{.495\textwidth} 
        \centering
        \includegraphics[width=1\linewidth]{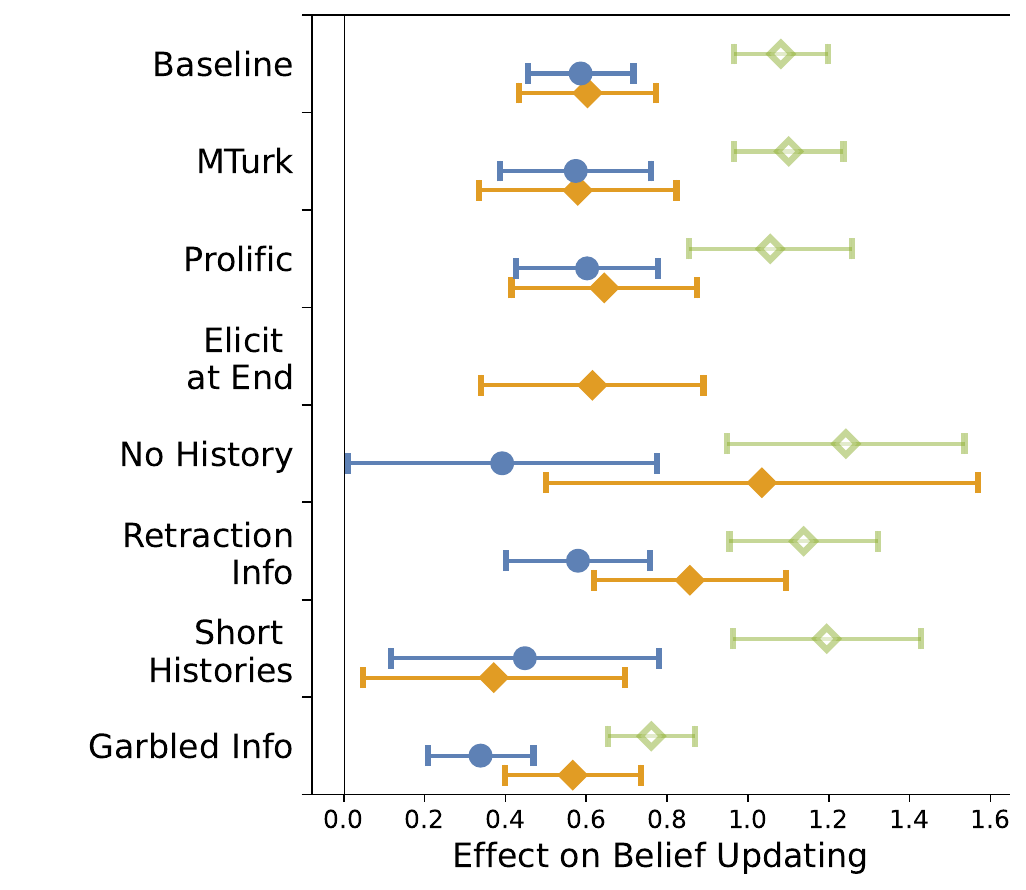}
        \vspace*{-2em}
        \caption{Beliefs}
    \end{subfigure}
    \begin{subfigure}{.495\textwidth}
        \centering
        \includegraphics[width=1\linewidth]{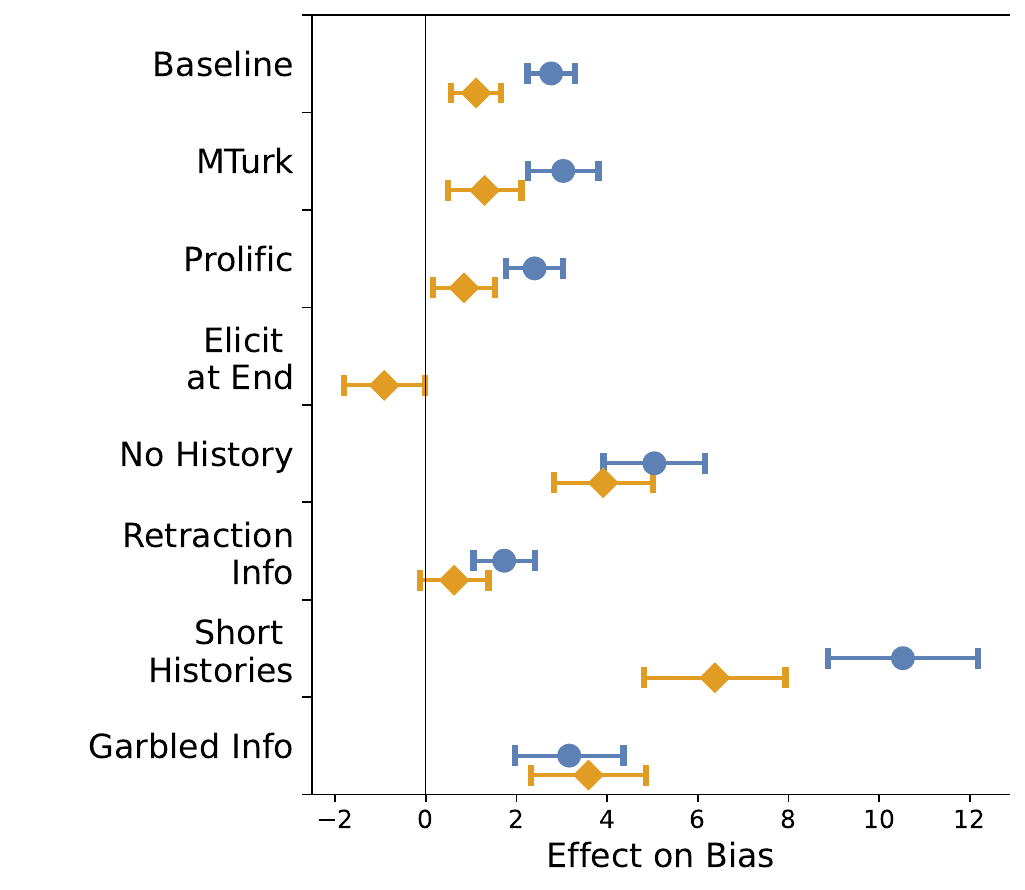}
        \vspace*{-2em}
        \caption{Accuracy}
    \end{subfigure}
    \begin{subfigure}{.495\textwidth} 
        \centering
        \includegraphics[width=1\linewidth]{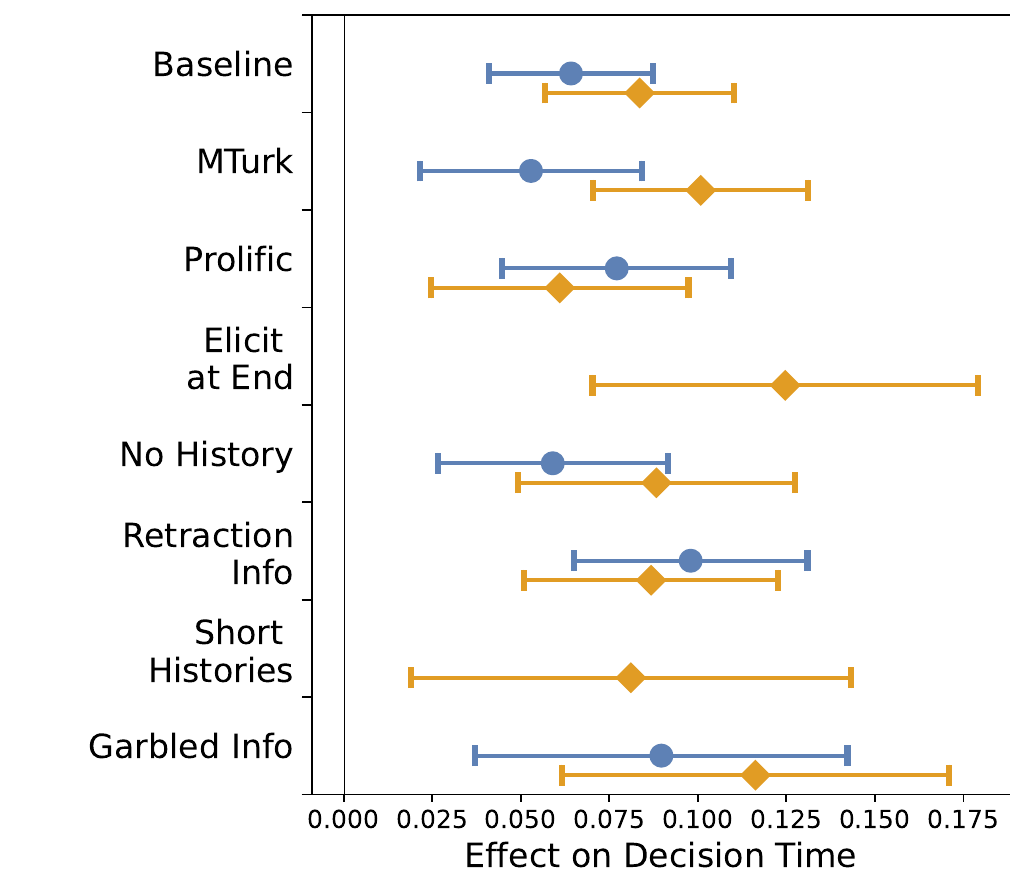}
        \vspace*{-2em}
        \caption{Speed}
    \end{subfigure}
    \begin{subfigure}{.495\textwidth}
        \centering
        \includegraphics[width=1\linewidth]{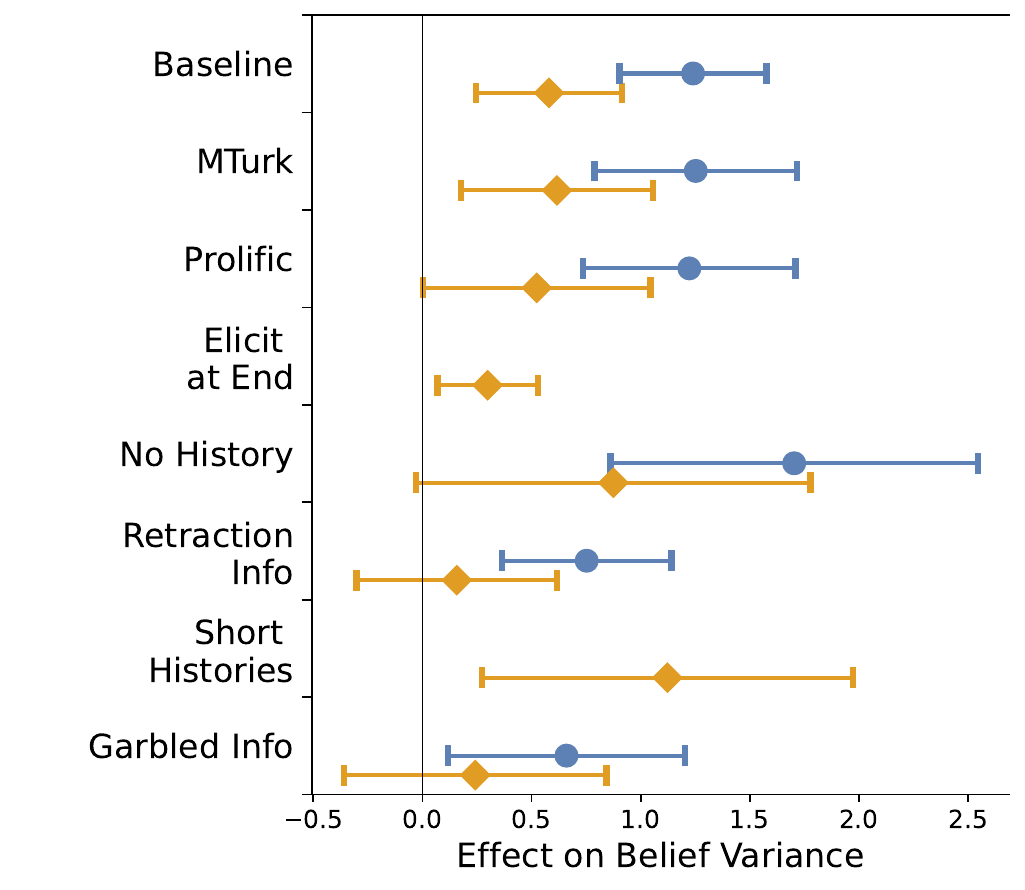}
        \vspace*{-2em}
        \caption{Variability}
    \end{subfigure}
    \begin{subfigure}{1\textwidth} 
        \centering
        \includegraphics[width=.8\linewidth]{Figures/figure-legend.pdf}
    \end{subfigure}
    \begin{minipage}{1\linewidth}
        \small
        \caption{Robustness 3: Variations on the Design}
        \label{figure-ate-robustness-design}
        \vspace*{-1.5em}
        \small
        \singlespacing \emph{Notes}: 
          This figure provides estimates of the effect of retractions on belief updating and on our three complexity measures across multiple variations on the baseline design. 
          ``Baseline'' is the pooled sample of our baseline treatment; 
          ``MTurk'' and ``Prolific'' split the sample by those platforms. 
          In the ``Elicit at End'' variant, beliefs are elicited only at the end of each round. 
          In ``No History'', participants were only shown the current observation, not the history of all observations in the current round. 
          In ``Short Histories'', there were only two periods per round, rather than four. 
          In ``Garbled Information'', truth balls were not fully informative. 
          Panel (a) displays the effect of retractions on belief updating, $\hat \ell_t$, under the same specification as for \hyref{figure-ate-retractions-beliefs}[Figure]. 
          ``Retraction Info'' is described in \hyref{figure-ate-robustness-understanding}[Figure] and included to facilitate comparison to other treatments.  
          Panels (b)--(d) display effects on our three complexity indicators---accuracy ($|\hat p_t-p_t|$), speed (ln(T$_t$)), and variability (Var($\hat \ell_t \mid h_t$))---under the same specifications as for \hyref{table-ate-complexity}[Table]. Plot whiskers represent 95\% confidence intervals.
     \end{minipage}
\end{figure}
\clearpage
}

\subsubsection{Variations on the Design}
We ran several variations of our baseline design as different treatments in our four experiments.
We discuss each of them, referring to our summary \hyref{figure-ate-robustness-design}[Figure] and additional analysis in \hyref{online-appendix:robustness:design}.

\paragraph{Elicit at the End}
An alternative explanation for the diminished effect of retractions is that it is difficult to disregard evidence that has been actively used, as might be suggested by explanations based on cognitive dissonance.  
We test whether this hypothesis could drive our results by contrasting updating from retractions when beliefs have already been elicited to when they have not.
In order to do so, we compare beliefs across our baseline---in which beliefs are elicited every period within a round---and the ``Elicit at End'' treatment in experiment A---in which beliefs are elicited only at the end of each round.\footnote{
    Specifically, the ``Elicit at End'' treatment consisted of a sequence of events identical to the baseline treatment, except for two differences:
    (1) beliefs are only elicited at the end of each round, rather than each period; 
    (2) with probability 1/3, the round ends in period two; with probability 2/3, the round ends in period three.
    The design ensures that, while we do not observe the \emph{entire} belief path, we can nevertheless observe beliefs after two draws, as well as in period 3, whether there is a third draw or a retraction. 
} 
The difference is null: having acted upon a piece of information or not does affect how much less one updates from retractions relative to equivalent new draws. 
Interestingly, accuracy in updating is lower, but a heterogeneity analysis reveals it to be only marginally significant (\hyref{table-hte-final}[Table]).
While this does not imply that retractions are as (in)effective when individuals act upon past information in other contexts, it does strengthen our conviction that our results are not due to design details.

\paragraph{No History of Past Draws}
It is often the case that, in real-world settings, past evidence remains available even if invalid, and retractions (e.g., of academic papers by journals or of news reports by media outlets) do not simply remove incorrect information but also describe what was corrected. 
Nevertheless, in many cases, the full history of past evidence may not be readily available either; it will necessarily be less salient and require being recalled. 
It is, therefore, natural to ask how omitting the history of past draws affects our baseline results. 
To speak to this, in our treatment ``No History'', the interface was kept exactly the same as in our baseline, except that the screen only showed the ball that had just been drawn and no other draws. 
When presenting retractions, we showed the retracted ball with the noise label, as in the original design, without any other draws. 
It was unclear if this would prompt participants to misinterpret retractions as evidence for the opposite state and therefore lead to treating retractions as \emph{more} informative than new draws and thus to updating more, not less, from retractions.

While removing the history does not result in statistically significant differences from our baseline in terms of how participants update from retractions relative to new draws (\hyref{table-hte-hist}[Table]), the data suggest that retractions become harder to interpret and that participants update even less from retractions relative to new draws. 
Interestingly, removing the history of draws leads to notably higher variability in beliefs and lower accuracy, resulting in less precise estimates for retraction effectiveness.
Note that we would not expect to find this effect if participants only paid attention to the last draw observed---the only piece of evidence necessary to update beliefs---suggesting that theoretically redundant past evidence plays a role in belief formation.

\paragraph{Short Histories}
In order to assess whether and how much our main finding that retractions entail diminished belief updating is due to the limited understanding of a complicated setup, we made the setup as simple as possible:
In a follow-up experiment, D, we presented participants with an updating task identical to our baseline, except that in this treatment---labelled ``Short Histories''---histories were shorter and ran for two periods only. 
Specifically, participants were provided one new draw in the first period, with the second signal being either a retraction or new draw. 
Our findings are robust even for short histories: participants infer less from retractions, take longer, and are more biased when updating from retractions than from equivalent new draws, and the variability of beliefs is also higher.
Although direct comparisons to our baseline are not well-founded, as these would partly reflect the documented heterogeneity of effects across histories (\hyref{section:different-biases}[Section]), 
we feel compelled to comment on the similarities and differences.
The effect of retractions on diminished belief updating and decision time is similar to that in our baseline.
In contrast, the ``Short Histories'' treatment features lower belief accuracy and greater belief variability, in line with suggestive evidence that these tend to be greater at histories leading to more moderate beliefs.

\paragraph{Garbled Information}
Our last design variation (experiment B, ``Garbled Info'' treatment) considered the case in which participants never perfectly learn $\theta$. 
Our goal was to allow participants to form nondegenerate beliefs about $\theta$ \emph{even} following an observation of a truth ball, thereby assessing robustness of our main results to an alternative specification of the information structure. 
As before, when a draw $s_t$ is labelled as noise ($n_t=1$), it is an independently drawn uniform $\epsilon_t$.
Unlike our baseline, however, even when labelled as a truth ball ($n_t=0$), $s_{t}$ matches $\theta$ with 80\% probability and is uniform noise with complementary probability.

The specific implementation of this design was as follows: 
    At the start of each round, a \emph{truth box} (instead of a truth ball) is chosen at random to be either ``mostly yellow'' or ``mostly blue'', each with equal probability. 
    A ``mostly yellow'' box has 9 yellow balls and 1 blue ball, and vice versa for a ``mostly blue'' box. 
    Participants could observe draws from the truth box or from a \emph{noise box} consisting of 5 yellow and 5 blue balls. 
    For periods 1 and 2, 
    a ball is drawn (with replacement) and shown to the participant; 
    with probability 1/2, the ball is from the noise box, and with probability 1/2 the ball is from the truth box. 
    In period 3, there is either a new draw or a ``fact-check'' (a slight variation in terminology relative to ``validation'' from the baseline design).
    In a fact-check, one of the prior draws is chosen uniformly at random, and the participant is told which box the ball is drawn from. 
In short, we simultaneously vary (i) the likelihood of new draws (from 3/2 to 7/3), (ii) the probability of drawing a noise ball, and (iii) the fact that now observing a ball from the truth box does not fully reveal the urn composition.

Despite the changes to the design, our results stand.
We again here find that participants update less from retractions than from direct evidence and behave as if it is more complex as per our indicators: they take longer and exhibit lower belief accuracy and greater variability.\footnote{
    Interestingly, they also update less from new draws---something in line with existing evidence that underinference from evidence is higher the greater its likelihood \citep[see][]{AugenblickEtAl}.
}

\section{Conclusion}
\label{section:conclusions}

This paper identifies and quantifies diminished updating from retractions and shows updating from retractions is revealed more complex. 
Our analyses distinguish diminished updating from retractions and other information-processing patterns that may not have been previously recognised as relevant to retraction effectiveness.
These findings provide insights into the design of interventions to address erroneous information. 
Specifically, we find that presenting direct evidence is more effective in correcting beliefs than retractions or corrections. 
Furthermore, corrections of erroneous evidence are more effective when they occur swiftly.

The minimality of our design facilitated a clear link between empirical results and their theoretical interpretation. 
But it certainly overlooks significant dimensions of real-world scenarios, where outcomes (e.g., citations) reflect factors other than probabilistic likelihood assessments, and domain-specific factors (e.g., memory frictions, motivated reasoning about health outcomes, etc.) may influence how individuals respond to retractions. 
However, the information structure in our study does approximate certain aspects of retractions in scientific articles, fact-checking, or other mechanisms of information correction.
Furthermore, we interpret the consistency of our results across variations of our baseline design as evidence for the external validity of our mechanism. 
As such, our contribution is to propose that the additional layers of complexity in updating from retractions are generically an important factor in explaining the diminished updating from retractions. 
To the extent that other factors are significant, our work suggests their impacts should be separately identified from---and interacted with---the effects of retractions on information processing analyzed here.

Our results point to several interesting potential directions for future work. 
Two strike us as particularly natural.

First, studying what makes indirect information more complex. 
Our experiment was designed to highlight how errors in information processing contribute to retraction failures. 
The richness afforded to us by variation in the design spoke to our proposed mechanism without altering the fundamental nature of the task at hand. 
Our findings suggest scope to further elucidate patterns in cognitive noise in indirect information. 
In particular, our results point toward the need for theoretical models of costly information processing to distinguish direct from indirect information. 
Additional research is necessary to document how belief updating depends on the degree of contingent reasoning involved. 
This agenda is not only of theoretical interest but also practical importance, as it aims to clarify how to correct misinformation and improve information transmission.

Second, exploring the implications of these patterns on optimal information design policies. 
In many settings---for example, interactions between politicians and the media, or firms and financial auditors---information receivers obtain results from strategic interplay between senders and third-party verification \citep[e.g.,][]{LevkunJMP}. 
While our results suggest receivers may be susceptible to err following certain kinds of information, we do not speak to how endogenous changes in information may influence belief-updating patterns. 
Furthermore, a broader implication of our work is that the way in which information is generated can influence its perception beyond the objective informational content. 
While work in information design commonly reduces information to posterior beliefs, such reductions may omit important economic forces that seem worth exploring. 
For instance, knowing that retractions are not fully effective in correcting beliefs, to what extent could an information designer (e.g., a partisan media outlet or a political campaign strategist) exploit under-reaction to retractions? 
How would our findings shape their information policy, and how should a third party design a verification or fact-checking policy to counter it?
If corrections but not validations are announced, will people correctly treat unretracted evidence as more reliable? 
When policies target evidence favoring a particular view, are the resulting corrections or fact-checks perceived as less informative? 
We believe answering these and related questions has substantial practical value.

\section{References}
\vspace*{-0em}\setlength{\bibhang}{0pt}
\bibliographystyle{aea}
{\setstretch{1.15}\setlength{\bibsep}{.0em plus .0ex}
\bibliography{retractions.bib}}

\newpage
\renewcommand{\thesection}{Appendix}
\renewcommand{\thesubsection}{Appendix \Alph{subsection}}
\renewcommand{\thesubsubsection}{\thesubsection.\arabic{subsubsection}}

\titleformat{\paragraph}[block]
{\normalfont\large\bfseries}{\theparagraph.}{.5em}{\large\bfseries}
\titlespacing*{\paragraph}{0pt}{*1}{*0}

\section{~}

\subsection{Additional Discussion of the Related Literature}
\label{appendix:related-literature}
In this appendix, we discuss existing domain-specific evidence of retraction ineffectiveness. We emphasize that this discussion focuses on the relationship between our design and those from past work, and is not meant to be a systematic survey or meta-analysis. As such, we mention broad themes that have been productively explored across a variety of research lines, but do not formally assess these papers or the state of these literatures. 

\noindent \emph{Political Information.} 
Perhaps the largest number of experiments in this literature have studied the correction of information in political settings. 
While interpreting magnitudes is sometimes difficult in these studies, most show retractions have diminished effectiveness in political contexts.\footnote{
    In the context of highly politically charged topics, retractions may in rare cases \emph{backfire}, leading participants to believe more strongly in the retracted information.
    \citet{NyhanReifler2010} noted the occurrence of backfiring in an experiment where they provided participants with information about the presence of weapons of mass destruction in Iraq during the early 2000s, and subsequently provided them with corrections. 
    This extreme form of retraction failure, for the most part, has not been replicated.
    See \citet{NyhanSurvey2021} for an authoritative discussion.
}
For instance, in the context of the 2016 US Presidential election \citep{SwireEtAl2017,Nyhanetal2019} and the 2017 French Presidential election \citep{Barreraetal2020}, fact-checking did improve factual knowledge, but was less effective than the original corrected information. \citet{GurievEtAl2023}, however, document relatively small impacts of fact-checking on perceived veracity in the context of the 2022 US Midterm elections. 
Many studies suggest motivated reasoning as the main explanation for the ineffectiveness of retractions in political contexts.\footnote{
    Various studies have articulated how motivated reasoning influences belief processing in political domains; for instance, see \citet{AngelucciPrat2020}, \citet{Thaler2020}, and \citet{TaberLodge2006}.
}  
Although it may indeed play a significant role, our results indicate that retractions fail even in the absence of motivated reasoning.

\medskip 

\textit{Fake News.} Prior literature on fake news across psychology, political science and economics has studied the effectiveness of fact checking in combating misinformation; \citet{PennycookRand2021} discuss several reasons for this apparent diminished effectiveness.  It is worth emphasizing that many papers in this literature vary the \emph{nature of the fact-check itself}, with the pattern of interest being whether some presentations of fact-checks are viewed as subjectively more informative; see \citet{EckerEtAl2020} for both an insightful discussion and an example.

\medskip

\noindent \emph{Financial Information.} 
Other work has focused on the effectiveness of retractions in financial settings, where designs tend to involve presentations of earnings reports or related financial statements and then instructions to disregard. 
The focus is typically less on beliefs themselves, but rather how the information is \emph{used} in assessments or investments.  
\citet{GrantEtAl}, \citet{TanandTan2009}, and \citet{TanandKoonce2011} run experiments using such designs, finding that retractions have diminished effectiveness in these domains, and discuss ways this can be combated.

\medskip

\noindent \emph{Jury Trials.} 
Jury trials often feature information which jurors are instructed to disregard. 
Experiments on this question tend to focus on whether the reason evidence should be disregarded matters. 
\citet{KassinSommers97}, \citet{ThompsonEtAl} and
\citet{FeinEtAl} conduct experiments documenting that juries do not always simply disregard information if instructed to do so. 
While these studies do show retracted information is not so easily disregarded, it is less clear that this reflects a departure from Bayesian rationality, since the retracted information is often meaningful.

\medskip

\noindent \emph{Academic Papers.} 
In addition to work studying society's beliefs in the association between vaccines and autism discussed in the introduction, other existing literature on retractions of scientific articles typically focuses on documenting the reasons why papers are retracted, as well as assessing the consequences for researchers. 
While fraud and academic misconduct are the main reasons behind retractions, error and failure to replicate constitute a significant fraction of the retraction notices \citep{Science2018,FangSteenCasadevall2012}---and it is important to note that many papers that do not replicate are not retracted \citep{Serra-GarciaGneezy2021SciAdv}. 
Among the academic community, there seems to be a significant penalty for researchers associated to retractions: a decrease in citations not only of the authors' prior work, but also of their collaborators', and, more generally, of work in related topics \citep{LuJinUzziJones2013,AzoulayFurmanKriegerMurray2015,HussingerPellens2019}. 
Of course, there are many reasons citations may be an imperfect proxy for retraction effectiveness (strategic citation motives, information about retractions not reaching the target audience, among others). 
Existing experimental evidence focusing on this setting suggests that retractions induce insufficient belief updating, even when the cited reason is fabricated data, and points to availability of a causal narrative as a possible reason \citep[see, e.g.,][]{Greitemeyer2014}. 
While these studies do show that retracted information is not so easily disregarded, relying on observational data is challenging. 
Indeed at least in some cases, the diminished updating from retractions may not reflect a misperception of its informational content and instead be consistent with Bayesian inference; for instance, a scientific article's retraction may involve a dispute with unclear implications, or follow-on work may find the retracted article made certain contributions which were accepted as valid (see \cite{FangSteenCasadevall2012} for examples).

\newpage

\subsection{Sample Characteristics}
\label{appendix:sample-characteristics}

\begin{table}[th!]\setstretch{1.1}
	\centering\footnotesize
	\begin{tabular}{l@{\extracolsep{4pt}}ccccccc@{}}
\hline\hline
& MTurk & Elicit & Garbled & Prolific & No History & Retraction & Short \\
&  & at End & Info &  &  & Info & Histories \\
& (1) & (2) & (3) & (4) & (5) & (6) & (7) \\
\hline
Total Subjects & 211 & 204 & 164 & 155 & 164 & 164 & 150 \\
\hline
Age & 37.7 & 39.5 & 38.9 & 38.5 & 37.8 & 37.6 & 35.5 \\
Female & 0.398 & 0.402 & 0.433 & 0.477 & 0.439 & 0.500 & 0.473 \\
High School & 0.900 & 0.887 & 0.927 & 0.865 & 0.860 & 0.884 & 0.880 \\
College Degree & 0.673 & 0.613 & 0.683 & 0.587 & 0.506 & 0.561 & 0.560 \\
Postgraduate & 0.180 & 0.201 & 0.189 & 0.148 & 0.177 & 0.177 & 0.153 \\
\hline
High Comprehension & 0.602 & 0.495 & 0.561 & 0.574 & 0.634 & 0.555 & 0.660 \\
High Quant & 0.275 & 0.294 & 0.171 & 0.290 & 0.317 & 0.317 & 0.307 \\
\hline
Experiment & A & A & B & C & C & C & D \\
Date & 2020-06 & 2020-06 & 2021-05 & 2024-01 & 2024-01 & 2024-01 & 2024-02 \\
Platform & MTurk & MTurk & MTurk & Prolific & Prolific & Prolific & Prolific \\
\hline\hline
\end{tabular}
    \begin{minipage}{1\linewidth}
        \footnotesize
        \vspace*{.5em}
        \caption{Sample Characteristics}
        \vspace*{-1.5em}
        \label{table-sample-characteristics}
        \singlespacing \emph{Notes}: 
        The table shows sample characteristics for each of our treatments in each of our experiments.  
        ``Age'' is measured in years; ``Female'' denotes the fraction of the sample that identifies as a woman; ``High School'', ``College Degree'', and ``Postgraduate Studies'' denote the fraction of the sample that has completed the respective level of education.
        ``Comprehension Correct'' shows the fraction of the sample that answered all comprehension questions correctly at first try; ``High Quant'' shows to the fraction of participants who answer all the quantitative questions correctly at on their first try.
        Finally, ``Date'' denotes when the data was collected, and ``Platform'' the venue used to recruit participants.
    \end{minipage}
\end{table}

\newpage

\subsection{Comparison of Recruitment Platforms}
\label{appendix:mturk-vs-prolific}

\begin{table}[h!]\setstretch{1.1}
	\centering\footnotesize
	\rotatebox{90}{\begin{tabular}{l@{\extracolsep{1pt}}cccccccc@{}}
\hline\hline
\multicolumn{1}{r}{Retraction vs.} &\multicolumn{4}{c}{No Retracted Draw}  &  \multicolumn{4}{c}{Equivalent New Draw} \\
\cline{2-5} \cline{6-9}
& (1) & (2) & (3) & (4) & (5) & (6) & (7) & (8) \\
& $\hat \ell_t$ & $|\hat p_t-p_t|$ & ln(T$_t$) & Var($\hat \ell_t \mid h_t$) & $\hat \ell_t$ & $|\hat p_t-p_t|$ & ln(T$_t$) & Var($\hat \ell_t \mid h_t$) \\
\hline
Retraction ($r_t$) & 0.009 & 2.454$^{***}$ & 0.061$^{***}$ & 1.146$^{***}$ & -0.031 & 0.797$^{**}$ & 0.080$^{***}$ & 0.486$^{*}$ \\
 & (0.027) & (0.306) & (0.016) & (0.232) & (0.030) & (0.331) & (0.016) & (0.251) \\ [.1em]
Retracted Draw ($r_t\cdot K(s_{\rho_t})$) & 0.600$^{***}$ & -- & -- & -- & 0.608$^{***}$ & -- & -- & -- \\
 & (0.090) &  &  &  & (0.104) &  &  &  \\ [.1em]
Retraction ($r_t$) x MTurk & 0.004 & 0.544 & 0.005 & 0.163 & 0.021 & 0.544 & 0.007 & 0.164 \\
 & (0.035) & (0.464) & (0.021) & (0.275) & (0.034) & (0.464) & (0.020) & (0.276) \\ [.1em]
Retracted Draw x MTurk & -0.024 & -- & -- & -- & -0.007 & -- & -- & -- \\
 & (0.131) &  &  &  & (0.131) &  &  &  \\ [.1em]

Mean Decision Time     &   &   & 8.830 &   &   &   & 8.830 &   \\
Compressed History FEs & Yes & Yes & Yes & Yes & No & No & No & No  \\
Sign History FEs & No & No & No & No & Yes & Yes & Yes & Yes  \\
\hline
R$^2$ & 0.27 & 0.08 & 0.18 & 0.03 & 0.27 & 0.08 & 0.18 & 0.03 \\
N & 39162 & 39162 & 39162 & 5236 & 39162 & 39162 & 39162 & 5236 \\
\hline\hline
\multicolumn{9}{l}{\footnotesize Clustered standard errors at the subject level in parentheses.}\\
\multicolumn{9}{l}{\footnotesize $^{*}$ \(p<0.1\), $^{**}$ \(p<0.05\), $^{***}$ \(p<0.01\)}\\
\end{tabular}}
    \begin{minipage}{1\linewidth}
        \footnotesize
        \vspace*{.5em}
        \caption{Treatment Effects
        across Recruitment Platform}
        \vspace*{-1.5em}
        \label{table-ate-mturk-vs-prolific}
        \singlespacing \emph{Notes}: 
        This table compares average treatment effects in our baseline treatment across experiments A (MTurk) and C (Prolific). 
        MTurk corresponds to an indicator variable that equals 1 when the observation is from our baseline treatment in experiment A. 
        There are two types of comparison: (a) updating from a retraction vs. without the retracted observation (Columns (1)-(4)) and (b) vs. an equivalent new draw (Columns (5)-(8)). 
        Columns (1) and (5) show effects on log-odds beliefs; (2) and (6) on the accuracy of belief updating; (3) and (7) on the speed of updating; (4) and (8) on the variability of updating. 
        The sample includes all observations of participants in the baseline treatments, excluding periods in which the truth ball is disclosed or in which there was a retraction in an earlier period. 
    \end{minipage}
\end{table}

\afterpage{\clearpage}
\newpage

\subsection{Comprehension Questionnaire}
\label{appendix:robustness:understanding:comp}

\begin{figure}[th!]
    \centering
    \includegraphics[width=0.48\linewidth,]{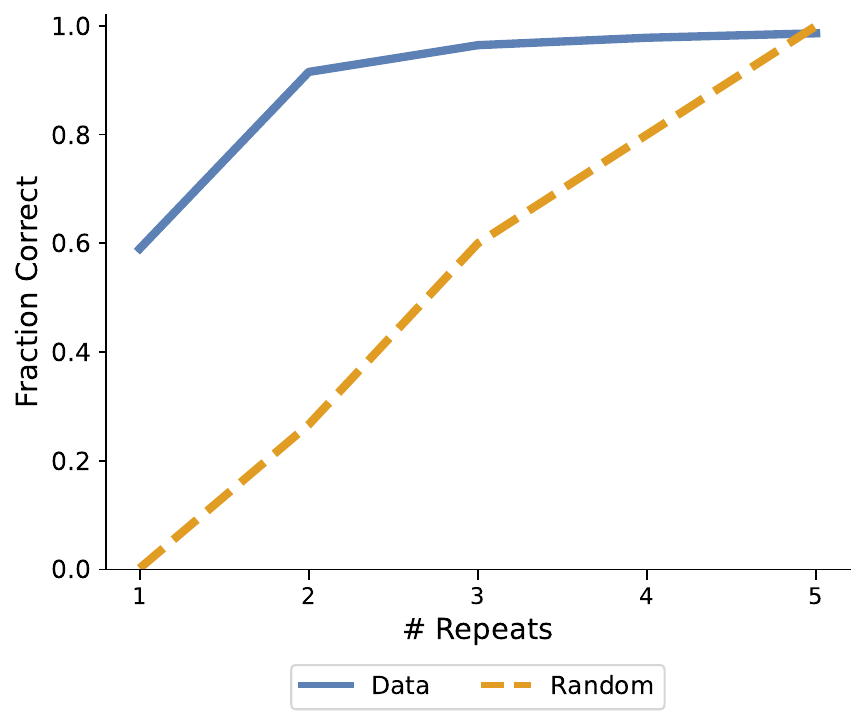}
    \begin{minipage}{1\linewidth}
        \footnotesize
        \caption{Comprehension Questions}
        \label{figure-repeats-comp}
        \vspace*{-1em}
        \singlespacing \emph{Notes}: 
        The comparison is to the case in which participants randomize uniformly over answers that were not previously tried and only in questions that were marked wrong.
    \end{minipage}
\end{figure} 

\afterpage{\clearpage}
\newpage

\newpage
\renewcommand{\thesection}{Online Appendix}
\renewcommand{\thesubsection}{Online Appendix \Alph{subsection}}
\renewcommand{\thesubsubsection}{\Alph{subsection}.\arabic{subsubsection}}
\renewcommand{\theparagraph}{\Alph{subsection}.\arabic{subsubsection}.\arabic{paragraph}}

\titleformat{\paragraph}[block]
{\normalfont\large\bfseries}{\theparagraph.}{.5em}{\large\bfseries}
\titlespacing*{\paragraph}{0pt}{*1}{*0}

\section{~}
\setcounter{subsection}{4}

\subsection{Comparing Beliefs to the Bayesian Posterior}
\label{online-appendix:comparison-to-bayes}

\begin{figure}[th!]
    \centering
    \begin{minipage}{1\linewidth}\centering
        \begin{subfigure}{0.48\linewidth}
            \centering \includegraphics[width=1\linewidth,]{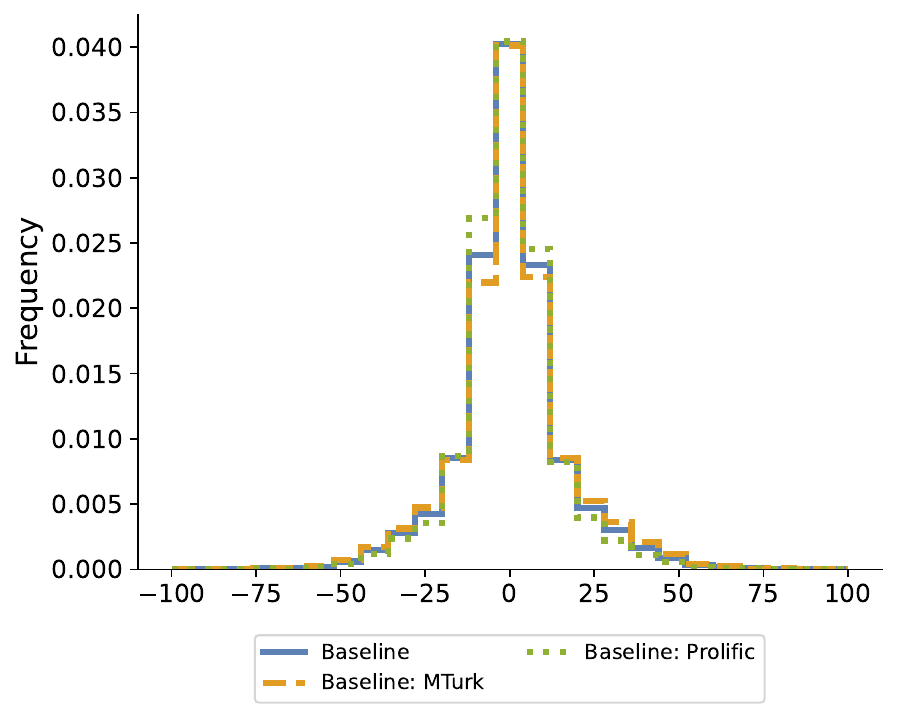}
            \caption{Difference to Bayesian Posteriors}
            \label{figure-difference-to-bayes}
        \end{subfigure}
        \begin{subfigure}{0.48\linewidth}
            \centering
            \includegraphics[width=1\linewidth,]{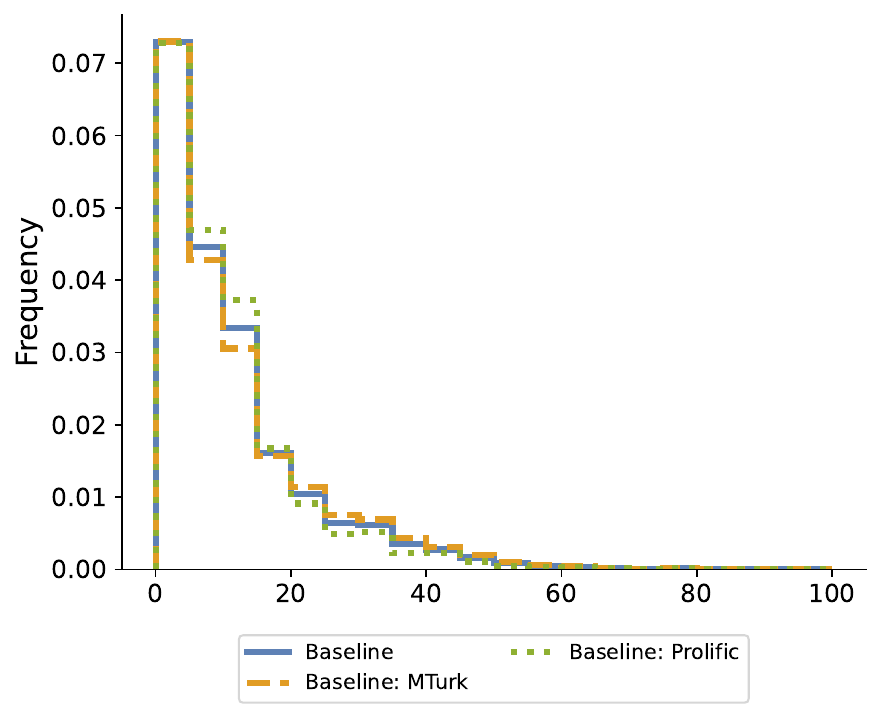}
            \caption{Distance from Bayesian Posteriors}
            \label{figure-distance-to-bayes}
        \end{subfigure}
    \end{minipage}
    
    \begin{minipage}{1\linewidth}
        \label{figure-comparison-to-bayes}
        \footnotesize
        \caption{Comparing Beliefs to Bayesian Posteriors}
        \vspace*{-1.5em}
        \singlespacing \emph{Notes}: 
        The figure shows (a) the average difference between beliefs and the Bayesian posterior and (b) the average distance between the beliefs and the Bayesian posterior. 
        The sample comprises the baseline treatments in both experiments A (MTurk; dashed orange line) and C (Prolific; dotted green line), as well as the pooled sample (solid blue line), and it includes all periods in which the history only include new draws.
    \end{minipage}
\end{figure} 

\afterpage{\clearpage}
\newpage

\subsection{Retraction Ineffectiveness}

In this section, we present results from \hyref{section:ineffectiveness}[Section] fully disaggregated by history. 
In \hyref{online-appendix:beliefs-by-history}[Section] we report beliefs (in levels) by history, first at histories that only include new draws, and then at histories in which the last observation is a retraction. 
In \hyref{online-appendix:table-hte-by-history-beliefs}[Section], we report treatment effects (in log-odds) of retraction ineffectiveness, corresponding to \hyref{table-retractions-beliefs}[Table], disaggregated by sign history for comparisons to no retracted draw (test (a)), and by compressed history for comparisons to equivalent new draws (test (b)). 

\newpage
\subsubsection{Beliefs Disaggregated by History}
\label{online-appendix:beliefs-by-history}

\paragraph{Beliefs Following New Draws}

\begin{table}[th!]\setstretch{1.1}
	\centering\footnotesize
	\begin{tabular}{l@{\extracolsep{4pt}}ccc@{}}
\hline\hline
Sign History & Bayesian Posterior & Mean Reported Belief & Obs \\
(1)  & (2)  & (3)  & (4) \\
\hline
$BBBB$ & 16.495 & 20.750 (1.773) & 208 \\
$BBB$ & 22.857 & 26.279 (0.865) & 785 \\
$BB$ & 30.769 & 35.435 (0.301) & 3027 \\
$BBBY$ & 30.769 & 28.268 (1.634) & 194 \\
$BBYB$ & 30.769 & 28.264 (1.708) & 170 \\
$YBBB$ & 30.769 & 31.357 (1.939) & 173 \\
$BYBB$ & 30.769 & 23.832 (1.498) & 165 \\
$B$ & 40.000 & 42.904 (0.188) & 5771 \\
$BBY$ & 40.000 & 40.982 (0.629) & 691 \\
$BYB$ & 40.000 & 38.142 (0.530) & 662 \\
$YBB$ & 40.000 & 43.105 (0.655) & 685 \\
$BBYY$ & 50.000 & 47.997 (1.117) & 171 \\
$BY$ & 50.000 & 50.127 (0.235) & 2744 \\
$BYBY$ & 50.000 & 48.888 (0.979) & 143 \\
$BYYB$ & 50.000 & 50.496 (0.926) & 154 \\
$YB$ & 50.000 & 51.800 (0.226) & 2765 \\
$YBBY$ & 50.000 & 50.232 (0.914) & 176 \\
$YBYB$ & 50.000 & 54.502 (1.146) & 163 \\
$YYBB$ & 50.000 & 53.827 (0.976) & 196 \\
$BYY$ & 60.000 & 61.127 (0.564) & 677 \\
$Y$ & 60.000 & 57.627 (0.173) & 5941 \\
$YBY$ & 60.000 & 61.725 (0.514) & 677 \\
$YYB$ & 60.000 & 58.987 (0.530) & 760 \\
$BYYY$ & 69.231 & 75.486 (1.598) & 183 \\
$YBYY$ & 69.231 & 74.454 (1.554) & 164 \\
$YY$ & 69.231 & 63.855 (0.284) & 3176 \\
$YYBY$ & 69.231 & 70.677 (1.714) & 172 \\
$YYYB$ & 69.231 & 72.480 (1.618) & 212 \\
$YYY$ & 77.143 & 75.432 (0.757) & 829 \\
$YYYY$ & 83.505 & 80.039 (1.545) & 230 \\
\hline\hline
\multicolumn{4}{l}{\footnotesize Standard errors in parentheses.}\\
\end{tabular}
    \begin{minipage}{1\linewidth}
        \footnotesize
        \vspace*{.5em}
        \caption{Beliefs Disaggregated by History: New Draws}
        \vspace*{-1.5em}
        \label{table-beliefs-by-history-baseline}
        \singlespacing \emph{Notes}: 
        The table shows, (1) for each history of draws, (2) the associated Bayesian Posterior, (3) the average beliefs for our baseline treatment (pooling experiments A and C), and (4) the number of observations.
        The sample only includes periods $t$ in which the history $\mathcal H_t$ only includes new draws. 
        Standard errors in parentheses.
    \end{minipage}
\end{table}

\afterpage{\clearpage}
\newpage  

\paragraph{Beliefs Following Retractions}

\begin{table}[h!]\setstretch{1.1}
	\centering\footnotesize
	\begin{tabular}{l@{\extracolsep{4pt}}ccc@{}}
\hline\hline
Sign History & Bayesian Posterior & Mean Reported Belief & Obs \\
(1)  & (2)  & (3)  & (4) \\
\hline
$BBBY$ & 30.769 & 27.219 (1.401) & 297 \\
$BBYB$ & 30.769 & 31.651 (1.770) & 92 \\
$YBBB$ & 30.769 & 37.160 (2.138) & 95 \\
$BYBB$ & 30.769 & 37.161 (2.049) & 89 \\
$BBY$ & 40.000 & 38.829 (0.490) & 1176 \\
$BYB$ & 40.000 & 42.965 (0.613) & 566 \\
$YBB$ & 40.000 & 47.019 (0.594) & 607 \\
$BBYY$ & 50.000 & 42.871 (1.124) & 199 \\
$BYBY$ & 50.000 & 44.870 (1.141) & 198 \\
$BYYB$ & 50.000 & 57.404 (1.261) & 178 \\
$YBBY$ & 50.000 & 48.191 (1.478) & 184 \\
$YBYB$ & 50.000 & 57.360 (0.923) & 193 \\
$YYBB$ & 50.000 & 57.132 (0.921) & 207 \\
$BYY$ & 60.000 & 55.819 (0.629) & 599 \\
$YBY$ & 60.000 & 58.012 (0.574) & 569 \\
$YYB$ & 60.000 & 60.423 (0.464) & 1242 \\
$BYYY$ & 69.231 & 66.010 (1.806) & 98 \\
$YBYY$ & 69.231 & 65.646 (1.485) & 102 \\
$YYBY$ & 69.231 & 67.820 (1.701) & 112 \\
$YYYB$ & 69.231 & 73.608 (1.334) & 295 \\
\hline\hline
\multicolumn{4}{l}{\footnotesize Standard errors in parentheses.}\\
\end{tabular}
    \begin{minipage}{1\linewidth}
        \footnotesize
        \vspace*{.5em}
        \caption{Beliefs Disaggregated by History: Retractions}
        \vspace*{-1.5em}
        \label{table-beliefs-by-history-baseline-retractions}
        \singlespacing \emph{Notes}: 
        The table shows, (1) for each history of draws, (2) the associated Bayesian Posterior, (3) the average beliefs for our baseline treatment (pooling experiments A and C), and (4) the number of observations.     
        The sample only includes periods $t$ in which the lagged history $\mathcal H_{t-1}$ only includes new draws and a retraction is observed in period $t$. 
        Standard errors in parentheses.
    \end{minipage}
\end{table}

\afterpage{\clearpage}
\newpage 
\subsubsection{Retraction Ineffectiveness: Treatment Effects by History}
\label{online-appendix:table-hte-by-history-beliefs}

\begin{table}[th!]\setstretch{1.1}
	\centering\footnotesize
	\begin{tabular}{l@{\extracolsep{4pt}}cccc@{}}
\hline\hline
Comparison to Retractions & Sign/Compressed History & Signal & Obs & ATE \\
\multicolumn{1}{c}{(1)} & \multicolumn{1}{c}{(2)} & \multicolumn{1}{c}{(3)} & \multicolumn{1}{c}{(4)} & \multicolumn{1}{c}{(5)}\\
\hline
Equivalent New Draw & $BBBY$ & +1/$Y$ & 491 & 0.462 (0.627)\\
Equivalent New Draw & $BBY$ & +1/$Y$ & 1867 & 0.326 (0.161)\\
Equivalent New Draw & $BBYB$ & -1/$B$ & 262 & 1.524 (0.668)\\
Equivalent New Draw & $BBYY$ & +1/$Y$ & 370 & 0.515 (0.261)\\
Equivalent New Draw & $BYB$ & -1/$B$ & 1228 & 0.526 (0.210)\\
Equivalent New Draw & $BYBB$ & -1/$B$ & 254 & 3.344 (0.661)\\
Equivalent New Draw & $BYBY$ & +1/$Y$ & 341 & 0.666 (0.298)\\
Equivalent New Draw & $BYY$ & +1/$Y$ & 1276 & 0.675 (0.168)\\
Equivalent New Draw & $BYYB$ & -1/$B$ & 332 & 1.085 (0.244)\\
Equivalent New Draw & $BYYY$ & +1/$Y$ & 281 & 2.921 (0.707)\\
Equivalent New Draw & $YBB$ & -1/$B$ & 1292 & 0.401 (0.158)\\
Equivalent New Draw & $YBBB$ & -1/$B$ & 268 & 1.947 (0.678)\\
Equivalent New Draw & $YBBY$ & +1/$Y$ & 360 & 0.331 (0.344)\\
Equivalent New Draw & $YBY$ & +1/$Y$ & 1246 & 0.384 (0.129)\\
Equivalent New Draw & $YBYB$ & -1/$B$ & 356 & 0.155 (0.247)\\
Equivalent New Draw & $YBYY$ & +1/$Y$ & 266 & 2.314 (0.689)\\
Equivalent New Draw & $YYB$ & -1/$B$ & 2002 & 0.178 (0.161)\\
Equivalent New Draw & $YYBB$ & -1/$B$ & 403 & 0.363 (0.238)\\
Equivalent New Draw & $YYBY$ & +1/$Y$ & 284 & 1.034 (0.606)\\
Equivalent New Draw & $YYYB$ & -1/$B$ & 507 & 0.121 (0.604)\\
No Retracted Draw & $B$ & -1/$B$ & 6944 & 0.311 (0.104)\\
No Retracted Draw & $B$ & +1/$Y$ & 6947 & 0.585 (0.101)\\
No Retracted Draw & $BB$ & -1/$B$ & 3303 & -0.212 (0.237)\\
No Retracted Draw & $BB$ & +1/$Y$ & 3324 & 3.033 (0.421)\\
No Retracted Draw & $BY$ & -1/$B$ & 3019 & 0.938 (0.186)\\
No Retracted Draw & $BY$ & +1/$Y$ & 3043 & 1.010 (0.191)\\
No Retracted Draw & $Y$ & -1/$B$ & 7183 & 0.373 (0.100)\\
No Retracted Draw & $Y$ & +1/$Y$ & 7109 & 0.112 (0.116)\\
No Retracted Draw & $YB$ & -1/$B$ & 3068 & 0.618 (0.130)\\
No Retracted Draw & $YB$ & +1/$Y$ & 3047 & 0.684 (0.196)\\
No Retracted Draw & $YY$ & -1/$B$ & 3471 & 3.247 (0.495)\\
No Retracted Draw & $YY$ & +1/$Y$ & 3488 & -0.717 (0.242)\\
\hline\hline
\multicolumn{5}{l}{\footnotesize Clustered standard errors at the subject level in parentheses.}\\
\end{tabular}
    \begin{minipage}{1\linewidth}
        \footnotesize
        \vspace*{.5em}
        \caption{Updating from Retractions: Disaggregated by History}
        \vspace*{-2em}
        \label{table-hte-by-history-beliefs}
        \singlespacing \emph{Notes}: 
        The table shows, the average treatment effect of a retraction on beliefs in log-odds ($\hat \ell_t$) for our baseline treatment (pooling experiments A and C) as estimated in \hyref{table-retractions-beliefs}[Table], but disaggregated 
        by compressed history ($C(\mathcal H_t)$) when comparing beliefs with a retraction and without the retracted observation (No Retracted Draw), 
        or 
        by sign history ($S(\mathcal H_t)$) when comparing beliefs with a retraction and with an equivalent new observation (Equivalent New Draw).
        Column (1) determines the comparison (`No Retracted Draw' or `Equivalent New Draw'), Column (2) the sign or compressed history, Column (3) the signal implied by the draw or the retraction, Column (4) the number of observations, and Column (5) the average treatment effect (ATE) with clustered standard errors in parentheses.
        As in \hyref{table-retractions-beliefs}[Table], the sample includes beliefs in log-odds ($\hat \ell_t$) of participants in the baseline treatments, excluding cases in which the truth ball is disclosed and in which there was a retraction in the past. 
        Clustered standard errors at the participant level in parentheses.
    \end{minipage}
\end{table}

\afterpage{\clearpage}
\newpage 
\subsection{Informational Complexity and Retraction Ineffectiveness}
\label{online-appendix:complexity}

This section supports \hyref{section:complexity}[Section] of the paper. 
In \hyref{online-appendix:variance-proof}[Section], we provide a short proof validating our measure of complexity based on the variability of belief reports, as referenced in \hyref{section:complexity:indicators}[Section]. 
In \hyref{online-appendix:complexity:regression-tables}[Section], we present tables corresponding to \hyref{figure-ate-complex}[Figure] of \hyref{section:complexity:variation}[Section].  

\subsubsection{Conditions for Higher Variability with Higher Complexity}
\label{online-appendix:variance-proof}
We recall that, in our model in \hyref{section:complexity:indicators}[Section],
posterior log-odds updates are given by
\[
    \hat \ell_t
    = 
    \hat \ell_{t-1} 
    + \beta K(E)+\beta\frac{\sigma_\zeta}{\sqrt{T}} \zeta,
\]
with $\beta=(1+\sigma_\zeta^2/(\sigma^2 T))^{-1}$ and $\zeta\sim\mathcal N(0,1)$. 
Further note that $\frac{\partial}{\partial \sigma_\zeta^2}\beta<0$ implies that $T'(\sigma_\zeta^2)\frac{\sigma_\zeta^2}{T(\sigma_\zeta^2)}<1$, i.e., that the elasticity of $T$ with respect to $\sigma_\zeta^2$ is lower than 1.
From the prediction that decision times increase with complexity, $T'(\sigma_\zeta^2)>0$, we obtain that $T$ is inelastic with respect to $\sigma_\zeta^2$.
Then, given $\hat\ell_{t-1}$ and $K(E)$, the variance of the posterior log-odds beliefs is 
$$\text{Var}(\hat \ell_t)=\beta^2 \frac{\sigma_\zeta^2}{T(\sigma_\zeta^2)} = \sigma^2\frac{ \frac{\sigma_\zeta^2}{T(\sigma_\zeta^2)}}{\left(\frac{\sigma_\zeta^2}{T(\sigma_\zeta^2)}+\sigma^2\right)^2}.$$
Hence, 
$$\frac{\partial}{\partial\sigma_\zeta^2}\text{Var}(\hat \ell_t)=\frac{\sigma^2 }{T(\sigma_\zeta^2) {\frac{\sigma_\zeta^2}{T(\sigma_\zeta^2)}+\sigma^2}^3}\left(\sigma^2-\frac{\sigma_\zeta^2}{T(\sigma_\zeta^2)}\right)\left(1-T'(\sigma_\zeta^2)\frac{\sigma_\zeta^2}{T(\sigma_\zeta^2)}\right)>0\Longrightarrow
\sigma^{2}/2>{(T(\sigma_\zeta^2)\sigma_\zeta^{-2} + \sigma^{-2})}^{-1}.$$

\afterpage{\clearpage}
\newpage 

\subsubsection{Validating and Varying Complexity}
\label{online-appendix:complexity:regression-tables}

We present tables for \hyref{figure-ate-complex}[Figure].

\begin{table}[th!]\setstretch{1.1}
	\centering\footnotesize
	\rotatebox{90}{\begin{tabular}{l@{\extracolsep{2pt}}cccccccc@{}}
\hline\hline
\multicolumn{1}{r}{Retraction vs.} &\multicolumn{4}{c}{No Retracted Draw}  &  \multicolumn{4}{c}{Equivalent New Draw} \\
\cline{2-5} \cline{6-9}
& (1) & (2) & (3) & (4) & (5) & (6) & (7) & (8) \\
& $\hat \ell_t$ & $|\hat p_t-p_t|$ & ln(T$_t$) & Var($\hat \ell_t \mid h_t$) & $\hat \ell_t$ & $|\hat p_t-p_t|$ & ln(T$_t$) & Var($\hat \ell_t \mid h_t$) \\
\hline
Retraction ($r_t$) & 0.006 & 2.244$^{***}$ & 0.038$^{**}$ & 0.676$^{***}$ & 0.014 & 1.286$^{***}$ & 0.065$^{***}$ & 0.288 \\
 & (0.027) & (0.284) & (0.015) & (0.239) & (0.038) & (0.336) & (0.018) & (0.223) \\ [.1em]
Retracted Draw ($r_t\cdot K(s_{\rho_t})$) & 0.392$^{***}$ & -- & -- & -- & 0.432$^{***}$ & -- & -- & -- \\
 & (0.083) &  &  &  & (0.096) &  &  &  \\ [.1em]

Mean Decision Time     &   &   & 8.774 &   &   &   & 8.774 &   \\
Compressed History FEs & Yes & Yes & Yes & Yes & No & No & No & No  \\
Sign History FEs & No & No & No & No & Yes & Yes & Yes & Yes  \\
\hline
R$^2$ & 0.27 & 0.08 & 0.01 & 0.03 & 0.28 & 0.08 & 0.01 & 0.03 \\
N & 35211 & 35211 & 35211 & 4643 & 35211 & 35211 & 35211 & 4643 \\
\hline\hline
\multicolumn{9}{l}{\footnotesize Clustered standard errors at the subject level in parentheses.}\\
\multicolumn{9}{l}{\footnotesize $^{*}$ \(p<0.1\), $^{**}$ \(p<0.05\), $^{***}$ \(p<0.01\)}\\
\end{tabular}}
    \begin{minipage}{1\linewidth}
        \footnotesize
        \vspace*{.5em}
        \caption{Treatment Effects: Retract Last Draw}
        \vspace*{-1.5em}
        \label{table-ate-last}
        \singlespacing \emph{Notes}: 
        This table reports the effect of retractions on updating and empirical complexity measures when the last ball is retracted, corresponding to ``Retract Last Draw'' of \hyref{figure-ate-complex}[Figure]. 
        Columns (1) and (5) are the regressions from \hyref{table-retractions-beliefs}[Table], but restricted to retractions of the last draw. 
        Similarly, Columns (2)-(4) and (6)-(8) are the regressions from \hyref{table-ate-complexity}[Table]. 
    \end{minipage}
\end{table}

\begin{table}[th!]\setstretch{1.1}
	\centering\footnotesize
	\rotatebox{90}{\begin{tabular}{l@{\extracolsep{4pt}}cccc@{}}
\hline\hline
\multicolumn{1}{r}{Retraction vs.}  &  \multicolumn{4}{c}{Equivalent New Draw} \\
\cline{2-5}
& (1) & (2) & (3) & (4) \\
& $\hat \ell_t$ & $|\hat p_t-p_t|$ & ln(T$_t$) & Var($\hat \ell_t \mid h_t$) \\
\hline
Retraction ($r_{t-1}$) & 0.026 & 2.270$^{***}$ & 0.082$^{***}$ & 0.803$^{**}$ \\
 & (0.057) & (0.455) & (0.021) & (0.390) \\ [.1em]
Retraction $\times$ Signal ($r_{t-1}\cdot K(s_{t})$) & 0.751$^{***}$ & -- & -- & -- \\
 & (0.162) &  &  &  \\ [.1em]

Mean Decision Time     &   &   & 8.794 &   \\
Sign History FEs & Yes & Yes & Yes & Yes  \\
\hline
R$^2$ & 0.26 & 0.08 & 0.01 & 0.03 \\
N & 39168 & 39168 & 39168 & 4494 \\
\hline\hline
\multicolumn{5}{l}{\footnotesize Clustered standard errors at the subject level in parentheses.}\\
\multicolumn{5}{l}{\footnotesize $^{*}$ \(p<0.1\), $^{**}$ \(p<0.05\), $^{***}$ \(p<0.01\)}\\
\end{tabular}}
    \begin{minipage}{1\linewidth}
        \footnotesize
        \vspace*{.5em}
        \caption{Treatment Effects: After Retractions}
        \vspace*{-1.5em}
        \label{table-ate-after}
        \singlespacing \emph{Notes}: 
        This table reports the effect of retractions on updating from subsequent signals, corresponding to ``After Retractions'' of \hyref{figure-ate-complex}[Figure].  
    \end{minipage}
\end{table}

We also exhibit the heterogeneous treatment effects for the retraction of the last draw, compared to retracting an earlier one.

\begin{table}[th!]\setstretch{1.1}
	\centering\footnotesize
	\rotatebox{90}{\begin{tabular}{l@{\extracolsep{4pt}}cccc@{}}
\hline\hline
\multicolumn{1}{r}{Retraction vs.} &\multicolumn{4}{c}{No Retracted Draw} \\
\cline{2-5}
& (1) & (2) & (3) & (4) \\
& $\hat \ell_t$ & $|\hat p_t-p_t|$ & ln(T$_t$) & Var($\hat \ell_t \mid h_t$)  \\
\hline
Retraction ($r_t$) & 0.014 & 3.098$^{***}$ & 0.080$^{***}$ & 1.030$^{***}$ \\
 & (0.024) & (0.313) & (0.014) & (0.168) \\ [.1em]
Retracted Draw ($r_t\cdot K(s_{\rho_t})$) & 0.740$^{***}$ & -- & -- & -- \\
 & (0.079) &  &  &  \\ [.1em]
Retraction of Last Draw ($1\{\rho_t=t-1\}$) & -0.008 & -0.760$^{***}$ & -0.037$^{**}$ & -0.231 \\
 & (0.037) & (0.284) & (0.017) & (0.219) \\ [.1em]
Retraction of Last Draw x Retracted Draw ($1\{\rho_t=t-1\}\,K(s_{\rho})$) & -0.348$^{***}$ & -- & -- & -- \\
 & (0.091) &  &  &  \\ [.1em]

Mean Decision Time     &   &   & 8.830 &   \\
Compressed History FEs & Yes & Yes & Yes & Yes  \\
\hline
R$^2$ & 0.27 & 0.08 & 0.01 & 0.02 \\
N & 39162 & 39162 & 39162 & 5709 \\
\hline\hline
\multicolumn{5}{l}{\footnotesize Clustered standard errors at the subject level in parentheses.}\\
\multicolumn{5}{l}{\footnotesize $^{*}$ \(p<0.1\), $^{**}$ \(p<0.05\), $^{***}$ \(p<0.01\)}\\
\end{tabular}}
    \begin{minipage}{1\linewidth}
        \footnotesize
        \vspace*{.5em}
        \caption{Heterogeneous Treatment Effects: Retract Last Draw}
        \vspace*{-1.5em}
        \label{table-hte-last}
        \singlespacing \emph{Notes}: 
        This table reports heterogeneous treatment effects of retractions on updating and empirical complexity measures, based on whether the last ball is retracted. 
        While \hyref{table-ate-last}[Table] considers the treatment effect restricted to when the last draw is retracted, here we use the full baseline sample and report treatment effect heterogeneity by it.
    \end{minipage}
\end{table}

\afterpage{\clearpage}
\newpage 

\subsection{Belief Updating Patterns under Retractions}
\label{online-appendix:different-biases}

This section supports \hyref{section:different-biases}[Section] of the paper. 
In \hyref{online-appendix:different-biases:regression-tables}[Section], we present the results from \hyref{table-grether}[Table], broken down by platform (MTurk vs. Prolific). 
In \hyref{online-appendix:different-biases:figures}[Section], we examine heterogeneous treatment effects on our measures of complexity across histories, following the same disaggregation as in \hyref{figure-ate-beliefs-sh}[Figure].

\subsubsection{Regression Tables}
\label{online-appendix:different-biases:regression-tables}

\begin{table}[th!]\setstretch{1.1}
	\centering\footnotesize
	\rotatebox{0}{\begin{tabular}{l@{\extracolsep{4pt}}cccccc@{}}
\hline\hline
&\multicolumn{2}{c}{Baseline} & \multicolumn{2}{c}{MTurk} & \multicolumn{2}{c}{Prolific} \\
\cline{2-3} \cline{4-5} \cline{6-7}
& (1) & (2) & (3) & (4) & (5) & (6) \\
& $\hat \ell_t$ & $\hat \ell_t$ & $\hat \ell_t$ & $\hat \ell_t$ & $\hat \ell_t$ & $\hat \ell_t$ \\
\hline
Signal ($K_t$) & 1.102$^{***}$ & 0.907$^{***}$ & 1.126$^{***}$ & 0.998$^{***}$ & 1.066$^{***}$ & 0.788$^{***}$ \\
 & (0.060) & (0.060) & (0.071) & (0.072) & (0.102) & (0.102) \\ [.1em]
Prior ($l_{t-1}$) & 0.801$^{***}$ & 0.747$^{***}$ & 0.834$^{***}$ & 0.800$^{***}$ & 0.742$^{***}$ & 0.664$^{***}$ \\
 & (0.032) & (0.032) & (0.037) & (0.037) & (0.052) & (0.044) \\ [.1em]
Confirmatory Signal ($K_t \cdot c_t$) & -- & 0.650$^{***}$ & -- & 0.417$^{***}$ & -- & 0.958$^{***}$ \\
 &  & (0.097) &  & (0.135) &  & (0.128) \\ [.1em]
\hline
R$^2$ & 0.38 & 0.38 & 0.41 & 0.41 & 0.33 & 0.34 \\
N & 32064 & 32064 & 18491 & 18491 & 13573 & 13573 \\
\hline\hline
\multicolumn{7}{l}{\footnotesize Clustered standard errors at the subject level in parentheses.}\\
\multicolumn{7}{l}{\footnotesize $^{*}$ \(p<0.1\), $^{**}$ \(p<0.05\), $^{***}$ \(p<0.01\)}\\
\end{tabular}}
    \begin{minipage}{1\linewidth}
        \footnotesize
        \vspace*{.5em}
        \caption{Robustness of Grether Regressions by Experimental Platform}
        \vspace*{-1.5em}
        \label{table-grether-byplatform}
        \singlespacing \emph{Notes}: 
        This table disaggregates the results from \hyref{table-grether}[Table] by experimental platform.
    \end{minipage}
\end{table}

\afterpage{\clearpage}
\newpage 

\subsubsection{Figures}
\label{online-appendix:different-biases:figures}

\begin{figure}[h!]\setstretch{1.1}
    \centering\footnotesize
    \begin{subfigure}{.495\textwidth}
        \centering
        \includegraphics[width=1\linewidth]{Figures/figure-ate-beliefs-sh.pdf}
        \vspace*{-2em}
        \caption{Beliefs}
    \end{subfigure}
    \begin{subfigure}{.495\textwidth} 
        \centering
        \includegraphics[width=1\linewidth]{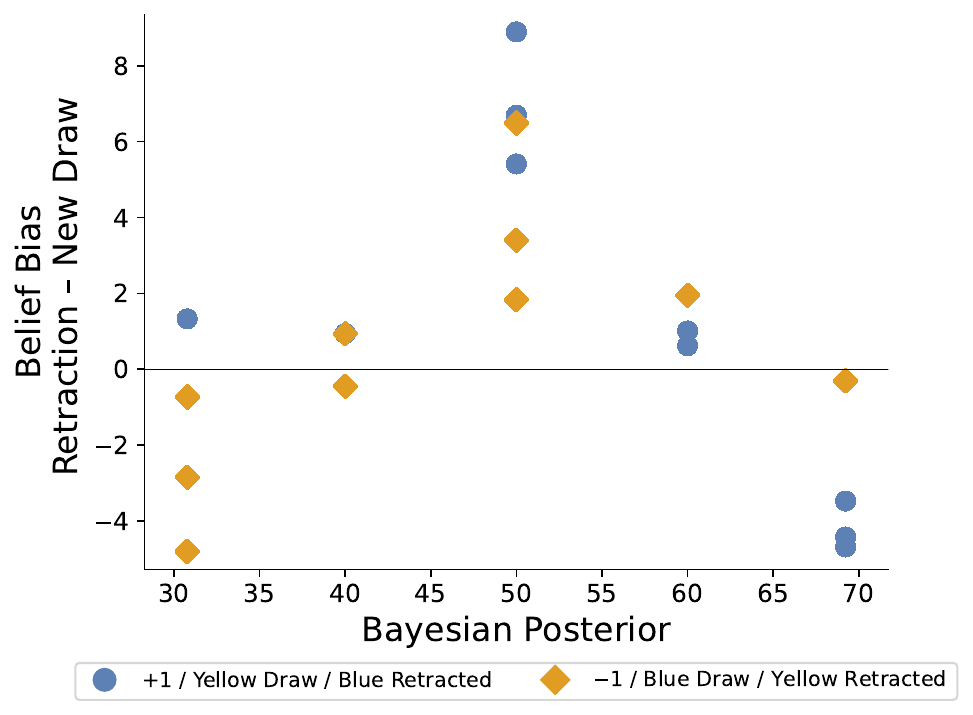}
        \vspace*{-2em}
        \caption{Accuracy}
    \end{subfigure}
    \begin{subfigure}{.495\textwidth}
        \centering
        \includegraphics[width=1\linewidth]{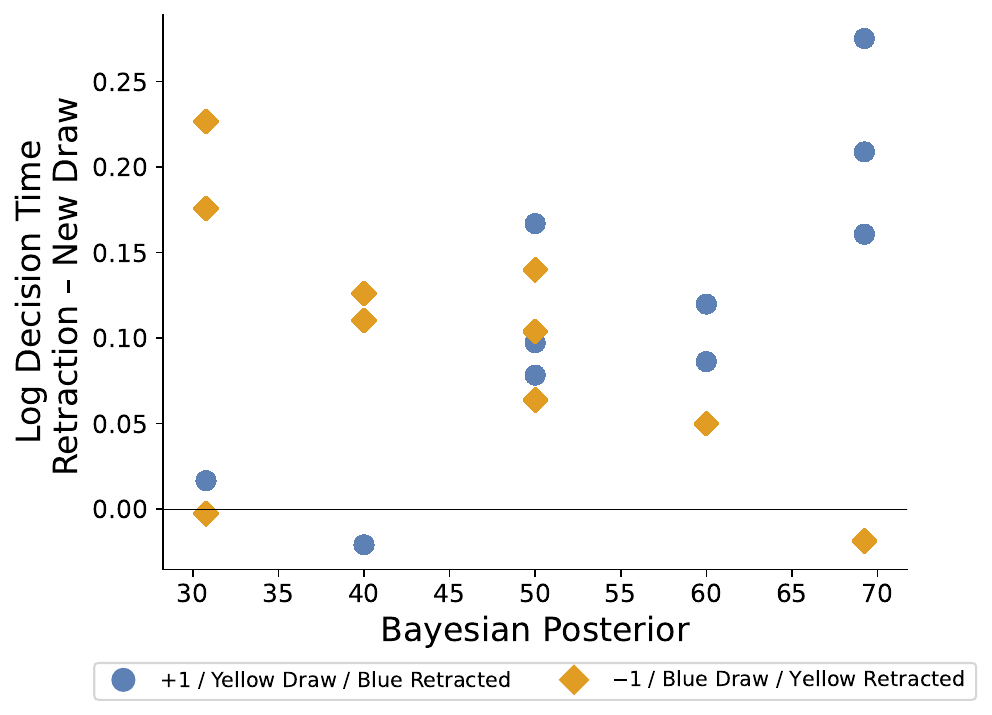}
        \vspace*{-2em}
        \caption{Speed}
    \end{subfigure}
    \begin{subfigure}{.495\textwidth} 
        \centering
        \includegraphics[width=1\linewidth]{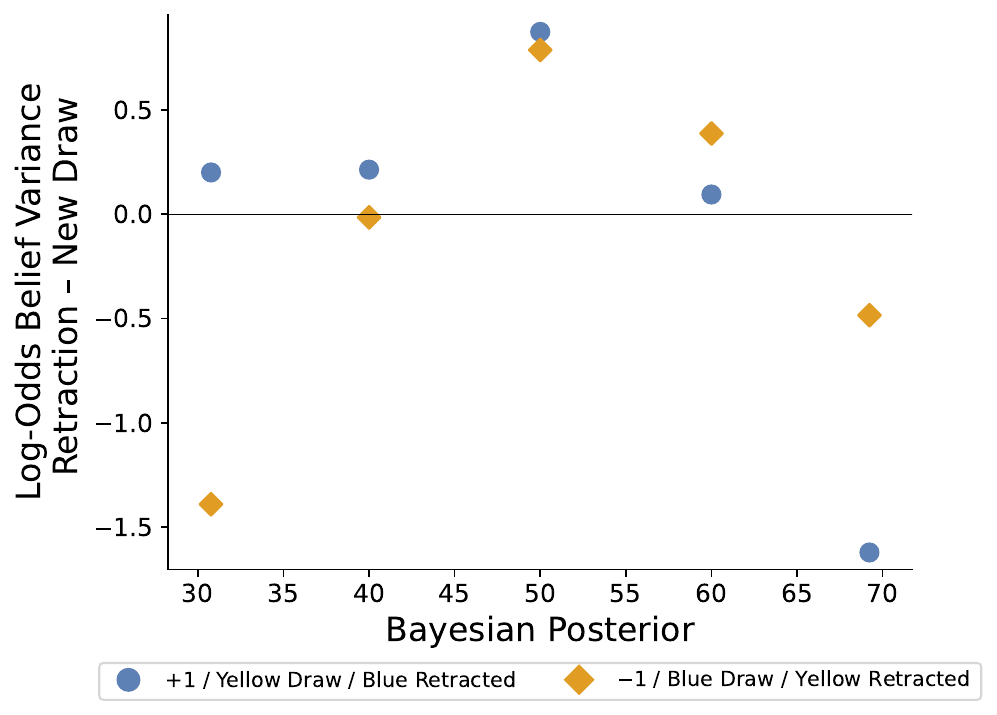}
        \vspace*{-2em}
        \caption{Variability}
    \end{subfigure}
    \begin{minipage}{1\linewidth}
        \footnotesize
        \vspace*{.5em}
        \caption{Updating from Retractions: Heterogeneity by History}
        \label{figure-ate-sh}
        \vspace*{-1em}
        \footnotesize
        \singlespacing \emph{Notes}: 
        This figure compares updating from retractions versus equivalent new draws, our test (b), disaggregated by sign history (i.e. corresponding to the disaggregation in \hyref{figure-ate-beliefs-sh}[Figure]). 
        Blue circles represent sign histories in which the last signal was a retraction of a blue draw, or a new yellow draw. 
        Orange diamonds represent those in which the last signal was a retraction of a yellow draw, or a new blue draw. 
        In both cases, the x-axis is the Bayesian posterior of the sign history. 
        Panel (a) displays the effect of retractions on belief updating, $\hat \ell_t$. 
        Panels (b)-(d) display effects on our three complexity indicators---accuracy ($|\hat p_t-p_t|$), speed (ln(T$_t$)), and variability (Var($\hat \ell_t \mid h_t$)). 
        The sample includes all observations of participants in the baseline treatment, excluding periods in which the truth ball is disclosed or in which there was a retraction in an earlier period. 
     \end{minipage}
\end{figure}

\afterpage{\clearpage}
\newpage

\subsection{Robustness 1: Participant Screening and Understanding}
\label{online-appendix:robustness:understanding}

This section supports \hyref{section:robustness:understanding}[Section] of the paper. 
In \hyref{online-appendix:robustness:understanding:regression-tables}[Section], we report regression tables corresponding to \hyref{figure-ate-robustness-understanding}[Figure]. 
In \hyref{online-appendix:robustness:understanding:confidence}[Section] we provide how confidence relates to the strength of updating to the kind of information participants are presented with. 
In \hyref{online-appendix:robustness:understanding:info}[Section], we disclose the estimates for heterogeneous treatment effects regarding the provision of additional information about retractions, comparing our baseline treatment and the ``Retraction Information'' treatment from experiment C.

\subsubsection{Regression Tables Corresponding to \hyref{figure-ate-robustness-understanding}[Figure]}

\label{online-appendix:robustness:understanding:regression-tables}

We include a table for each of the robustness checks presented in the figure, each of which presents our main results on updating and empirical complexity measures but restricting the sample considering various dimensions of demonstrated understanding.

\newpage

\paragraph{Comprehension Questionnaire Correct at 1st Try}
\label{online-appendix:robustness:understanding:regression-tables:comp}

\begin{table}[th!]\setstretch{1.1}
	\centering\footnotesize
	\rotatebox{90}{\begin{tabular}{l@{\extracolsep{4pt}}cccccccc@{}}
\hline\hline
\multicolumn{1}{r}{Retraction vs.} &\multicolumn{4}{c}{No Retracted Draw}  &  \multicolumn{4}{c}{Equivalent New Draw} \\
\cline{2-5} \cline{6-9}
& (1) & (2) & (3) & (4) & (5) & (6) & (7) & (8) \\
& $\hat \ell_t$ & $|\hat p_t-p_t|$ & ln(T$_t$) & Var($\hat \ell_t \mid h_t$) & $\hat \ell_t$ & $|\hat p_t-p_t|$ & ln(T$_t$) & Var($\hat \ell_t \mid h_t$) \\
\hline
Retraction ($r_t$) & -0.004 & 2.635$^{***}$ & 0.068$^{***}$ & 1.014$^{***}$ & -0.036$^{*}$ & 1.643$^{***}$ & 0.090$^{***}$ & 0.638$^{***}$ \\
 & (0.016) & (0.320) & (0.015) & (0.175) & (0.022) & (0.340) & (0.017) & (0.190) \\ [.1em]
Retracted Draw ($r_t\cdot K(s_{\rho_t})$) & 0.596$^{***}$ & -- & -- & -- & 0.642$^{***}$ & -- & -- & -- \\
 & (0.076) &  &  &  & (0.086) &  &  &  \\ [.1em]

Mean Decision Time     &   &   & 8.688 &   &   &   & 8.688 &   \\
Compressed History FEs & Yes & Yes & Yes & Yes & No & No & No & No  \\
Sign History FEs & No & No & No & No & Yes & Yes & Yes & Yes  \\
\hline
R$^2$ & 0.33 & 0.09 & 0.01 & 0.05 & 0.34 & 0.09 & 0.01 & 0.06 \\
N & 23110 & 23110 & 23110 & 3079 & 23110 & 23110 & 23110 & 3079 \\
\hline\hline
\multicolumn{9}{l}{\footnotesize Clustered standard errors at the subject level in parentheses.}\\
\multicolumn{9}{l}{\footnotesize $^{*}$ \(p<0.1\), $^{**}$ \(p<0.05\), $^{***}$ \(p<0.01\)}\\
\end{tabular}}
    \begin{minipage}{1\linewidth}
        \footnotesize
        \vspace*{.5em}
        \caption{Treatment Effects: Comprehension Correct}
        \vspace*{-1.5em}
        \label{table-ate-comp}
        \singlespacing \emph{Notes}: 
        This table reports the effect of retractions on updating and empirical complexity measures when restricting the baseline sample to participants who answered all experimental comprehension questions, corresponding to ``Comprehension Correct'' of \hyref{figure-ate-robustness-understanding}[Figure]. 
        Columns (1) and (5) are the regressions from \hyref{table-retractions-beliefs}[Table], but restricted to this sample. 
        Similarly, Columns (2)-(4) and (6)-(8) are the regressions from \hyref{table-ate-complexity}[Table]. 
    \end{minipage}
\end{table}

\afterpage{\clearpage}
\newpage 
\paragraph{Understands Disclosure of Truth Ball}
\label{online-appendix:robustness:understanding:regression-tables:revelation}

\begin{table}[th!]\setstretch{1.1}
	\centering\footnotesize
	\rotatebox{90}{\begin{tabular}{l@{\extracolsep{4pt}}cccccccc@{}}
\hline\hline
\multicolumn{1}{r}{Retraction vs.} &\multicolumn{4}{c}{No Retracted Draw}  &  \multicolumn{4}{c}{Equivalent New Draw} \\
\cline{2-5} \cline{6-9}
& (1) & (2) & (3) & (4) & (5) & (6) & (7) & (8) \\
& $\hat \ell_t$ & $|\hat p_t-p_t|$ & ln(T$_t$) & Var($\hat \ell_t \mid h_t$) & $\hat \ell_t$ & $|\hat p_t-p_t|$ & ln(T$_t$) & Var($\hat \ell_t \mid h_t$) \\
\hline
Retraction ($r_t$) & -0.007 & 2.450$^{***}$ & 0.078$^{***}$ & 0.856$^{***}$ & -0.034 & 1.798$^{***}$ & 0.098$^{***}$ & 0.458$^{**}$ \\
 & (0.017) & (0.355) & (0.017) & (0.191) & (0.024) & (0.379) & (0.020) & (0.214) \\ [.1em]
Retracted Draw ($r_t\cdot K(s_{\rho_t})$) & 0.541$^{***}$ & -- & -- & -- & 0.552$^{***}$ & -- & -- & -- \\
 & (0.079) &  &  &  & (0.088) &  &  &  \\ [.1em]

Mean Decision Time     &   &   & 8.413 &   &   &   & 8.413 &   \\
Compressed History FEs & Yes & Yes & Yes & Yes & No & No & No & No  \\
Sign History FEs & No & No & No & No & Yes & Yes & Yes & Yes  \\
\hline
R$^2$ & 0.39 & 0.11 & 0.01 & 0.04 & 0.40 & 0.11 & 0.02 & 0.05 \\
N & 17958 & 17958 & 17958 & 2402 & 17958 & 17958 & 17958 & 2402 \\
\hline\hline
\multicolumn{9}{l}{\footnotesize Clustered standard errors at the subject level in parentheses.}\\
\multicolumn{9}{l}{\footnotesize $^{*}$ \(p<0.1\), $^{**}$ \(p<0.05\), $^{***}$ \(p<0.01\)}\\
\end{tabular}}
    \begin{minipage}{1\linewidth}
        \footnotesize
        \vspace*{.5em}
        \caption{Treatment Effects: Understand Disclosure}
        \vspace*{-1.5em}
        \label{table-ate-revelation}
        \singlespacing \emph{Notes}: 
        This table reports the effect of retractions on updating and empirical complexity measures when restricting the baseline sample to participants who, when the state is revealed, correctly report that they know the state, corresponding to ``Understand Disclosure'' of \hyref{figure-ate-robustness-understanding}[Figure]. 
        Columns (1) and (5) are the regressions from \hyref{table-retractions-beliefs}[Table], but restricted to this sample. 
        Similarly, Columns (2)-(4) and (6)-(8) are the regressions from \hyref{table-ate-complexity}[Table]. 
    \end{minipage}
\end{table}

\afterpage{\clearpage}
\newpage 
\paragraph{Few Mistakes}
\label{online-appendix:robustness:understanding:regression-tables:mistakes}

\begin{table}[th!]\setstretch{1.1}
	\centering\footnotesize
	\rotatebox{90}{\begin{tabular}{l@{\extracolsep{4pt}}cccccccc@{}}
\hline\hline
\multicolumn{1}{r}{Retraction vs.} &\multicolumn{4}{c}{No Retracted Draw}  &  \multicolumn{4}{c}{Equivalent New Draw} \\
\cline{2-5} \cline{6-9}
& (1) & (2) & (3) & (4) & (5) & (6) & (7) & (8) \\
& $\hat \ell_t$ & $|\hat p_t-p_t|$ & ln(T$_t$) & Var($\hat \ell_t \mid h_t$) & $\hat \ell_t$ & $|\hat p_t-p_t|$ & ln(T$_t$) & Var($\hat \ell_t \mid h_t$) \\
\hline
Retraction ($r_t$) & -0.044$^{**}$ & 2.981$^{***}$ & 0.072$^{***}$ & 0.944$^{***}$ & -0.047$^{*}$ & 1.650$^{***}$ & 0.100$^{***}$ & 0.708$^{***}$ \\
 & (0.018) & (0.392) & (0.018) & (0.190) & (0.027) & (0.394) & (0.021) & (0.191) \\ [.1em]
Retracted Draw ($r_t\cdot K(s_{\rho_t})$) & 0.674$^{***}$ & -- & -- & -- & 0.763$^{***}$ & -- & -- & -- \\
 & (0.090) &  &  &  & (0.108) &  &  &  \\ [.1em]

Mean Decision Time     &   &   & 8.449 &   &   &   & 8.449 &   \\
Compressed History FEs & Yes & Yes & Yes & Yes & No & No & No & No  \\
Sign History FEs & No & No & No & No & Yes & Yes & Yes & Yes  \\
\hline
R$^2$ & 0.44 & 0.13 & 0.01 & 0.04 & 0.44 & 0.14 & 0.02 & 0.04 \\
N & 17357 & 17357 & 17357 & 2332 & 17357 & 17357 & 17357 & 2332 \\
\hline\hline
\multicolumn{9}{l}{\footnotesize Clustered standard errors at the subject level in parentheses.}\\
\multicolumn{9}{l}{\footnotesize $^{*}$ \(p<0.1\), $^{**}$ \(p<0.05\), $^{***}$ \(p<0.01\)}\\
\end{tabular}}
    \begin{minipage}{1\linewidth}
        \footnotesize
        \vspace*{.5em}
        \caption{Treatment Effects: Few Mistakes}
        \vspace*{-1.5em}
        \label{table-ate-mistakes}
        \singlespacing \emph{Notes}: 
       This table reports the effect of retractions on updating and empirical complexity measures when restricting the baseline sample to participants who update in the opposite direction to the signal more than 10\% of the time, corresponding to ``Few Mistakes'' of \hyref{figure-ate-robustness-understanding}[Figure]. 
       Columns (1) and (5) are the regressions from \hyref{table-retractions-beliefs}[Table], but restricted to this sample. 
       Similarly, Columns (2)-(4) and (6)-(8) are the regressions from \hyref{table-ate-complexity}[Table]. 
    \end{minipage}
\end{table}

\afterpage{\clearpage}
\newpage 
\paragraph{Understands Replacement}
\label{online-appendix:robustness:understanding:regression-tables:replacement}

\begin{table}[th!]\setstretch{1.1}
	\centering\footnotesize
	\rotatebox{90}{\begin{tabular}{l@{\extracolsep{4pt}}cccccccc@{}}
\hline\hline
\multicolumn{1}{r}{Retraction vs.} &\multicolumn{4}{c}{No Retracted Draw}  &  \multicolumn{4}{c}{Equivalent New Draw} \\
\cline{2-5} \cline{6-9}
& (1) & (2) & (3) & (4) & (5) & (6) & (7) & (8) \\
& $\hat \ell_t$ & $|\hat p_t-p_t|$ & ln(T$_t$) & Var($\hat \ell_t \mid h_t$) & $\hat \ell_t$ & $|\hat p_t-p_t|$ & ln(T$_t$) & Var($\hat \ell_t \mid h_t$) \\
\hline
Retraction ($r_t$) & -0.000 & 2.551$^{***}$ & 0.057$^{***}$ & 0.964$^{***}$ & -0.027 & 1.311$^{***}$ & 0.079$^{***}$ & 0.405$^{***}$ \\
 & (0.017) & (0.268) & (0.012) & (0.164) & (0.025) & (0.299) & (0.014) & (0.155) \\ [.1em]
Retracted Draw ($r_t\cdot K(s_{\rho_t})$) & 0.482$^{***}$ & -- & -- & -- & 0.507$^{***}$ & -- & -- & -- \\
 & (0.063) &  &  &  & (0.085) &  &  &  \\ [.1em]

Mean Decision Time     &   &   & 8.925 &   &   &   & 8.925 &   \\
Compressed History FEs & Yes & Yes & Yes & Yes & No & No & No & No  \\
Sign History FEs & No & No & No & No & Yes & Yes & Yes & Yes  \\
\hline
R$^2$ & 0.23 & 0.06 & 0.01 & 0.03 & 0.24 & 0.07 & 0.01 & 0.04 \\
N & 36059 & 36059 & 36059 & 4820 & 36059 & 36059 & 36059 & 4820 \\
\hline\hline
\multicolumn{9}{l}{\footnotesize Clustered standard errors at the subject level in parentheses.}\\
\multicolumn{9}{l}{\footnotesize $^{*}$ \(p<0.1\), $^{**}$ \(p<0.05\), $^{***}$ \(p<0.01\)}\\
\end{tabular}}
    \begin{minipage}{1\linewidth}
        \footnotesize
        \vspace*{.5em}
        \caption{Treatment Effects: Understands Replacement}
        \vspace*{-1.5em}
        \label{table-ate-replacement}
        \singlespacing \emph{Notes}: 
        This table reports the effect of retractions on updating and empirical complexity measures when excluding participants from the baseline sample excludes participants who could be mistaking sampling with and without replacement, corresponding to ``Understands Replacement'' of \hyref{figure-ate-robustness-understanding}[Figure]. 
        Columns (1) and (5) are the regressions from \hyref{table-retractions-beliefs}[Table], but restricted to this sample. 
        Similarly, Columns (2)-(4) and (6)-(8) are the regressions from \hyref{table-ate-complexity}[Table]
    \end{minipage}
\end{table}

\afterpage{\clearpage}
\newpage 
\paragraph{Retraction Information}
\label{online-appendix:robustness:understanding:regression-tables:info}

\begin{table}[th!]\setstretch{1.1}
	\centering\footnotesize
	\rotatebox{90}{\begin{tabular}{l@{\extracolsep{4pt}}cccccccc@{}}
\hline\hline
\multicolumn{1}{r}{Retraction vs.} &\multicolumn{4}{c}{No Retracted Draw}  &  \multicolumn{4}{c}{Equivalent New Draw} \\
\cline{2-5} \cline{6-9}
& (1) & (2) & (3) & (4) & (5) & (6) & (7) & (8) \\
& $\hat \ell_t$ & $|\hat p_t-p_t|$ & ln(T$_t$) & Var($\hat \ell_t \mid h_t$) & $\hat \ell_t$ & $|\hat p_t-p_t|$ & ln(T$_t$) & Var($\hat \ell_t \mid h_t$) \\
\hline
Retraction ($r_t$) & 0.002 & 1.731$^{***}$ & 0.098$^{***}$ & 0.752$^{***}$ & 0.015 & 0.625 & 0.087$^{***}$ & 0.157 \\
 & (0.022) & (0.344) & (0.017) & (0.199) & (0.030) & (0.386) & (0.018) & (0.234) \\ [.1em]
Retracted Draw ($r_t\cdot K(s_{\rho_t})$) & 0.580$^{***}$ & -- & -- & -- & 0.856$^{***}$ & -- & -- & -- \\
 & (0.091) &  &  &  & (0.121) &  &  &  \\ [.1em]

Mean Decision Time     &   &   & 12.256 &   &   &   & 12.256 &   \\
Compressed History FEs & Yes & Yes & Yes & Yes & No & No & No & No  \\
Sign History FEs & No & No & No & No & Yes & Yes & Yes & Yes  \\
\hline
R$^2$ & 0.30 & 0.09 & 0.01 & 0.04 & 0.29 & 0.09 & 0.01 & 0.04 \\
N & 17553 & 17553 & 17553 & 2338 & 17553 & 17553 & 17553 & 2338 \\
\hline\hline
\multicolumn{9}{l}{\footnotesize Clustered standard errors at the subject level in parentheses.}\\
\multicolumn{9}{l}{\footnotesize $^{*}$ \(p<0.1\), $^{**}$ \(p<0.05\), $^{***}$ \(p<0.01\)}\\
\end{tabular}}
    \begin{minipage}{1\linewidth}
        \footnotesize
        \vspace*{.5em}
        \caption{Treatment Effects: Understands Replacement}
        \vspace*{-1.5em}
        \label{table-ate-info}
        \singlespacing \emph{Notes}: 
        This table reports the effect of retractions on updating and empirical complexity measures in the treatment conveying additional retraction information in experiment C, corresponding to ``Retraction Info'' of \hyref{figure-ate-robustness-understanding}[Figure]. 
        Columns (1) and (5) are the regressions from \hyref{table-retractions-beliefs}[Table], but restricted to this sample. 
        Similarly, Columns (2)-(4) and (6)-(8) are the regressions from \hyref{table-ate-complexity}[Table]
    \end{minipage}
\end{table}

\afterpage{\clearpage}
\newpage 
\subsubsection{Participant Confidence}
\label{online-appendix:robustness:understanding:confidence}

\begin{table}[th!]\setstretch{1.1}
	\centering\footnotesize
    \rotatebox{0}{\begin{tabular}{l@{\extracolsep{1pt}}cc@{}}
\hline\hline
& (1) & (2) \\
& $\hat \ell_t$ & $\hat \ell_t$ \\
\hline
Signal ($K(s_t)$) & 1.008$^{***}$ & 0.801$^{***}$ \\
 & (0.172) & (0.186) \\ [.1em]
Prior ($l_{t-1}$) & 0.672$^{***}$ & 0.620$^{***}$ \\
 & (0.060) & (0.047) \\ [.1em]
Confirmatory Signal ($K(s_t) \cdot c_t$) & -- & 0.700$^{***}$ \\
 &  & (0.171) \\ [.1em]
High Confidence x Signal ($K(s_t)$) & 0.114 & -0.023 \\
 & (0.200) & (0.213) \\ [.1em]
High Confidence x Prior ($l_{t-1}$) & 0.153$^{*}$ & 0.097 \\
 & (0.082) & (0.075) \\ [.1em]
High Confidence x Confirmatory Signal ($K(s_t) \cdot c_t$) & -- & 0.513$^{*}$ \\
 &  & (0.266) \\ [.1em]

\hline
R$^2$ & 0.33 & 0.34 \\
N & 13573 & 13573 \\
\hline\hline
\multicolumn{3}{l}{\footnotesize Clustered standard errors at the subject level in parentheses.}\\
\multicolumn{3}{l}{\footnotesize $^{*}$ \(p<0.1\), $^{**}$ \(p<0.05\), $^{***}$ \(p<0.01\)}\\
\end{tabular}}
    \begin{minipage}{1\linewidth}
        \footnotesize
        \vspace*{.5em}
        \caption{Belief Updating Patterns and Between-Participant Confidence: Grether Regressions}
        \vspace*{-1.5em}
        \label{table-grether-hte-chigh}
        \singlespacing \emph{Notes}: 
        This table examines how between-participant heterogeneity relates to patterns in belief updating from new draws. 
        It reports estimates of \hyref{equation:grether-retractions}[Equation] interacting the independent variables with whether or not the participant reports, on average, a higher confidence level than that of the median participant. 
        The sample includes all observations of participants in the baseline treatment in experiment C, in which we collect such confidence level measure, excluding periods in which the truth ball is disclosed, a retraction occurs, or in which there was a retraction in an earlier period. 
    \end{minipage}
\end{table}

In our experiment C, in each period we elicit how confident participants are about their answer, as described in \hyref{section:robustness:understanding}[Section].

In \hyref{table-grether-hte-chigh}[Table], we estimate the standard \citet{Grether1980QJE} log-odds regression as per \hyref{equation:grether-baseline}[Equation], interacted with an indicator variable `High Confidence' that equals 1 for participants whose average reported confidence level is higher than that of the median participant.
This corresponds to a between-participant analysis of how confidence relates to the belief updating patterns.
We find that participants who are more confident tend to update more, but especially so from confirmatory signals.
They also tend to exhibit lower base-rate neglect.
However, these patterns are not clear---they are, at best, significantly at a 10\% significance level.

\hyref{table-grether-hte-cstd}[Table] exhibits the same regression but now interacted with the measure of confidence, `Confidence', standardized within-participant.
That is, for each participant, we subtract to the measure the mean reported confidence for that participant, and divide by the within-participant standard deviation of their reported level of confidence.
We find beliefs react more strongly to information when participants are more confident, but this is especially true for confirmatory information.
Furthermore, base-rate neglect is attenuated in instances in which participants are more confident.

\begin{table}[t]\setstretch{1.1}
	\centering\footnotesize
    \rotatebox{0}{\begin{tabular}{l@{\extracolsep{1pt}}cc@{}}
\hline\hline
& (1) & (2) \\
& $\hat \ell_t$ & $\hat \ell_t$ \\
\hline
Signal ($K(s_t)$) & 1.097$^{***}$ & 0.771$^{***}$ \\
 & (0.101) & (0.110) \\ [.1em]
Prior ($l_{t-1}$) & 0.656$^{***}$ & 0.582$^{***}$ \\
 & (0.033) & (0.032) \\ [.1em]
Confirmatory Signal ($K(s_t) \cdot c_t$) & -- & 0.999$^{***}$ \\
 &  & (0.141) \\ [.1em]
Confidence x Signal ($K(s_t)$) & 0.506$^{***}$ & 0.106$^{*}$ \\
 & (0.088) & (0.060) \\ [.1em]
Confidence x Prior ($l_{t-1}$) & 0.176$^{***}$ & 0.102$^{***}$ \\
 & (0.027) & (0.028) \\ [.1em]
Confidence x Confirmatory Signal ($K(s_t) \cdot c_t$) & -- & 1.186$^{***}$ \\
 &  & (0.184) \\ [.1em]

\hline
R$^2$ & 0.36 & 0.39 \\
N & 13316 & 13316 \\
\hline\hline
\multicolumn{3}{l}{\footnotesize Clustered standard errors at the subject level in parentheses.}\\
\multicolumn{3}{l}{\footnotesize $^{*}$ \(p<0.1\), $^{**}$ \(p<0.05\), $^{***}$ \(p<0.01\)}\\
\end{tabular}}
    \begin{minipage}{1\linewidth}
        \footnotesize
        \vspace*{.5em}
        \caption{Belief Updating Patterns and Within-Participant Confidence: Grether Regressions}
        \vspace*{-1.5em}
        \label{table-grether-hte-cstd}
        \singlespacing \emph{Notes}: 
        This table examines how within-participant heterogeneity relates to patterns in belief updating from new draws. 
        It reports estimates of \hyref{equation:grether-retractions}[Equation] interacting the independent variables with reported confidence, standardized within-participant. 
        The sample includes all observations of participants in the baseline treatment in experiment C, in which we collect such confidence level measure, excluding periods in which the truth ball is disclosed, a retraction occurs, or in which there was a retraction in an earlier period. 
    \end{minipage}
\end{table}

\hyref{table-c-ret}[Table] examines if retractions affect confidence levels.
It shows that indeed confidence is lower when participants update from a retraction, but this effect is small and not robustly significant.

\begin{table}[th!]\setstretch{1.1}
	\centering\footnotesize
    \rotatebox{0}{\begin{tabular}{l@{\extracolsep{1pt}}cccc@{}}
\hline\hline
\multicolumn{1}{r}{Retraction vs.} &\multicolumn{2}{c}{No Retracted Draw}  &  \multicolumn{2}{c}{Equivalent New Draw} \\
& (1) & (2) & (3) & (4) \\
& Confidence & Confidence & Confidence & Confidence \\
& (Levels) & (Standardized) & (Levels) & (Standardized) \\
\hline
Retraction ($r_t$) & -0.878 & -0.045 & -1.928$^{***}$ & -0.091$^{***}$ \\
 & (0.988) & (0.047) & (0.675) & (0.031) \\ [.1em]

Compressed History FEs & Yes  & Yes  & No  &  No \\
Sign History FEs       & No   & No   & Yes  & Yes \\
\hline
R$^2$ & 0.02 & 0.04 & 0.02 & 0.04 \\
N & 16267 & 16267 & 16267 & 16267 \\
\hline\hline
\multicolumn{5}{l}{\footnotesize Clustered standard errors at the subject level in parentheses.}\\
\multicolumn{5}{l}{\footnotesize $^{*}$ \(p<0.1\), $^{**}$ \(p<0.05\), $^{***}$ \(p<0.01\)}\\
\end{tabular}}
    \begin{minipage}{1\linewidth}
        \footnotesize
        \vspace*{.5em}
        \caption{Effect of Retractions on Confidence}
        \vspace*{-1.5em}
        \label{table-c-ret}
        \singlespacing \emph{Notes}: 
        This table provides estimates of the effect of retractions on confidence, following  \hyref{equation:retractions-time}[Equation].
        There are two types of comparison: (a) updating from a retraction vs. without the retracted observation (Columns (1)-(2)) and (b) updating from a retraction vs. an equivalent new draw (Columns (3)-(4)). 
        Columns (1) and (3) refer to reported confidence levels, while Columns (2) and (4) use confidence levels standardized for each participant.  
        The sample includes all observations of participants in the baseline treatment in experiment C, in which we collect such confidence level measure, excluding periods in which the truth ball is disclosed or in which there was a retraction in an earlier period. 
    \end{minipage}
\end{table}

\afterpage{\clearpage}
\newpage 
\subsubsection{Additional Retraction Information}
\label{online-appendix:robustness:understanding:info}

\begin{table}[th!]\setstretch{1.1}
	\centering\footnotesize
    \rotatebox{0}{\begin{tabular}{l@{\extracolsep{1pt}}cccc@{}}
\hline\hline
\multicolumn{1}{r}{Retraction vs.} &\multicolumn{2}{c}{No Retracted Draw}  &  \multicolumn{2}{c}{Equivalent New Draw} \\
& (1) & (2) & (3) & (4) \\
& Confidence & Confidence & Confidence & Confidence \\
& (Levels) & (Standardized) & (Levels) & (Standardized) \\
\hline
Retraction ($r_t$) & -0.840 & -0.044 & -2.061$^{***}$ & -0.102$^{***}$ \\
 & (0.875) & (0.041) & (0.713) & (0.032) \\ [.1em]
Retraction Info & 1.590 & -0.005 & 1.602 & -0.004 \\
 & (2.246) & (0.016) & (2.247) & (0.016) \\ [.1em]
Retraction Info x Retraction ($r_t$) & 0.797 & 0.021 & 0.760 & 0.019 \\
 & (0.932) & (0.045) & (0.927) & (0.045) \\ [.1em]

Compressed History FEs & Yes  & Yes  & No  &  No \\
Sign History FEs       & No   & No   & Yes  & Yes \\
\hline
R$^2$ & 0.02 & 0.04 & 0.02 & 0.04 \\
N & 33820 & 33820 & 33820 & 33820 \\
\hline\hline
\multicolumn{5}{l}{\footnotesize Clustered standard errors at the subject level in parentheses.}\\
\multicolumn{5}{l}{\footnotesize $^{*}$ \(p<0.1\), $^{**}$ \(p<0.05\), $^{***}$ \(p<0.01\)}\\
\end{tabular}}
    \begin{minipage}{1\linewidth}
        \footnotesize
        \vspace*{.5em}
        \caption{Effect of Retractions on Confidence: Additional Retraction Information}
        \vspace*{-1.5em}
        \label{table-c-ret-info}
        \singlespacing \emph{Notes}: 
        This table provides estimates on heterogeneity regarding the effect of retractions on confidence, following  \hyref{equation:retractions-time}[Equation], interacting the main explanatory variable with an indicator, 
        ``Retraction Info'', which equals 1 for participants assigned to this treatment and is 0 for participants assigned to our baseline treatment. 
        There are two types of comparison: (a) updating from a retraction vs. without the retracted observation (Columns (1)-(2)) and (b) updating from a retraction vs. an equivalent new draw (Columns (3)-(4)). 
        Columns (1) and (3) refer to reported confidence levels, while Columns (2) and (4) use confidence levels standardized for each participant.  
        The sample includes all observations of participants in the baseline and ``Retraction Info'' treatments in experiment C, in which we collect such confidence level measure, excluding periods in which the truth ball is disclosed or in which there was a retraction in an earlier period. 
    \end{minipage}
\end{table}

\hyref{table-c-ret-info}[Table] shows that providing additional information about retractions leads to participants reporting higher confidence levels when updating from retractions, even if only marginally so and not in a statistically significant manner.

\begin{table}[th!]\setstretch{1.1}
	\centering\footnotesize
    \rotatebox{90}{\begin{tabular}{l@{\extracolsep{1pt}}cccccccc@{}}
\hline\hline
\multicolumn{1}{r}{Retraction vs.} &\multicolumn{4}{c}{No Retracted Draw}  &  \multicolumn{4}{c}{Equivalent New Draw} \\
\cline{2-5} \cline{6-9}
& (1) & (2) & (3) & (4) & (5) & (6) & (7) & (8) \\
& $\hat \ell_t$ & $|\hat p_t-p_t|$ & ln(T$_t$) & Var($\hat \ell_t \mid h_t$) & $\hat \ell_t$ & $|\hat p_t-p_t|$ & ln(T$_t$) & Var($\hat \ell_t \mid h_t$) \\
\hline
Retraction ($r_t$) & 0.007 & 2.271$^{***}$ & 0.075$^{***}$ & 1.173$^{***}$ & -0.018 & 0.934$^{***}$ & 0.063$^{***}$ & 0.542$^{**}$ \\
 & (0.027) & (0.304) & (0.016) & (0.232) & (0.030) & (0.321) & (0.017) & (0.241) \\ [.1em]
Retracted Draw ($r_t\cdot K(s_{\rho_t})$) & 0.602$^{***}$ & -- & -- & -- & 0.751$^{***}$ & -- & -- & -- \\
 & (0.090) &  &  &  & (0.105) &  &  &  \\ [.1em]
Retraction ($r_t$) x Ret Info & 0.007 & -0.409 & 0.025 & -0.383 & 0.031 & -0.405 & 0.025 & -0.400 \\
 & (0.034) & (0.425) & (0.021) & (0.285) & (0.033) & (0.425) & (0.021) & (0.287) \\ [.1em]
Retracted Draw x Ret Info & -0.021 & -- & -- & -- & -0.006 & -- & -- & -- \\
 & (0.128) &  &  &  & (0.128) &  &  &  \\ [.1em]

Mean Decision Time     &   &   & 12.018 &   &   &   & 12.018 &   \\
Compressed History FEs & Yes & Yes & Yes & Yes & No & No & No & No  \\
Sign History FEs & No & No & No & No & Yes & Yes & Yes & Yes  \\
\hline
R$^2$ & 0.29 & 0.10 & 0.01 & 0.04 & 0.29 & 0.10 & 0.01 & 0.05 \\
N & 34137 & 34137 & 34137 & 4544 & 34137 & 34137 & 34137 & 4544 \\
\hline\hline
\multicolumn{9}{l}{\footnotesize Clustered standard errors at the subject level in parentheses.}\\
\multicolumn{9}{l}{\footnotesize $^{*}$ \(p<0.1\), $^{**}$ \(p<0.05\), $^{***}$ \(p<0.01\)}\\
\end{tabular}}
    \begin{minipage}{1\linewidth}
        \footnotesize
        \vspace*{.5em}
        \caption{Heterogenous Treatment Effects: Retraction Info}
        \vspace*{-1.5em}
        \label{table-hte-info}
        \singlespacing \emph{Notes}: 
        This table examines how treatment effects are affected by the provision of additional information about retractions.
        ``Ret Info'' corresponds to an indicator variable that equals 1 when the observation is from our treatment with additional retraction information in experiment C, and equals 0 when it is from our baseline treatment in experiment C. 
        There are two types of comparison: (a) updating from a retraction vs. without the retracted observation (Columns (1)-(4)) and (b) vs. an equivalent new draw (Columns (5)-(8)). 
        Columns (1) and (5) show effects on log-odds beliefs; (2) and (6) on the accuracy of belief updating; (3) and (7) on the speed of updating; (4) and (8) on the variability of updating.
        The sample includes all observations of participants in the baseline and ``Retraction Info'' treatments in experiment C, in which we collect such confidence level measure, excluding periods in which the truth ball is disclosed or in which there was a retraction in an earlier period. 
    \end{minipage}
\end{table}

\hyref{table-hte-info}[Table] reports on heterogeneous treatment effects regarding the provision of additional information about retractions.
No significant differences are detected.

\afterpage{\clearpage}
\newpage 
\subsection{Robustness 2: Consistency Across Heterogeneity}
\label{online-appendix:robustness:heterogeneity}

This section supports \hyref{section:robustness:heterogeneity}[Section] of the paper. 
In \hyref{online-appendix:robustness:heterogeneity:regression-tables}[Section], we report regression tables corresponding to \hyref{figure-ate-robustness-heterogeneity}[Figure]. 
In \hyref{online-appendix:robustness:heterogeneity:individual}[Section] we investigate heterogeneity in retraction ineffectiveness by participant, reporting summary statics on the participant-level estimates of the coefficient of interest (Retracted draw) in \hyref{table-retractions-beliefs}[Table].

\subsubsection{Regression Tables Corresponding to \hyref{figure-ate-robustness-heterogeneity}[Figure]}
\label{online-appendix:robustness:heterogeneity:regression-tables}

We include tables for each of the robustness checks presented in the figure, which explore heterogeneity in updating from retractions across multiple dimensions. 
For each dimension, we present two tables. 
The first re-estimates our main specification but with the sample restricted to the group in question (corresponding to the figure), while the second expands our main specification with interaction terms to account for heterogeneity, estimating the resulting regression on the full Baseline sample.

\newpage

\paragraph{High Quantitative Ability}
\label{online-appendix:robustness:heterogeneity:regression-tables:quantall}

\begin{table}[th!]\setstretch{1.1}
	\centering\footnotesize
	\rotatebox{90}{\begin{tabular}{l@{\extracolsep{4pt}}cccccccc@{}}
\hline\hline
\multicolumn{1}{r}{Retraction vs.} &\multicolumn{4}{c}{No Retracted Draw}  &  \multicolumn{4}{c}{Equivalent New Draw} \\
\cline{2-5} \cline{6-9}
& (1) & (2) & (3) & (4) & (5) & (6) & (7) & (8) \\
& $\hat \ell_t$ & $|\hat p_t-p_t|$ & ln(T$_t$) & Var($\hat \ell_t \mid h_t$) & $\hat \ell_t$ & $|\hat p_t-p_t|$ & ln(T$_t$) & Var($\hat \ell_t \mid h_t$) \\
\hline
Retraction ($r_t$) & -0.014 & 2.019$^{***}$ & 0.097$^{***}$ & 0.868$^{***}$ & -0.033 & 1.384$^{***}$ & 0.109$^{***}$ & 0.281 \\
 & (0.022) & (0.469) & (0.021) & (0.232) & (0.032) & (0.461) & (0.026) & (0.274) \\ [.1em]
Retracted Draw ($r_t\cdot K(s_{\rho_t})$) & 0.487$^{***}$ & -- & -- & -- & 0.563$^{***}$ & -- & -- & -- \\
 & (0.092) &  &  &  & (0.098) &  &  &  \\ [.1em]

Mean Decision Time     &   &   & 8.299 &   &   &   & 8.299 &   \\
Compressed History FEs & Yes & Yes & Yes & Yes & No & No & No & No  \\
Sign History FEs & No & No & No & No & Yes & Yes & Yes & Yes  \\
\hline
R$^2$ & 0.40 & 0.13 & 0.02 & 0.02 & 0.41 & 0.14 & 0.02 & 0.02 \\
N & 11044 & 11044 & 11044 & 1470 & 11044 & 11044 & 11044 & 1470 \\
\hline\hline
\multicolumn{9}{l}{\footnotesize Clustered standard errors at the subject level in parentheses.}\\
\multicolumn{9}{l}{\footnotesize $^{*}$ \(p<0.1\), $^{**}$ \(p<0.05\), $^{***}$ \(p<0.01\)}\\
\end{tabular}}
    \begin{minipage}{1\linewidth}
        \footnotesize
        \vspace*{.5em}
        \caption{Treatment Effects: High Quant Ability}
        \vspace*{-1.5em}
        \label{table-ate-quantall}
        \singlespacing \emph{Notes}: 
        This table reports the effect of retractions on updating and empirical complexity measures when restricting the baseline sample to participants with above median score on a quantitative test in the experiment, corresponding to ``High Quant Ability'' of \hyref{figure-ate-robustness-heterogeneity}[Figure]. 
        Columns (1) and (5) are the regressions from \hyref{table-retractions-beliefs}[Table], but restricted to this sample. 
        Similarly, Columns (2)-(4) and (6)-(8) are the regressions from \hyref{table-ate-complexity}[Table]. 
    \end{minipage}
\end{table}

\begin{table}[th!]\setstretch{1.1}
	\centering\footnotesize
	\rotatebox{90}{\begin{tabular}{l@{\extracolsep{1pt}}cccccccc@{}}
\hline\hline
\multicolumn{1}{r}{Retraction vs.} &\multicolumn{4}{c}{No Retracted Draw}  &  \multicolumn{4}{c}{Equivalent New Draw} \\
\cline{2-5} \cline{6-9}
& (1) & (2) & (3) & (4) & (5) & (6) & (7) & (8) \\
& $\hat \ell_t$ & $|\hat p_t-p_t|$ & ln(T$_t$) & Var($\hat \ell_t \mid h_t$) & $\hat \ell_t$ & $|\hat p_t-p_t|$ & ln(T$_t$) & Var($\hat \ell_t \mid h_t$) \\
\hline
Retraction ($r_t$) & 0.015 & 2.899$^{***}$ & 0.051$^{***}$ & 1.300$^{***}$ & -0.017 & 1.208$^{***}$ & 0.072$^{***}$ & 0.639$^{***}$ \\
 & (0.023) & (0.316) & (0.014) & (0.205) & (0.028) & (0.343) & (0.015) & (0.205) \\ [.1em]
Retracted Draw ($r_t\cdot K(s_{\rho_t})$) & 0.628$^{***}$ & -- & -- & -- & 0.644$^{***}$ & -- & -- & -- \\
 & (0.086) &  &  &  & (0.103) &  &  &  \\ [.1em]
Retraction ($r_t$) x High Quant & -0.015 & -0.461 & 0.045$^{*}$ & -0.213 & -0.008 & -0.483 & 0.044$^{*}$ & -0.213 \\
 & (0.031) & (0.487) & (0.024) & (0.235) & (0.030) & (0.488) & (0.023) & (0.234) \\ [.1em]
Retracted Draw x High Quant & -0.149 & -- & -- & -- & -0.143 & -- & -- & -- \\
 & (0.126) &  &  &  & (0.125) &  &  &  \\ [.1em]

Mean Decision Time     &   &   & 8.830 &   &   &   & 8.830 &   \\
Compressed History FEs & Yes & Yes & Yes & Yes & No & No & No & No  \\
Sign History FEs & No & No & No & No & Yes & Yes & Yes & Yes  \\
\hline
R$^2$ & 0.27 & 0.12 & 0.01 & 0.03 & 0.27 & 0.12 & 0.01 & 0.04 \\
N & 39162 & 39162 & 39162 & 5236 & 39162 & 39162 & 39162 & 5236 \\
\hline\hline
\multicolumn{9}{l}{\footnotesize Clustered standard errors at the subject level in parentheses.}\\
\multicolumn{9}{l}{\footnotesize $^{*}$ \(p<0.1\), $^{**}$ \(p<0.05\), $^{***}$ \(p<0.01\)}\\
\end{tabular}}
    \begin{minipage}{1\linewidth}
        \footnotesize
        \vspace*{.5em}
        \caption{Heterogeneous Treatment Effects: Quant Ability}
        \vspace*{-1.5em}
        \label{table-hte-quantall}
        \singlespacing \emph{Notes}: 
        This table examines how treatment effects are affected by the participants' quantitative ability.
        `High Quant Ability' corresponds to an indicator variable that equals 1 when the observation is from participants with above median score on a quantitative test in the experiment. 
        There are two types of comparison: (a) updating from a retraction vs. without the retracted observation (Columns (1)-(4)) and (b) vs. an equivalent new draw (Columns (5)-(8)). 
        Columns (1) and (5) show effects on log-odds beliefs; (2) and (6) on the accuracy of belief updating; (3) and (7) on the speed of updating; (4) and (8) on the variability of updating.
        The sample includes all observations of participants in the baseline treatment, excluding periods in which the truth ball is disclosed or in which there was a retraction in an earlier period. 
    \end{minipage}
\end{table}

\afterpage{\clearpage}
\newpage 
\paragraph{High Confidence}
\label{online-appendix:robustness:heterogeneity:regression-tables:chigh}

\begin{table}[th!]\setstretch{1.1}
	\centering\footnotesize
	\rotatebox{90}{\begin{tabular}{l@{\extracolsep{4pt}}cccccccc@{}}
\hline\hline
\multicolumn{1}{r}{Retraction vs.} &\multicolumn{4}{c}{No Retracted Draw}  &  \multicolumn{4}{c}{Equivalent New Draw} \\
\cline{2-5} \cline{6-9}
& (1) & (2) & (3) & (4) & (5) & (6) & (7) & (8) \\
& $\hat \ell_t$ & $|\hat p_t-p_t|$ & ln(T$_t$) & Var($\hat \ell_t \mid h_t$) & $\hat \ell_t$ & $|\hat p_t-p_t|$ & ln(T$_t$) & Var($\hat \ell_t \mid h_t$) \\
\hline
Retraction ($r_t$) & -0.010 & 2.560$^{***}$ & 0.110$^{***}$ & 1.257$^{***}$ & -0.094$^{*}$ & 0.561 & 0.089$^{***}$ & 0.537 \\
 & (0.050) & (0.447) & (0.022) & (0.343) & (0.054) & (0.528) & (0.029) & (0.358) \\ [.1em]
Retracted Draw ($r_t\cdot K(s_{\rho_t})$) & 0.554$^{***}$ & -- & -- & -- & 0.548$^{***}$ & -- & -- & -- \\
 & (0.126) &  &  &  & (0.185) &  &  &  \\ [.1em]

Mean Decision Time     &   &   & 12.105 &   &   &   & 12.105 &   \\
Compressed History FEs & Yes & Yes & Yes & Yes & No & No & No & No  \\
Sign History FEs & No & No & No & No & Yes & Yes & Yes & Yes  \\
\hline
R$^2$ & 0.34 & 0.10 & 0.01 & 0.06 & 0.35 & 0.11 & 0.01 & 0.07 \\
N & 8130 & 8130 & 8130 & 1079 & 8130 & 8130 & 8130 & 1079 \\
\hline\hline
\multicolumn{9}{l}{\footnotesize Clustered standard errors at the subject level in parentheses.}\\
\multicolumn{9}{l}{\footnotesize $^{*}$ \(p<0.1\), $^{**}$ \(p<0.05\), $^{***}$ \(p<0.01\)}\\
\end{tabular}}
    \begin{minipage}{1\linewidth}
        \footnotesize
        \vspace*{.5em}
        \caption{Treatment Effects: High Confidence}
        \vspace*{-1.5em}
        \label{table-ate-chigh}
        \singlespacing \emph{Notes}: 
        This table reports the effect of retractions on updating and empirical complexity measures when restricting the baseline sample to participants with above median confidence in their beliefs, corresponding to ``High Confidence'' of \hyref{figure-ate-robustness-heterogeneity}[Figure]. 
        Columns (1) and (5) are the regressions from \hyref{table-retractions-beliefs}[Table], but restricted to this sample. 
        Similarly, Columns (2)-(4) and (6)-(8) are the regressions from \hyref{table-ate-complexity}[Table]. 
    \end{minipage}
\end{table}

\begin{table}[th!]\setstretch{1.1}
	\centering\footnotesize
	\rotatebox{90}{\begin{tabular}{l@{\extracolsep{1pt}}cccccccc@{}}
\hline\hline
\multicolumn{1}{r}{Retraction vs.} &\multicolumn{4}{c}{No Retracted Draw}  &  \multicolumn{4}{c}{Equivalent New Draw} \\
\cline{2-5} \cline{6-9}
& (1) & (2) & (3) & (4) & (5) & (6) & (7) & (8) \\
& $\hat \ell_t$ & $|\hat p_t-p_t|$ & ln(T$_t$) & Var($\hat \ell_t \mid h_t$) & $\hat \ell_t$ & $|\hat p_t-p_t|$ & ln(T$_t$) & Var($\hat \ell_t \mid h_t$) \\
\hline
Retraction ($r_t$) & 0.057$^{**}$ & 2.258$^{***}$ & 0.047$^{**}$ & 1.302$^{***}$ & 0.014 & 0.717 & 0.029 & 0.611 \\
 & (0.027) & (0.441) & (0.023) & (0.361) & (0.038) & (0.451) & (0.022) & (0.378) \\ [.1em]
Retracted Draw ($r_t\cdot K(s_{\rho_t})$) & 0.644$^{***}$ & -- & -- & -- & 0.718$^{***}$ & -- & -- & -- \\
 & (0.131) &  &  &  & (0.147) &  &  &  \\ [.1em]
Retraction ($r_t$) x Confident & -0.077 & 0.295 & 0.062$^{**}$ & -0.161 & -0.074 & 0.264 & 0.066$^{**}$ & -0.180 \\
 & (0.053) & (0.580) & (0.029) & (0.428) & (0.051) & (0.582) & (0.029) & (0.425) \\ [.1em]
Retracted Draw x Confident & -0.085 & -- & -- & -- & -0.155 & -- & -- & -- \\
 & (0.183) &  &  &  & (0.183) &  &  &  \\ [.1em]

Mean Decision Time     &   &   & 11.765 &   &   &   & 11.765 &   \\
Compressed History FEs & Yes & Yes & Yes & Yes & No & No & No & No  \\
Sign History FEs & No & No & No & No & Yes & Yes & Yes & Yes  \\
\hline
R$^2$ & 0.28 & 0.11 & 0.01 & 0.05 & 0.29 & 0.11 & 0.01 & 0.06 \\
N & 16584 & 16584 & 16584 & 2206 & 16584 & 16584 & 16584 & 2206 \\
\hline\hline
\multicolumn{9}{l}{\footnotesize Clustered standard errors at the subject level in parentheses.}\\
\multicolumn{9}{l}{\footnotesize $^{*}$ \(p<0.1\), $^{**}$ \(p<0.05\), $^{***}$ \(p<0.01\)}\\
\end{tabular}}
    \begin{minipage}{1\linewidth}
        \footnotesize
        \vspace*{.5em}
        \caption{Heterogeneous Treatment Effects: Confidence}
        \vspace*{-1.5em}
        \label{table-hte-chigh}
        \singlespacing \emph{Notes}: 
        This table examines how treatment effects are affected by the participants' confidence.
        ``High Quant Ability'' corresponds to an indicator variable that equals 1 when the observation is from participants with above median confidence in their beliefs. 
        There are two types of comparison: (a) updating from a retraction vs. without the retracted observation (Columns (1)-(4)) and (b) vs. an equivalent new draw (Columns (5)-(8)). 
        Columns (1) and (5) show effects on log-odds beliefs; (2) and (6) on the accuracy of belief updating; (3) and (7) on the speed of updating; (4) and (8) on the variability of updating.
        The sample includes all observations of participants in the baseline treatment in experiment C, for which we collect confidence data, excluding periods in which the truth ball is disclosed or in which there was a retraction in an earlier period. 
    \end{minipage}
\end{table}

\afterpage{\clearpage}
\newpage 

\paragraph{More Bayesian}
\label{online-appendix:robustness:heterogeneity:regression-tables:morebayes}

\begin{table}[th!]\setstretch{1.1}
	\centering\footnotesize
	\rotatebox{90}{\begin{tabular}{l@{\extracolsep{4pt}}cccccccc@{}}
\hline\hline
\multicolumn{1}{r}{Retraction vs.} &\multicolumn{4}{c}{No Retracted Draw}  &  \multicolumn{4}{c}{Equivalent New Draw} \\
\cline{2-5} \cline{6-9}
& (1) & (2) & (3) & (4) & (5) & (6) & (7) & (8) \\
& $\hat \ell_t$ & $|\hat p_t-p_t|$ & ln(T$_t$) & Var($\hat \ell_t \mid h_t$) & $\hat \ell_t$ & $|\hat p_t-p_t|$ & ln(T$_t$) & Var($\hat \ell_t \mid h_t$) \\
\hline
Retraction ($r_t$) & 0.014 & 2.828$^{***}$ & 0.075$^{***}$ & 0.794$^{***}$ & 0.003 & 1.835$^{***}$ & 0.109$^{***}$ & 0.465$^{***}$ \\
 & (0.015) & (0.318) & (0.017) & (0.148) & (0.020) & (0.296) & (0.019) & (0.147) \\ [.1em]
Retracted Draw ($r_t\cdot K(s_{\rho_t})$) & 0.571$^{***}$ & -- & -- & -- & 0.651$^{***}$ & -- & -- & -- \\
 & (0.057) &  &  &  & (0.067) &  &  &  \\ [.1em]

Mean Decision Time     &   &   & 8.736 &   &   &   & 8.736 &   \\
Compressed History FEs & Yes & Yes & Yes & Yes & No & No & No & No  \\
Sign History FEs & No & No & No & No & Yes & Yes & Yes & Yes  \\
\hline
R$^2$ & 0.52 & 0.16 & 0.01 & 0.05 & 0.53 & 0.17 & 0.02 & 0.06 \\
N & 20972 & 20972 & 20972 & 2791 & 20972 & 20972 & 20972 & 2791 \\
\hline\hline
\multicolumn{9}{l}{\footnotesize Clustered standard errors at the subject level in parentheses.}\\
\multicolumn{9}{l}{\footnotesize $^{*}$ \(p<0.1\), $^{**}$ \(p<0.05\), $^{***}$ \(p<0.01\)}\\
\end{tabular}}
    \begin{minipage}{1\linewidth}
        \footnotesize
        \vspace*{.5em}
        \caption{Treatment Effects: More Bayesian}
        \vspace*{-1.5em}
        \label{table-ate-morebayes}
        \singlespacing \emph{Notes}: 
        This table reports the effect of retractions on updating and empirical complexity measures when restricting the baseline sample to participants who have, on average, lower than median distance to the Bayesian posterior when updating from new draws, corresponding to ``More Bayesian'' of \hyref{figure-ate-robustness-heterogeneity}[Figure]. 
        Columns (1) and (5) are the regressions from \hyref{table-retractions-beliefs}[Table], but restricted to this sample. 
        Similarly, Columns (2)-(4) and (6)-(8) are the regressions from \hyref{table-ate-complexity}[Table]. 
    \end{minipage}
\end{table}

\begin{table}[th!]\setstretch{1.1}
	\centering\footnotesize
	\rotatebox{90}{\begin{tabular}{l@{\extracolsep{1pt}}cccccccc@{}}
\hline\hline
\multicolumn{1}{r}{Retraction vs.} &\multicolumn{4}{c}{No Retracted Draw}  &  \multicolumn{4}{c}{Equivalent New Draw} \\
\cline{2-5} \cline{6-9}
& (1) & (2) & (3) & (4) & (5) & (6) & (7) & (8) \\
& $\hat \ell_t$ & $|\hat p_t-p_t|$ & ln(T$_t$) & Var($\hat \ell_t \mid h_t$) & $\hat \ell_t$ & $|\hat p_t-p_t|$ & ln(T$_t$) & Var($\hat \ell_t \mid h_t$) \\
\hline
Retraction ($r_t$) & -0.003 & 2.285$^{***}$ & 0.046$^{***}$ & 1.481$^{***}$ & -0.028 & 0.542 & 0.065$^{***}$ & 0.819$^{***}$ \\
 & (0.034) & (0.404) & (0.016) & (0.276) & (0.038) & (0.400) & (0.018) & (0.274) \\ [.1em]
Retracted Draw ($r_t\cdot K(s_{\rho_t})$) & 0.611$^{***}$ & -- & -- & -- & 0.618$^{***}$ & -- & -- & -- \\
 & (0.128) &  &  &  & (0.141) &  &  &  \\ [.1em]
Retraction ($r_t$) x More Bayesian & 0.026 & 0.877$^{*}$ & 0.034 & -0.461$^{*}$ & 0.017 & 0.925$^{*}$ & 0.036$^{*}$ & -0.458 \\
 & (0.037) & (0.485) & (0.022) & (0.279) & (0.036) & (0.484) & (0.021) & (0.279) \\ [.1em]
Retracted Draw x More Bayesian & -0.045 & -- & -- & -- & -0.025 & -- & -- & -- \\
 & (0.140) &  &  &  & (0.139) &  &  &  \\ [.1em]

Mean Decision Time     &   &   & 8.830 &   &   &   & 8.830 &   \\
Compressed History FEs & Yes & Yes & Yes & Yes & No & No & No & No  \\
Sign History FEs & No & No & No & No & Yes & Yes & Yes & Yes  \\
\hline
R$^2$ & 0.27 & 0.27 & 0.01 & 0.05 & 0.27 & 0.27 & 0.01 & 0.06 \\
N & 39162 & 39162 & 39162 & 5236 & 39162 & 39162 & 39162 & 5236 \\
\hline\hline
\multicolumn{9}{l}{\footnotesize Clustered standard errors at the subject level in parentheses.}\\
\multicolumn{9}{l}{\footnotesize $^{*}$ \(p<0.1\), $^{**}$ \(p<0.05\), $^{***}$ \(p<0.01\)}\\
\end{tabular}}
    \begin{minipage}{1\linewidth}
        \footnotesize
        \vspace*{.5em}
        \caption{Heterogenous Treatment Effects: Bayesian}
        \vspace*{-1.5em}
        \label{table-hte-morebayes}
        \singlespacing \emph{Notes}: 
        This table examines how treatment effects are affected by how correctly participants update beliefs from new draws.
        ``More Bayesian'' corresponds to an indicator variable that equals 1 when the observation is from participants more Bayesian than the median participant when updating from new draws (i.e., have on average lower than median distance to the Bayesian posterior). 
        There are two types of comparison: (a) updating from a retraction vs. without the retracted observation (Columns (1)-(4)) and (b) vs. an equivalent new draw (Columns (5)-(8)). 
        Columns (1) and (5) show effects on log-odds beliefs; (2) and (6) on the accuracy of belief updating; (3) and (7) on the speed of updating; (4) and (8) on the variability of updating.
        The sample includes all observations of participants in the baseline treatment, excluding periods in which the truth ball is disclosed or in which there was a retraction in an earlier period. 
    \end{minipage}
\end{table}

\afterpage{\clearpage}
\newpage 
\paragraph{More Experienced (2nd Half)}
\label{online-appendix:robustness:heterogeneity:regression-tables:2half}

\begin{table}[th!]\setstretch{1.1}
	\centering\footnotesize
	\rotatebox{90}{\begin{tabular}{l@{\extracolsep{4pt}}cccccccc@{}}
\hline\hline
\multicolumn{1}{r}{Retraction vs.} &\multicolumn{4}{c}{No Retracted Draw}  &  \multicolumn{4}{c}{Equivalent New Draw} \\
\cline{2-5} \cline{6-9}
& (1) & (2) & (3) & (4) & (5) & (6) & (7) & (8) \\
& $\hat \ell_t$ & $|\hat p_t-p_t|$ & ln(T$_t$) & Var($\hat \ell_t \mid h_t$) & $\hat \ell_t$ & $|\hat p_t-p_t|$ & ln(T$_t$) & Var($\hat \ell_t \mid h_t$) \\
\hline
Retraction ($r_t$) & 0.039$^{*}$ & 2.743$^{***}$ & 0.052$^{***}$ & 0.997$^{***}$ & -0.038 & 1.206$^{***}$ & 0.059$^{***}$ & 0.712$^{***}$ \\
 & (0.020) & (0.309) & (0.013) & (0.258) & (0.029) & (0.352) & (0.016) & (0.233) \\ [.1em]
Retracted Draw ($r_t\cdot K(s_{\rho_t})$) & 0.585$^{***}$ & -- & -- & -- & 0.550$^{***}$ & -- & -- & -- \\
 & (0.080) &  &  &  & (0.106) &  &  &  \\ [.1em]

Mean Decision Time     &   &   & 7.414 &   &   &   & 7.414 &   \\
Compressed History FEs & Yes & Yes & Yes & Yes & No & No & No & No  \\
Sign History FEs & No & No & No & No & Yes & Yes & Yes & Yes  \\
\hline
R$^2$ & 0.28 & 0.07 & 0.01 & 0.02 & 0.28 & 0.07 & 0.01 & 0.02 \\
N & 20834 & 20834 & 20834 & 4054 & 20834 & 20834 & 20834 & 4054 \\
\hline\hline
\multicolumn{9}{l}{\footnotesize Clustered standard errors at the subject level in parentheses.}\\
\multicolumn{9}{l}{\footnotesize $^{*}$ \(p<0.1\), $^{**}$ \(p<0.05\), $^{***}$ \(p<0.01\)}\\
\end{tabular}}
    \begin{minipage}{1\linewidth}
        \footnotesize
        \vspace*{.5em}
        \caption{Treatment Effects: Experienced}
        \vspace*{-1.5em}
        \label{table-ate-2half}
        \singlespacing \emph{Notes}: 
        This table reports the effect of retractions on updating and empirical complexity measures when restricting the baseline sample to the second half of rounds for each participant, corresponding to ``Experienced'' of \hyref{figure-ate-robustness-heterogeneity}[Figure]. 
        Columns (1) and (5) are the regressions from \hyref{table-retractions-beliefs}[Table], but restricted to this sample. 
        Similarly, Columns (2)-(4) and (6)-(8) are the regressions from \hyref{table-ate-complexity}[Table]. 
    \end{minipage}
\end{table}

\begin{table}[th!]\setstretch{1.1}
	\centering\footnotesize
	\rotatebox{90}{\begin{tabular}{l@{\extracolsep{1pt}}cccccccc@{}}
\hline\hline
\multicolumn{1}{r}{Retraction vs.} &\multicolumn{4}{c}{No Retracted Draw}  &  \multicolumn{4}{c}{Equivalent New Draw} \\
\cline{2-5} \cline{6-9}
& (1) & (2) & (3) & (4) & (5) & (6) & (7) & (8) \\
& $\hat \ell_t$ & $|\hat p_t-p_t|$ & ln(T$_t$) & Var($\hat \ell_t \mid h_t$) & $\hat \ell_t$ & $|\hat p_t-p_t|$ & ln(T$_t$) & Var($\hat \ell_t \mid h_t$) \\
\hline
Retraction ($r_t$) & -0.007 & 2.675$^{***}$ & 0.080$^{***}$ & 0.817$^{***}$ & -0.038 & 1.020$^{***}$ & 0.098$^{***}$ & 0.541$^{***}$ \\
 & (0.026) & (0.287) & (0.016) & (0.144) & (0.031) & (0.302) & (0.017) & (0.136) \\ [.1em]
Retracted Draw ($r_t\cdot K(s_{\rho_t})$) & 0.587$^{***}$ & -- & -- & -- & 0.608$^{***}$ & -- & -- & -- \\
 & (0.079) &  &  &  & (0.094) &  &  &  \\ [.1em]
Retraction ($r_t$) x Experienced & 0.034 & 0.169 & -0.032$^{*}$ & 0.235 & 0.037 & 0.163 & -0.032$^{**}$ & 0.240 \\
 & (0.029) & (0.272) & (0.016) & (0.236) & (0.029) & (0.270) & (0.016) & (0.232) \\ [.1em]
Retracted Draw x Experienced & -0.001 & -- & -- & -- & -0.008 & -- & -- & -- \\
 & (0.086) &  &  &  & (0.084) &  &  &  \\ [.1em]

Mean Decision Time     &   &   & 8.830 &   &   &   & 8.830 &   \\
Compressed History FEs & Yes & Yes & Yes & Yes & No & No & No & No  \\
Sign History FEs & No & No & No & No & Yes & Yes & Yes & Yes  \\
\hline
R$^2$ & 0.26 & 0.08 & 0.03 & 0.02 & 0.27 & 0.08 & 0.04 & 0.02 \\
N & 39162 & 39162 & 39162 & 7822 & 39162 & 39162 & 39162 & 7822 \\
\hline\hline
\multicolumn{9}{l}{\footnotesize Clustered standard errors at the subject level in parentheses.}\\
\multicolumn{9}{l}{\footnotesize $^{*}$ \(p<0.1\), $^{**}$ \(p<0.05\), $^{***}$ \(p<0.01\)}\\
\end{tabular}}
    \begin{minipage}{1\linewidth}
        \footnotesize
        \vspace*{.5em}
        \caption{Heterogenous Treatment Effects: Experience}
        \vspace*{-1.5em}
        \label{table-hte-2half}
        \singlespacing \emph{Notes}: 
        This table examines how treatment effects are affected by experience.
        ``Experienced'' corresponds to an indicator variable that equals 1 when the observation is from the second half of the rounds. 
        There are two types of comparison: (a) updating from a retraction vs. without the retracted observation (Columns (1)-(4)) and (b) vs. an equivalent new draw (Columns (5)-(8)). 
        Columns (1) and (5) show effects on log-odds beliefs; (2) and (6) on the accuracy of belief updating; (3) and (7) on the speed of updating; (4) and (8) on the variability of updating.
        The sample includes all observations of participants in the baseline treatment, excluding periods in which the truth ball is disclosed or in which there was a retraction in an earlier period. 
    \end{minipage}
\end{table}

\afterpage{\clearpage}
\newpage 

\subsubsection{Individual Heterogeneity}
\label{online-appendix:robustness:heterogeneity:individual}

\begin{table}[th!]\setstretch{1.1}
	\centering\footnotesize
	\rotatebox{0}{\begin{tabular}{l@{\extracolsep{4pt}}cc@{}}
\hline\hline
\multicolumn{1}{r}{Retraction vs.} &\multicolumn{1}{c}{No Retracted Draw}  &  \multicolumn{1}{c}{Equivalent New Draw} \\
\cline{2-2} \cline{3-3}
& (1) & (2) \\
& $\hat \ell_t$ & $\hat \ell_t$ \\
\hline
Mean Subject-level Effect & 0.628 & 0.564\\
    & (0.080) & (0.106) \\
Median Subject-level Effect & 0.340 & 0.297\\
    & (0.059) & (0.058) \\
Fraction $\beta_1>0$ & 0.720 & 0.683\\
Mean Std Error & 0.462 & 0.502\\
Median Std Error & 0.268 & 0.274\\

Compressed History FEs & Yes & No \\
Sign History FEs       & No  & Yes \\
\hline
\hline\hline
\multicolumn{3}{l}{\footnotesize Bootstrapped standard errors in parentheses.}\\
\end{tabular}}
    \begin{minipage}{1\linewidth}
        \footnotesize
        \vspace*{.5em}
        \caption{Updating from Retractions (\hyref{hypothesis:retractions-beliefs}[Hypothesis]): Participant-level Estimates}
        \vspace*{-1.5em}
        \label{table-retractions-beliefs-individual-hetero}
        \singlespacing \emph{Notes}: 
        This table provides summary statistics on distribution of participant-level estimates of the coefficient of interest in the main specification of interest in this paper. 
        We investigate the existence of individual-level heterogeneity by estimating the specifications in \hyref{table-retractions-beliefs}[Table] for each participant.
        The sample includes all observations of participants in the baseline treatment, excluding periods in which the truth ball is disclosed or in which there was a retraction in an earlier period. 
    \end{minipage}
\end{table}

\afterpage{\clearpage}
\newpage 
\subsection{Robustness 3: Variations on the Design}
\label{online-appendix:robustness:design}

This section supports \hyref{section:robustness:design}[Section] of the paper. 
In \hyref{online-appendix:robustness:design:regression-tables}[Section], we report regression tables corresponding to \hyref{figure-ate-robustness-design}[Figure]. 

\subsubsection{Regression Tables Corresponding to \hyref{figure-ate-robustness-design}[Figure]}
\label{online-appendix:robustness:design:regression-tables}

We include tables for each of the robustness checks presented in the figure, which explore the robustness of our results to variations in the experimental design. 
For each variation, we re-estimate our main specification using the results from the variation of the design (corresponding to the figure). 
The first and second variants were randomized against the baseline variant, at the individual participant level. 
Hence, for those variants, we also present results on heterogeneous treatment effects, directly using an interaction term.

\paragraph{Elicit at End}
\label{online-appendix:robustness:design:regression-tables:final}

\begin{table}[th!]\setstretch{1.1}
	\centering\footnotesize
	\rotatebox{0}{\begin{tabular}{l@{\extracolsep{4pt}}cccc@{}}
\hline\hline
\multicolumn{1}{r}{Retraction vs.}  &  \multicolumn{4}{c}{Equivalent New Draw} \\
\cline{2-5}
& (1) & (2) & (3) & (4) \\
& $\hat \ell_t$ & $|\hat p_t-p_t|$ & ln(T$_t$) & Var($\hat \ell_t \mid h_t$) \\
\hline
Retraction ($r_t$) & 0.034 & -0.916$^{**}$ & 0.125$^{***}$ & 0.299$^{**}$ \\
 & (0.041) & (0.456) & (0.028) & (0.117) \\ [.1em]
Retracted Draw ($r_t\cdot K(s_{\rho_t})$) & 0.615$^{***}$ & -- & -- & -- \\
 & (0.140) &  &  &  \\ [.1em]

Mean Decision Time     &   &   & 9.769 &   \\
Sign History FEs & Yes & Yes & Yes & Yes  \\
\hline
R$^2$ & 0.42 & 0.03 & 0.01 & 0.02 \\
N & 6093 & 6093 & 6093 & 1436 \\
\hline\hline
\multicolumn{5}{l}{\footnotesize Clustered standard errors at the subject level in parentheses.}\\
\multicolumn{5}{l}{\footnotesize $^{*}$ \(p<0.1\), $^{**}$ \(p<0.05\), $^{***}$ \(p<0.01\)}\\
\end{tabular}}
    \begin{minipage}{1\linewidth}
        \footnotesize
        \vspace*{.5em}
        \caption{Treatment Effects: Elicit at End}
        \vspace*{-1.5em}
        \label{table-ate-final}
        \singlespacing \emph{Notes}:   
        This table reports the effect of retractions on updating and empirical complexity measures in a variant of the experiment in which beliefs are elicited only at the end of each round, corresponding to ``Elicit at End'' of \hyref{figure-ate-robustness-design}[Figure]. 
        Column (1) is the equivalent regression from \hyref{table-retractions-beliefs}[Table]. 
        Similarly, Columns (2)-(4) are the regressions from \hyref{table-ate-complexity}[Table]. 
        We cannot compare to `No Retracted Draw' as beliefs are only elicited in the last period of each round. 
    \end{minipage}
\end{table}

\begin{table}[th!]\setstretch{1.1}
	\centering\footnotesize
	\rotatebox{90}{\begin{tabular}{l@{\extracolsep{1pt}}cccccccc@{}}
\hline\hline
\multicolumn{1}{r}{Retraction vs.} &\multicolumn{4}{c}{No Retracted Draw}  &  \multicolumn{4}{c}{Equivalent New Draw} \\
\cline{2-5} \cline{6-9}
& (1) & (2) & (3) & (4) & (5) & (6) & (7) & (8) \\
& $\hat \ell_t$ & $|\hat p_t-p_t|$ & ln(T$_t$) & Var($\hat \ell_t \mid h_t$) & $\hat \ell_t$ & $|\hat p_t-p_t|$ & ln(T$_t$) & Var($\hat \ell_t \mid h_t$) \\
\hline
Retraction ($r_t$) & 0.008 & 3.050$^{***}$ & 0.053$^{***}$ & 1.176$^{***}$ & -0.016 & 0.825$^{**}$ & 0.109$^{***}$ & 0.593$^{***}$ \\
 & (0.023) & (0.391) & (0.016) & (0.240) & (0.032) & (0.406) & (0.015) & (0.201) \\ [.1em]
Retracted Draw ($r_t\cdot K(s_{\rho_t})$) & 0.572$^{***}$ & -- & -- & -- & 0.584$^{***}$ & -- & -- & -- \\
 & (0.095) &  &  &  & (0.113) &  &  &  \\ [.1em]
Retraction ($r_t$) x Elicit End & 0.040 & -1.053$^{*}$ & -0.054$^{*}$ & -0.317 & 0.046 & -0.650 & -0.006 & -0.243 \\
 & (0.051) & (0.552) & (0.028) & (0.204) & (0.051) & (0.538) & (0.028) & (0.199) \\ [.1em]
Retracted Draw x Elicit End & -0.071 & -- & -- & -- & 0.062 & -- & -- & -- \\
 & (0.152) &  &  &  & (0.149) &  &  &  \\ [.1em]

Mean Decision Time     &   &   & 7.332 &   &   &   & 7.332 &   \\
Compressed History FEs & Yes & Yes & Yes & Yes & No & No & No & No  \\
Sign History FEs & No & No & No & No & Yes & Yes & Yes & Yes  \\
\hline
R$^2$ & 0.29 & 0.05 & 0.07 & 0.02 & 0.29 & 0.05 & 0.07 & 0.03 \\
N & 28671 & 28671 & 28671 & 4466 & 28671 & 28671 & 28671 & 4466 \\
\hline\hline
\multicolumn{9}{l}{\footnotesize Clustered standard errors at the subject level in parentheses.}\\
\multicolumn{9}{l}{\footnotesize $^{*}$ \(p<0.1\), $^{**}$ \(p<0.05\), $^{***}$ \(p<0.01\)}\\
\end{tabular}}
    \begin{minipage}{1\linewidth}
        \footnotesize
        \vspace*{.5em}
        \caption{Heterogenous Treatment Effects: Elicit at End}
        \vspace*{-1.5em}
        \label{table-hte-final}
        \singlespacing \emph{Notes}: 
        This table examines how treatment effects are affected by eliciting beliefs only at the end of the round.
        ``Elicit at End'' corresponds to an indicator variable that equals 1 when the observation is from that treatment in experiment A, and equals 0 when it is from the baseline treatment in experiment A. 
        There are two types of comparison: (a) updating from a retraction vs. without the retracted observation (Columns (1)-(4)) and (b) vs. an equivalent new draw (Columns (5)-(8)). 
        Columns (1) and (5) show effects on log-odds beliefs; (2) and (6) on the accuracy of belief updating; (3) and (7) on the speed of updating; (4) and (8) on the variability of updating.
        The sample includes all observations of participants in the baseline and `Elicit at the End' treatments of experiment A, excluding periods in which the truth ball is disclosed or in which there was a retraction in an earlier period. 
    \end{minipage}
\end{table}

\afterpage{\clearpage}
\newpage 
\paragraph{No History}
\label{online-appendix:robustness:design:regression-tables:hist}

\begin{table}[th!]\setstretch{1.1}
	\centering\footnotesize
	\rotatebox{90}{\begin{tabular}{l@{\extracolsep{4pt}}cccccccc@{}}
\hline\hline
\multicolumn{1}{r}{Retraction vs.} &\multicolumn{4}{c}{No Retracted Draw}  &  \multicolumn{4}{c}{Equivalent New Draw} \\
\cline{2-5} \cline{6-9}
& (1) & (2) & (3) & (4) & (5) & (6) & (7) & (8) \\
& $\hat \ell_t$ & $|\hat p_t-p_t|$ & ln(T$_t$) & Var($\hat \ell_t \mid h_t$) & $\hat \ell_t$ & $|\hat p_t-p_t|$ & ln(T$_t$) & Var($\hat \ell_t \mid h_t$) \\
\hline
Retraction ($r_t$) & 0.018 & 5.043$^{***}$ & 0.059$^{***}$ & 1.703$^{***}$ & 0.029 & 3.915$^{***}$ & 0.088$^{***}$ & 0.875$^{*}$ \\
 & (0.059) & (0.572) & (0.017) & (0.429) & (0.056) & (0.558) & (0.020) & (0.461) \\ [.1em]
Retracted Draw ($r_t\cdot K(s_{\rho_t})$) & 0.392$^{**}$ & -- & -- & -- & 1.034$^{***}$ & -- & -- & -- \\
 & (0.195) &  &  &  & (0.273) &  &  &  \\ [.1em]

Mean Decision Time     &   &   & 12.000 &   &   &   & 12.000 &   \\
Compressed History FEs & Yes & Yes & Yes & Yes & No & No & No & No  \\
Sign History FEs & No & No & No & No & Yes & Yes & Yes & Yes  \\
\hline
R$^2$ & 0.16 & 0.04 & 0.01 & 0.02 & 0.16 & 0.04 & 0.01 & 0.02 \\
N & 17642 & 17642 & 17642 & 2342 & 17642 & 17642 & 17642 & 2342 \\
\hline\hline
\multicolumn{9}{l}{\footnotesize Clustered standard errors at the subject level in parentheses.}\\
\multicolumn{9}{l}{\footnotesize $^{*}$ \(p<0.1\), $^{**}$ \(p<0.05\), $^{***}$ \(p<0.01\)}\\
\end{tabular}}
    \begin{minipage}{1\linewidth}
        \footnotesize
        \vspace*{.5em}
        \caption{Treatment Effects: No History}
        \vspace*{-1.5em}
        \label{table-ate-hist}
        \singlespacing \emph{Notes}: 
        This table reports the effect of retractions on updating and empirical complexity measures in a variant of the experiment in which participants were only shown the current observation, not the history of all observations in the current round, corresponding to ``No History'' of \hyref{figure-ate-robustness-design}[Figure]. 
        Columns (1) and (5) are the equivalent regressions from \hyref{table-retractions-beliefs}[Table]. 
        Similarly, Columns (2)-(4) and (6)-(8) are the regressions from \hyref{table-ate-complexity}[Table].  
    \end{minipage}
\end{table}

\begin{table}[th!]\setstretch{1.1}
	\centering\footnotesize
	\rotatebox{90}{\begin{tabular}{l@{\extracolsep{1pt}}cccccccc@{}}
\hline\hline
\multicolumn{1}{r}{Retraction vs.} &\multicolumn{4}{c}{No Retracted Draw}  &  \multicolumn{4}{c}{Equivalent New Draw} \\
\cline{2-5} \cline{6-9}
& (1) & (2) & (3) & (4) & (5) & (6) & (7) & (8) \\
& $\hat \ell_t$ & $|\hat p_t-p_t|$ & ln(T$_t$) & Var($\hat \ell_t \mid h_t$) & $\hat \ell_t$ & $|\hat p_t-p_t|$ & ln(T$_t$) & Var($\hat \ell_t \mid h_t$) \\
\hline
Retraction ($r_t$) & 0.005 & 2.253$^{***}$ & 0.070$^{***}$ & 1.028$^{***}$ & -0.012 & 0.907$^{***}$ & 0.078$^{***}$ & 0.258 \\
 & (0.027) & (0.302) & (0.016) & (0.227) & (0.031) & (0.322) & (0.017) & (0.258) \\ [.1em]
Retracted Draw ($r_t\cdot K(s_{\rho_t})$) & 0.601$^{***}$ & -- & -- & -- & 0.954$^{***}$ & -- & -- & -- \\
 & (0.090) &  &  &  & (0.124) &  &  &  \\ [.1em]
Retraction ($r_t$) x No Hist & 0.026 & 2.918$^{***}$ & -0.004 & 0.853$^{*}$ & 0.032 & 2.918$^{***}$ & -0.005 & 0.853$^{*}$ \\
 & (0.062) & (0.622) & (0.021) & (0.453) & (0.062) & (0.621) & (0.021) & (0.452) \\ [.1em]
Retracted Draw x No Hist & -0.209 & -- & -- & -- & -0.203 & -- & -- & -- \\
 & (0.215) &  &  &  & (0.215) &  &  &  \\ [.1em]

Mean Decision Time     &   &   & 11.886 &   &   &   & 11.886 &   \\
Compressed History FEs & Yes & Yes & Yes & Yes & No & No & No & No  \\
Sign History FEs & No & No & No & No & Yes & Yes & Yes & Yes  \\
\hline
R$^2$ & 0.21 & 0.07 & 0.01 & 0.02 & 0.21 & 0.07 & 0.01 & 0.02 \\
N & 34226 & 34226 & 34226 & 4548 & 34226 & 34226 & 34226 & 4548 \\
\hline\hline
\multicolumn{9}{l}{\footnotesize Clustered standard errors at the subject level in parentheses.}\\
\multicolumn{9}{l}{\footnotesize $^{*}$ \(p<0.1\), $^{**}$ \(p<0.05\), $^{***}$ \(p<0.01\)}\\
\end{tabular}}
    \begin{minipage}{1\linewidth}
        \footnotesize
        \vspace*{.5em}
        \caption{Heterogenous Treatment Effects: No History}
        \vspace*{-1.5em}
        \label{table-hte-hist}
        \singlespacing \emph{Notes}: 
        This table examines how treatment effects are affected by not showing the history of past signals.
        ``No History'' corresponds to an indicator variable that equals 1 when the observation is from that treatment in experiment C, and equals 0 when it is from the baseline treatment in experiment C. 
        There are two types of comparison: (a) updating from a retraction vs. without the retracted observation (Columns (1)-(4)) and (b) vs. an equivalent new draw (Columns (5)-(8)). 
        Columns (1) and (5) show effects on log-odds beliefs; (2) and (6) on the accuracy of belief updating; (3) and (7) on the speed of updating; (4) and (8) on the variability of updating.
        The sample includes all observations of participants in the baseline and `No History' treatments of experiment C, excluding periods in which the truth ball is disclosed or in which there was a retraction in an earlier period. 
    \end{minipage}
\end{table}

\afterpage{\clearpage}
\newpage 
\paragraph{Short Histories}
\label{online-appendix:robustness:design:regression-tables:2draw}

\begin{table}[th!]\setstretch{1.1}
	\centering\footnotesize
	\rotatebox{90}{\begin{tabular}{l@{\extracolsep{4pt}}cccccc@{}}
\hline\hline
\multicolumn{1}{r}{Retraction vs.} &\multicolumn{2}{c}{No Retracted Draw}  &  \multicolumn{4}{c}{Equivalent New Draw} \\
\cline{2-3} \cline{4-7}
& (1) & (2) & (3) & (4) & (5) & (6) \\
& $\hat \ell_t$ & $|\hat p_t-p_t|$ & $\hat \ell_t$ & $|\hat p_t-p_t|$ & ln(T$_t$) & Var($\hat \ell_t \mid h_t$) \\
\hline
Retraction ($r_t$) & 0.001 & 1.525$^{**}$ & -0.013 & 6.378$^{***}$ & 0.081$^{**}$ & 1.123$^{***}$ \\
 & (0.025) & (0.697) & (0.027) & (0.796) & (0.032) & (0.434) \\ [.1em]
Retracted Draw ($r_t\cdot K(s_{\rho_t})$) & 0.448$^{***}$ & -- & 0.371$^{**}$ & -- & -- & -- \\
 & (0.169) &  & (0.166) &  &  &  \\ [.1em]

Mean Decision Time     &   &   &   &   & 8.131 &   \\
Compressed History FEs & Yes & Yes & No & No & No & No  \\
Sign History FEs & No & No & Yes & Yes & Yes & Yes  \\
\hline
R$^2$ & 0.01 & 0.00 & 0.21 & 0.08 & 0.00 & 0.02 \\
N & 13938 & 13938 & 9138 & 9138 & 9138 & 882 \\
\hline\hline
\multicolumn{7}{l}{\footnotesize Clustered standard errors at the subject level in parentheses.}\\
\multicolumn{7}{l}{\footnotesize $^{*}$ \(p<0.1\), $^{**}$ \(p<0.05\), $^{***}$ \(p<0.01\)}\\
\end{tabular}}
    \begin{minipage}{1\linewidth}
        \footnotesize
        \vspace*{.5em}
        \caption{Treatment Effects: Short Histories}
        \vspace*{-1.5em}
        \label{table-ate-2draw}
        \singlespacing \emph{Notes}: 
        This table reports the effect of retractions on updating and empirical complexity measures in a variant of the experiment in which there were only two periods per round, rather than four, corresponding to ``Short Histories'' of \hyref{figure-ate-robustness-design}[Figure]. 
        Columns (1) and (3) are the equivalent regressions from \hyref{table-retractions-beliefs}[Table]. 
        Similarly, Columns (2) and (4)-(6) are the regressions from \hyref{table-ate-complexity}[Table]. 
        There are no treatment effects on decision times and variability for the comparison to `No Retracted Draw', because, given there are only two periods, the comparison is to the prior at period 0 (i.e. before any observations) which we did not elicit and assume here to be 0.5.
    \end{minipage}
\end{table}

\afterpage{\clearpage}
\newpage 
\paragraph{Garbled Information}
\label{online-appendix:robustness:design:regression-tables:box}

\begin{table}[th!]\setstretch{1.1}
	\centering\footnotesize
	\rotatebox{90}{\begin{tabular}{l@{\extracolsep{4pt}}cccccccc@{}}
\hline\hline
\multicolumn{1}{r}{Retraction vs.} &\multicolumn{4}{c}{No Retracted Draw}  &  \multicolumn{4}{c}{Equivalent New Draw} \\
\cline{2-5} \cline{6-9}
& (1) & (2) & (3) & (4) & (5) & (6) & (7) & (8) \\
& $\hat \ell_t$ & $|\hat p_t-p_t|$ & ln(T$_t$) & Var($\hat \ell_t \mid h_t$) & $\hat \ell_t$ & $|\hat p_t-p_t|$ & ln(T$_t$) & Var($\hat \ell_t \mid h_t$) \\
\hline
Retraction ($r_t$) & 0.001 & 3.164$^{***}$ & 0.090$^{***}$ & 0.660$^{**}$ & -0.007 & 3.591$^{***}$ & 0.116$^{***}$ & 0.242 \\
 & (0.026) & (0.610) & (0.027) & (0.277) & (0.040) & (0.647) & (0.028) & (0.307) \\ [.1em]
Retracted Draw ($r_t\cdot K(s_{\rho_t})$) & 0.338$^{***}$ & -- & -- & -- & 0.567$^{***}$ & -- & -- & -- \\
 & (0.067) &  &  &  & (0.086) &  &  &  \\ [.1em]

Mean Decision Time     &   &   & 8.761 &   &   &   & 8.761 &   \\
Compressed History FEs & Yes & Yes & Yes & Yes & No & No & No & No  \\
Sign History FEs & No & No & No & No & Yes & Yes & Yes & Yes  \\
\hline
R$^2$ & 0.33 & 0.04 & 0.01 & 0.01 & 0.33 & 0.04 & 0.01 & 0.01 \\
N & 14427 & 14427 & 14344 & 1703 & 14427 & 14427 & 14344 & 1703 \\
\hline\hline
\multicolumn{9}{l}{\footnotesize Clustered standard errors at the subject level in parentheses.}\\
\multicolumn{9}{l}{\footnotesize $^{*}$ \(p<0.1\), $^{**}$ \(p<0.05\), $^{***}$ \(p<0.01\)}\\
\end{tabular}}
    \begin{minipage}{1\linewidth}
        \footnotesize
        \vspace*{.5em}
        \caption{Treatment Effects: Garbled Information}
        \vspace*{-1.5em}
        \label{table-ate-box}
        \singlespacing \emph{Notes}: 
        This table reports the effect of retractions on updating and empirical complexity measures in a variant of the experiment in which truth balls were not fully informative, corresponding to ``Garbled Info'' of \hyref{figure-ate-robustness-design}[Figure]. 
        Columns (1) and (5) are the equivalent regressions from \hyref{table-retractions-beliefs}[Table]. 
        Similarly, Columns (2)-(4) and (6)-(8) are the regressions from \hyref{table-ate-complexity}[Table]. 
    \end{minipage}
\end{table}

\afterpage{\clearpage}
\newpage

\subsection{Instructions and Screenshots} 
\label{online-appendix:instructions}

All our treatments followed small variations of our baseline instructions presented here.

\subsubsection{Start Screen and Instructions}
Below are screenshots of the start screen and the instructions as presented to the participants.
\begin{figure}[H]
    \centering
    \includegraphics[width=.8\textwidth]{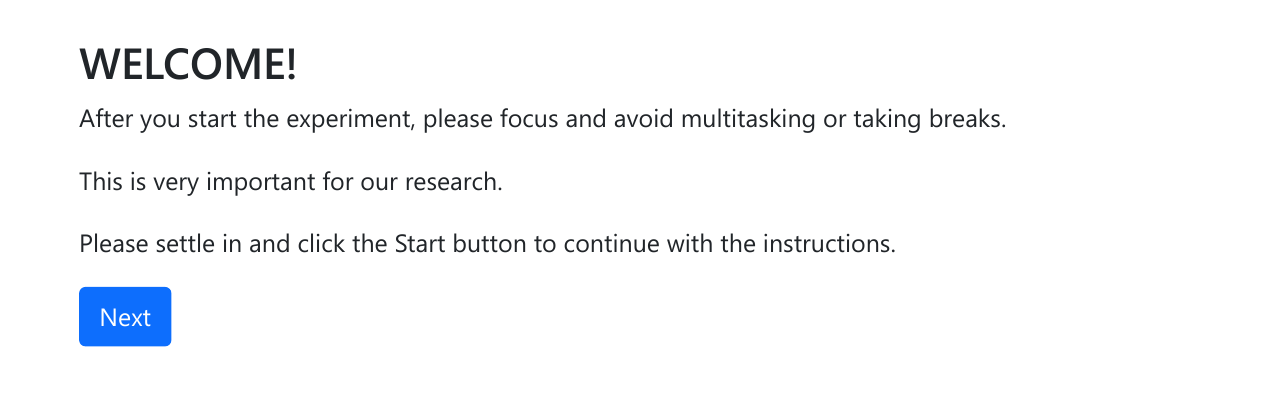}
\end{figure}
\begin{figure}[H]
    \centering
    \includegraphics[width=\textwidth]{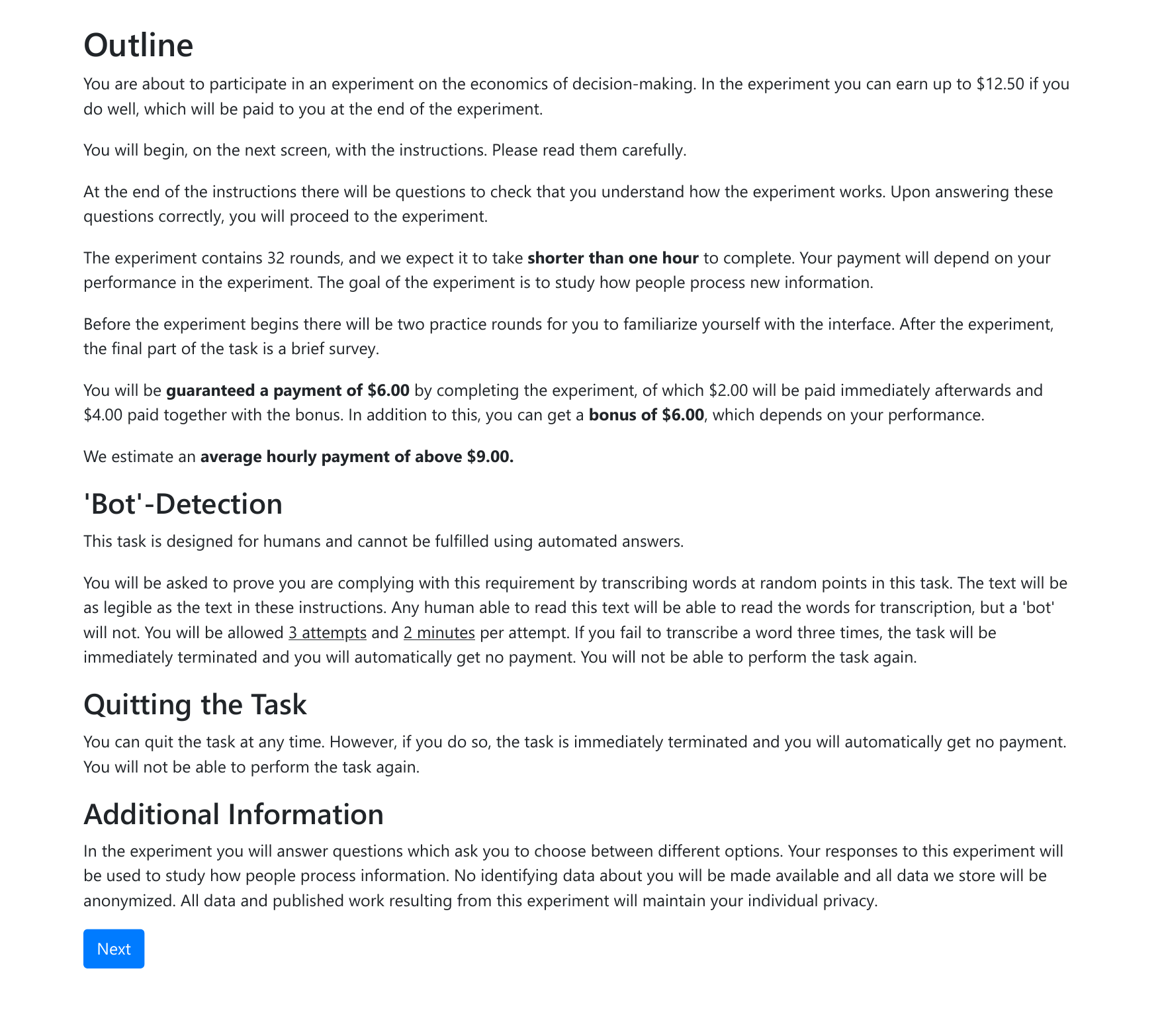}
\end{figure}
\begin{figure}[H]
    \centering
    \includegraphics[page=1,width=\textwidth]{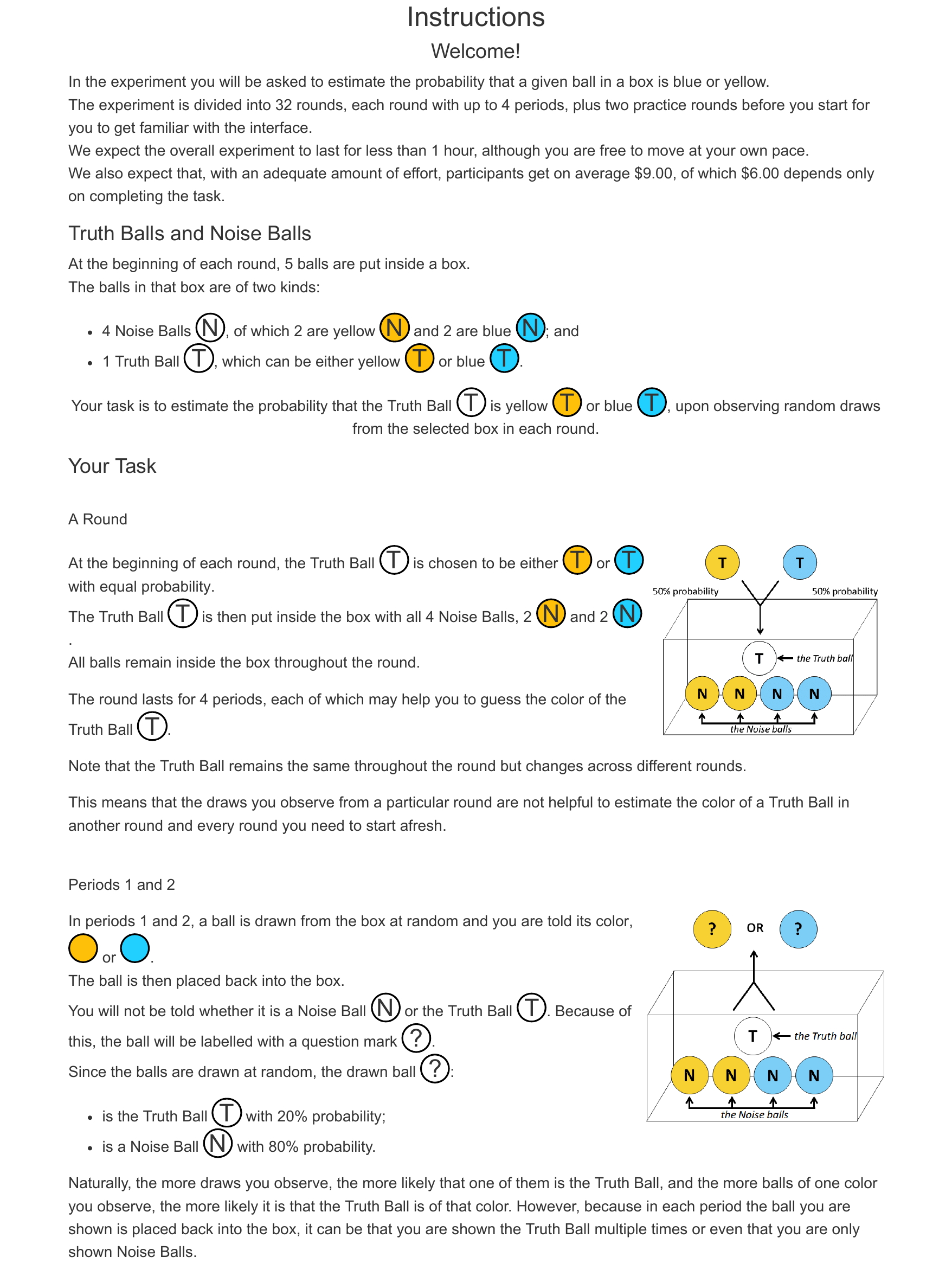}
\end{figure}

\begin{figure}[H]
    \centering
    \includegraphics[page=2,width=\textwidth]{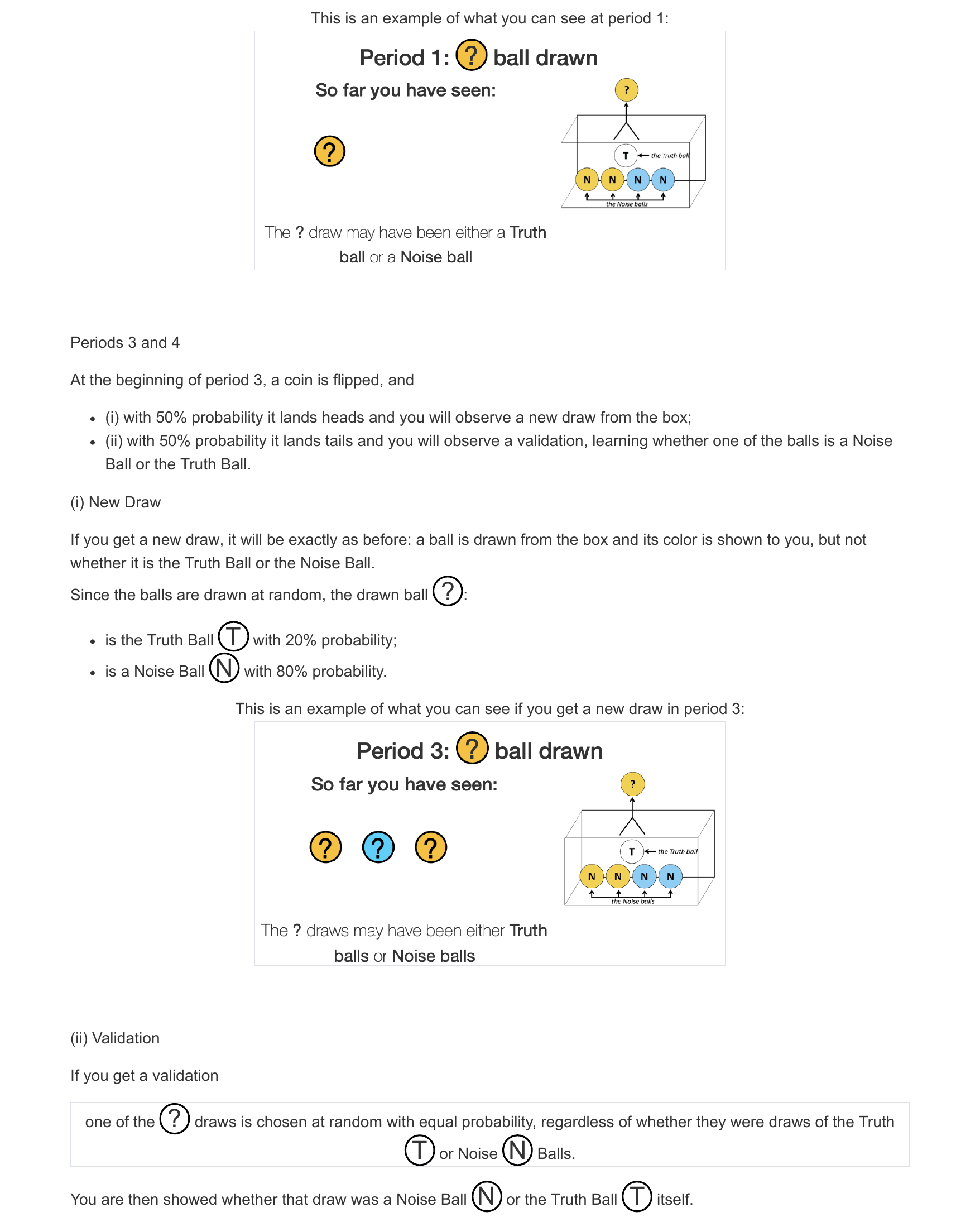}
\end{figure}

\begin{figure}[H]
    \centering
    \includegraphics[page=3,width=\textwidth]{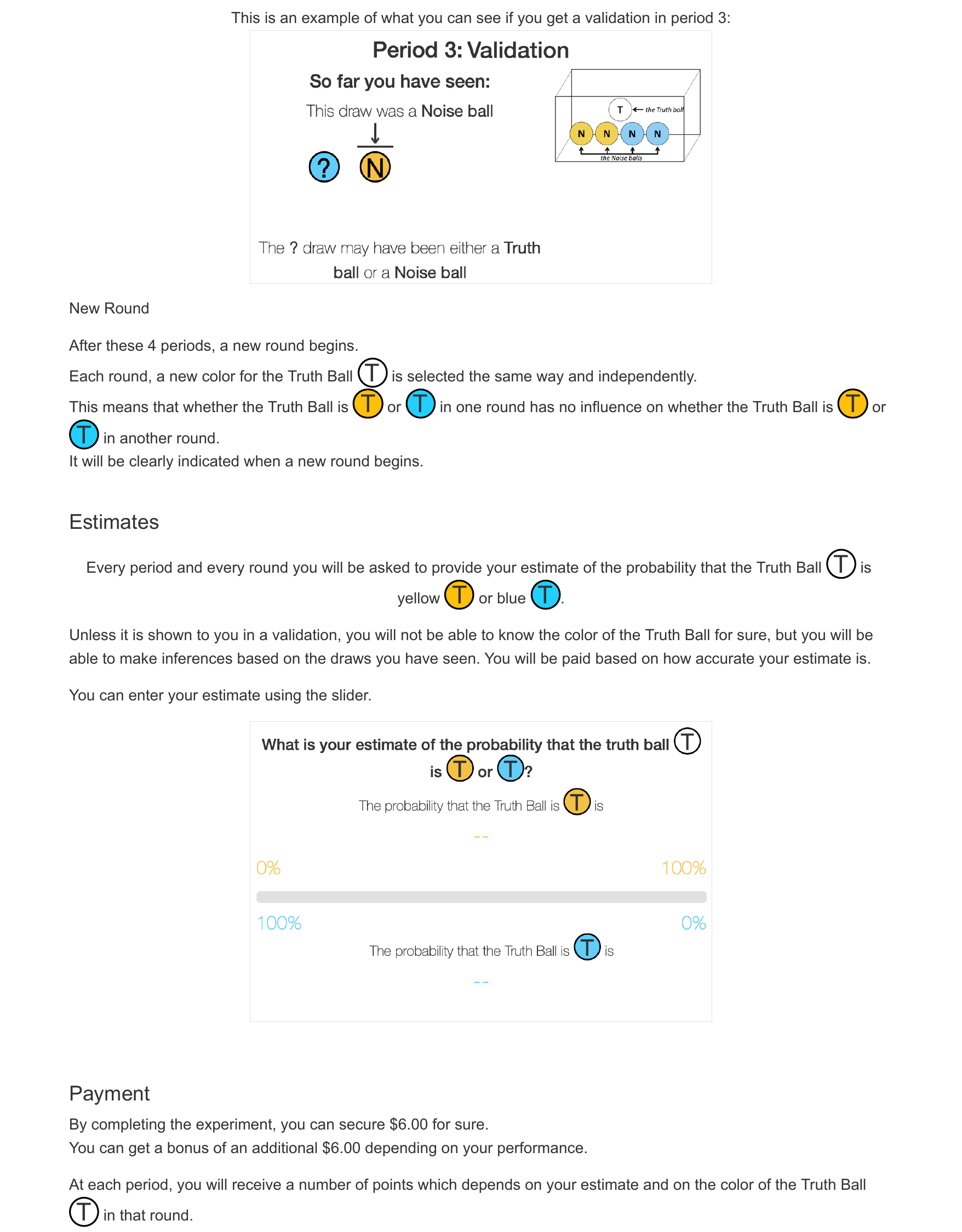}
\end{figure}

\begin{figure}[H]
    \centering
    \includegraphics[page=4,width=\textwidth]{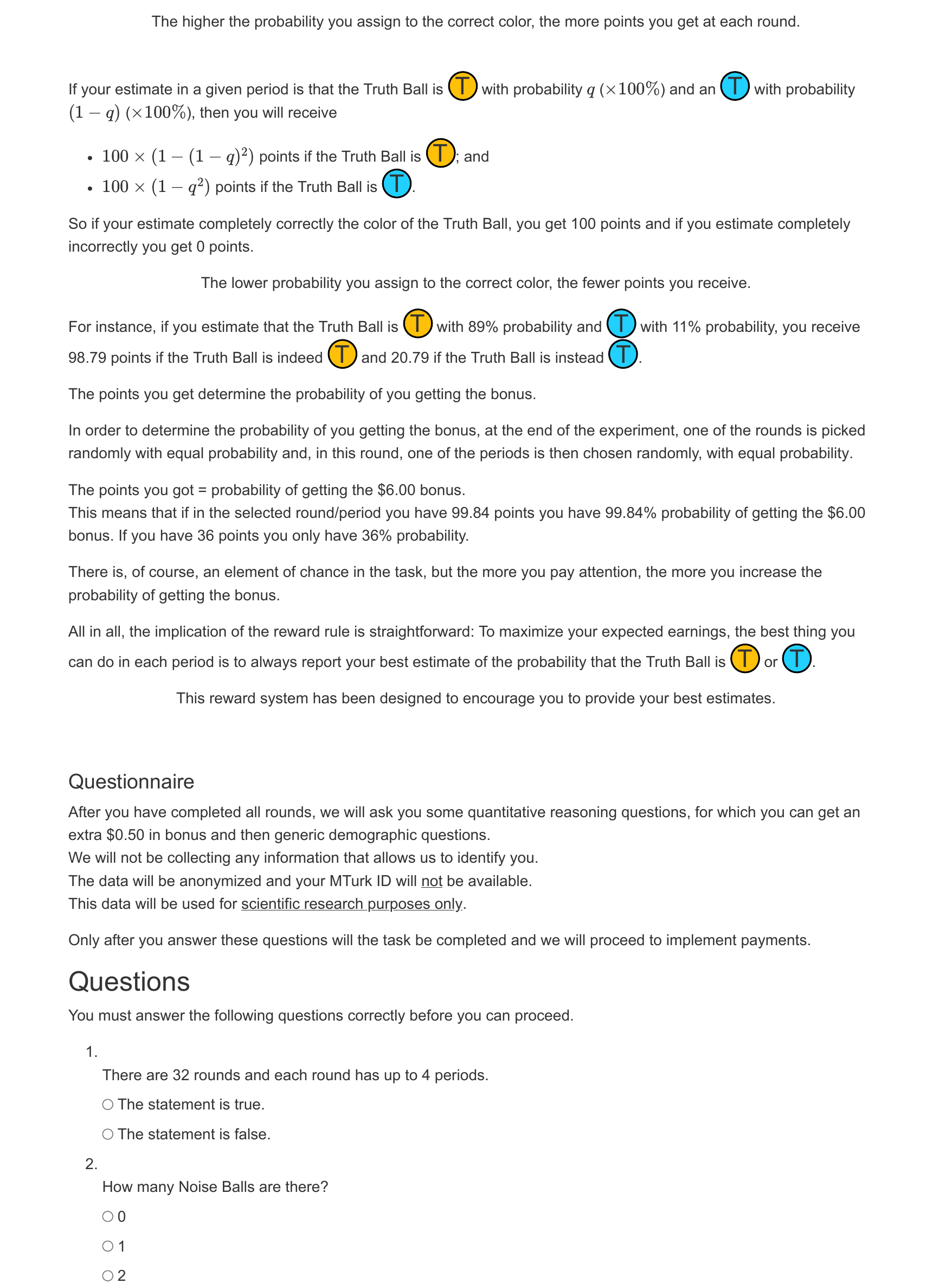}
\end{figure}

\begin{figure}[H]
    \centering
    \includegraphics[page=5,width=\textwidth]{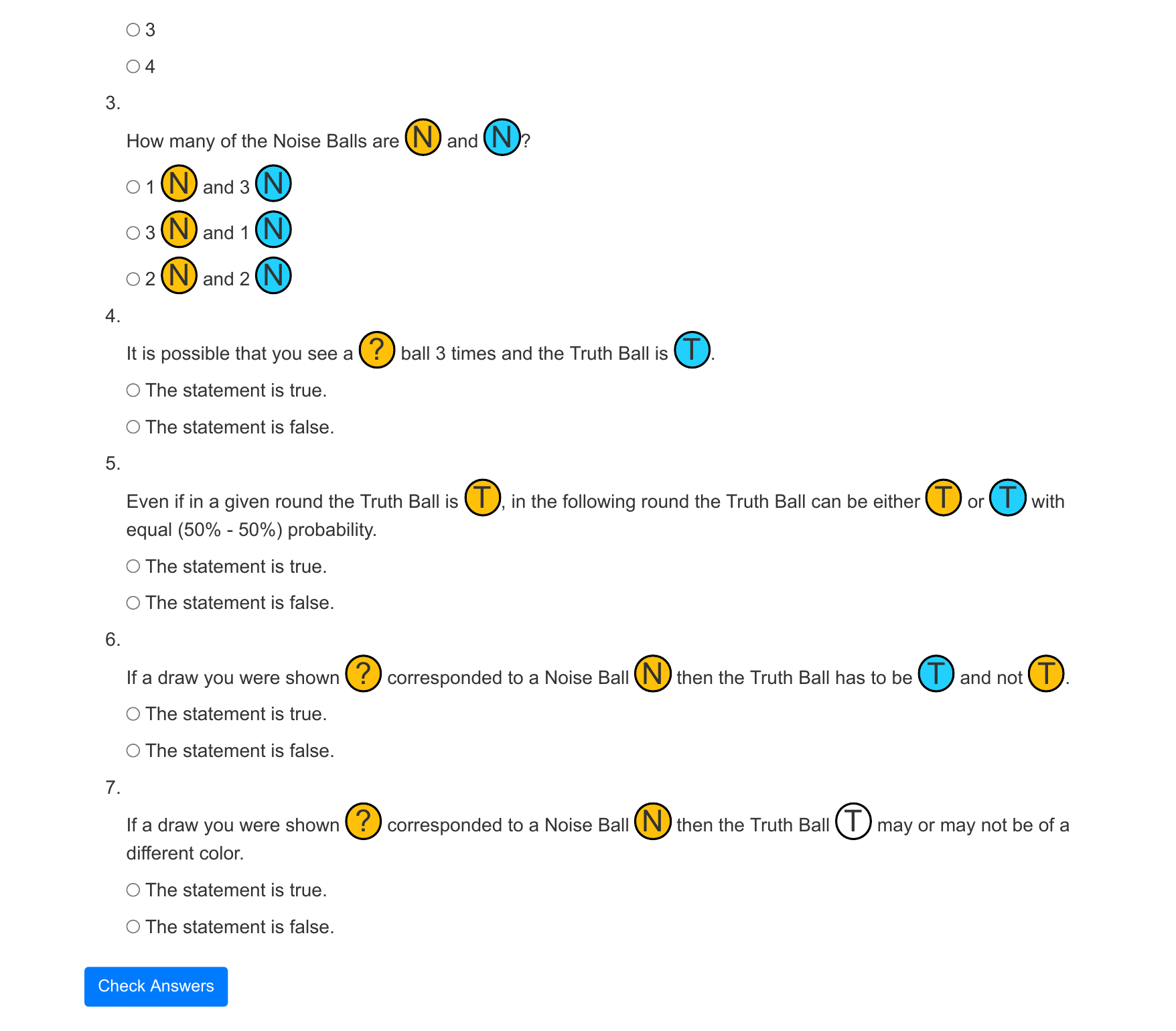}
\end{figure}

\subsubsection{Practice Round}
Participants played had two practice rounds before starting the task.
It was explicitly mentioned that these would not count toward their payment.
\begin{figure}[H]
    \centering
    \includegraphics[width=\textwidth]{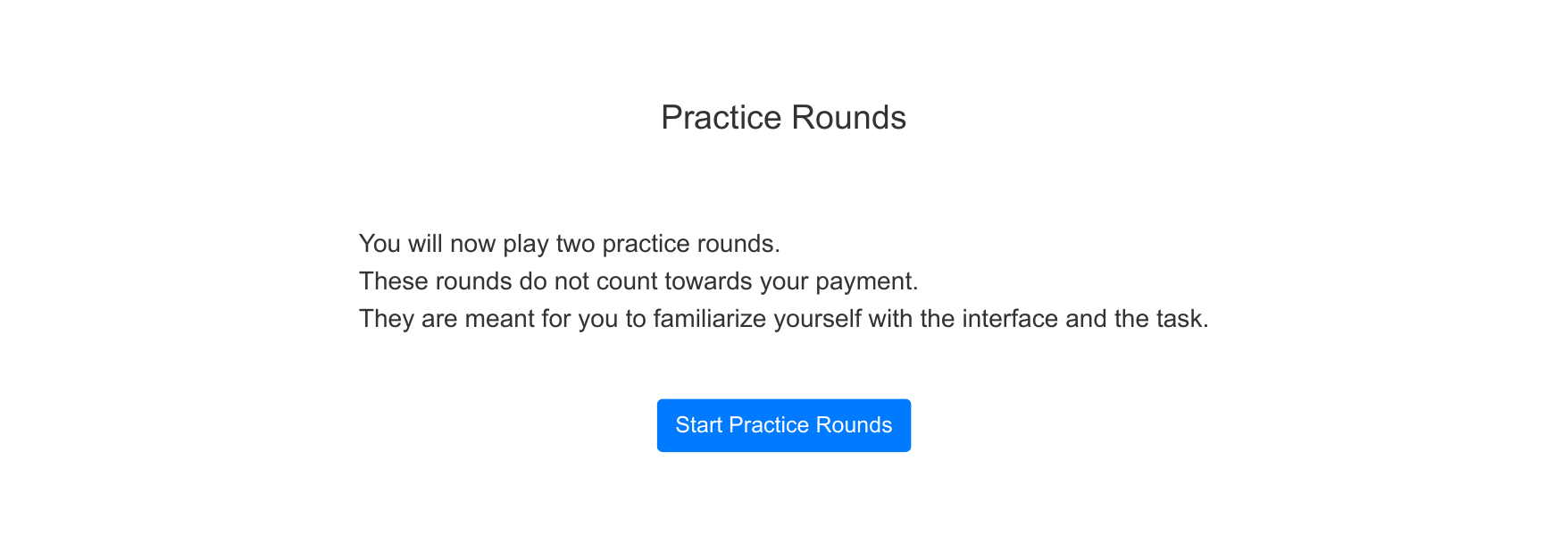}
\end{figure}
\begin{figure}[H]
    \centering
    \includegraphics[width=\textwidth]{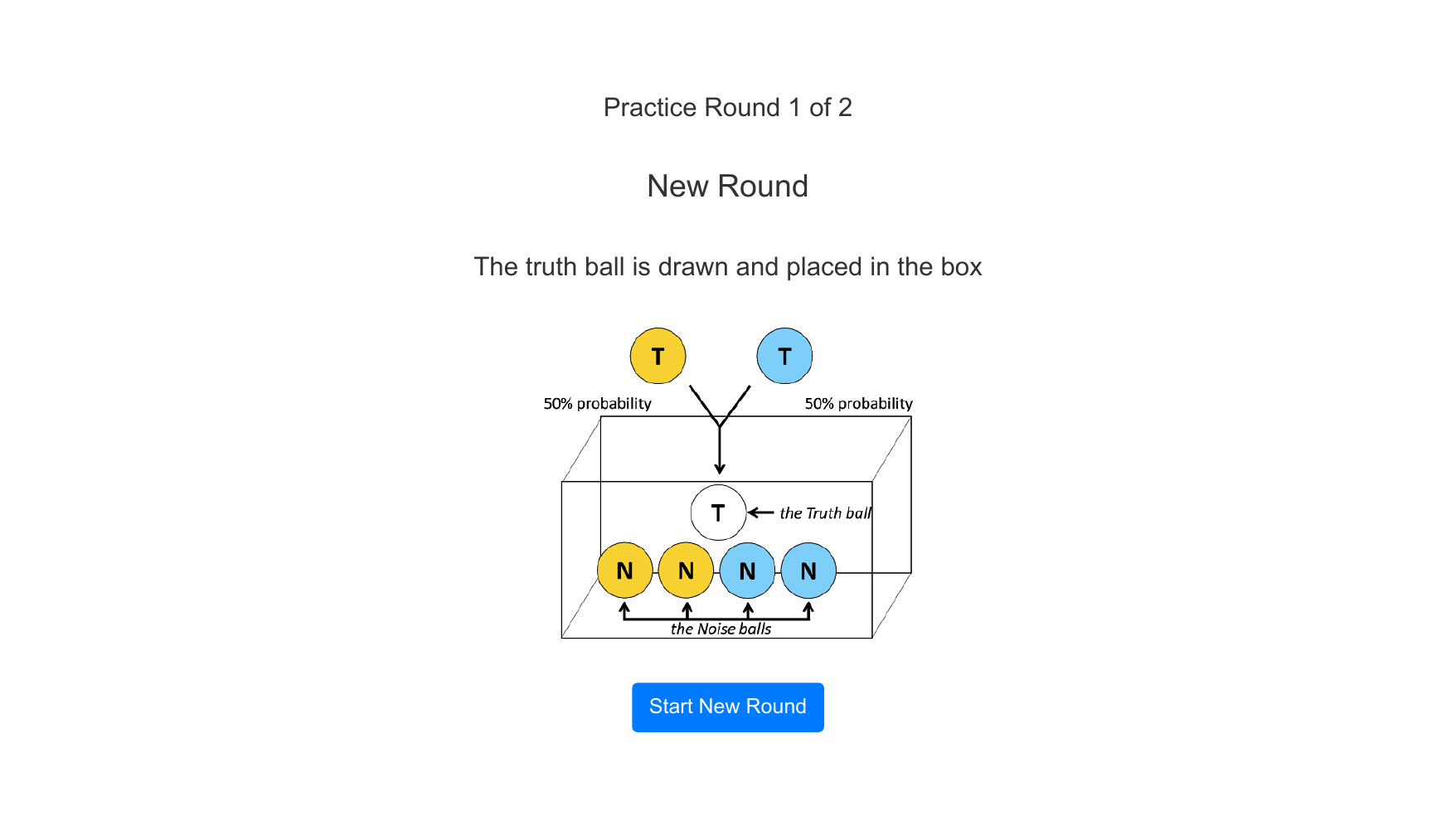}
\end{figure}

\newpage
One the page loaded, the slider was blank and only activated once the participants clicked on it.
\begin{figure}[H]
    \centering
    \includegraphics[width=\textwidth]{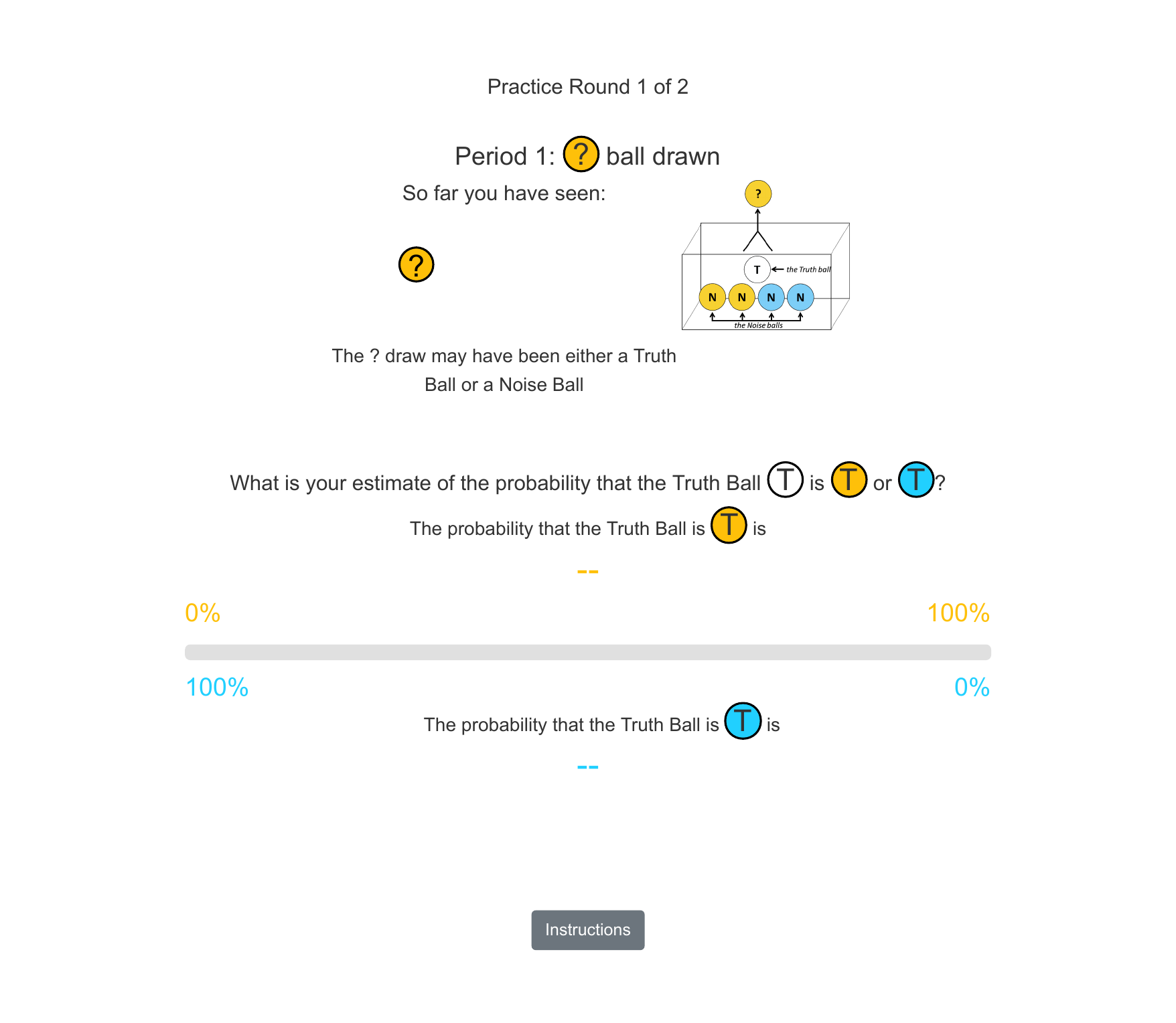}
\end{figure}
\begin{figure}[H]
    \centering
    \includegraphics[width=\textwidth]{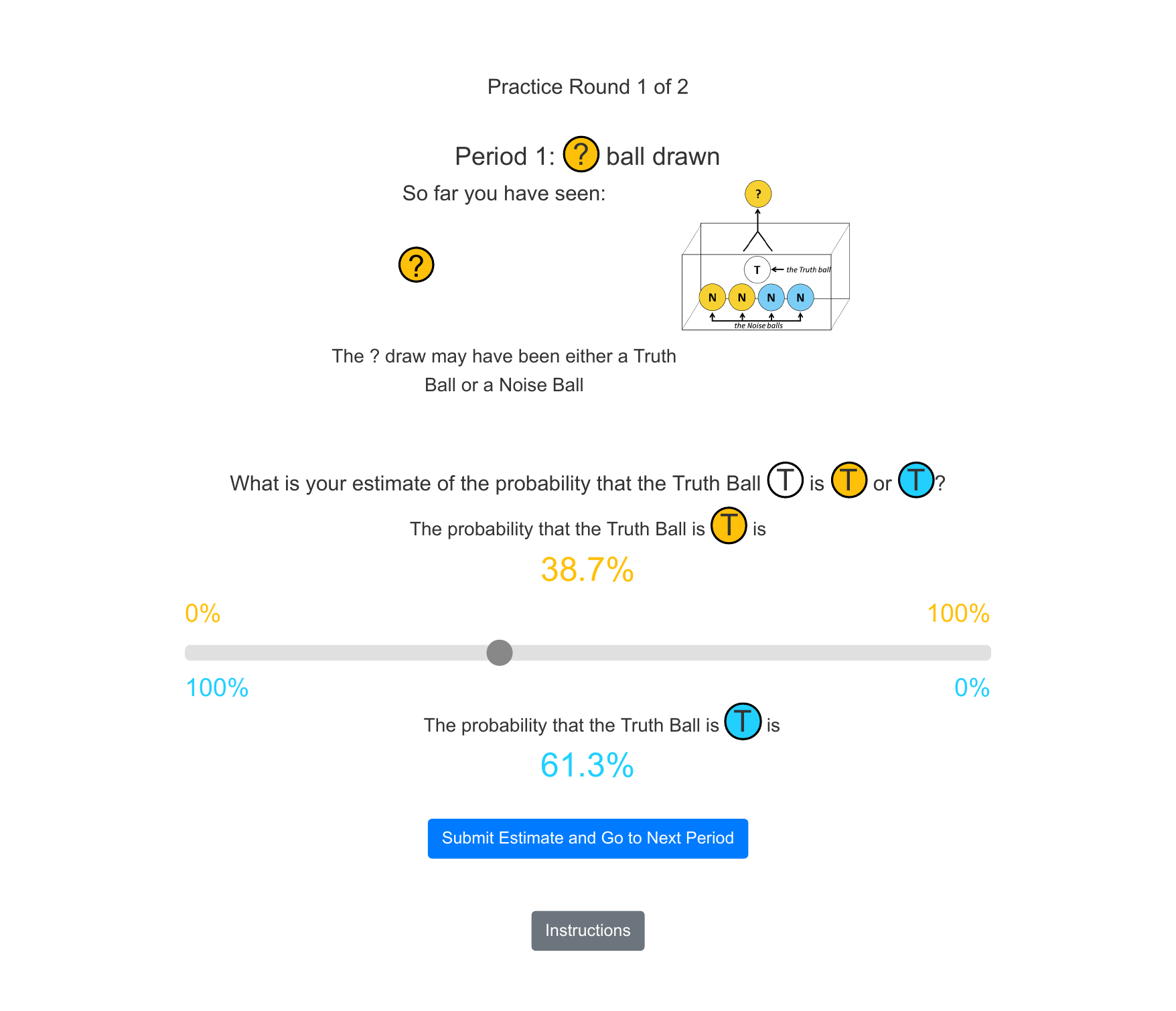}
\end{figure}

\subsubsection{Captchas}
Participants face five different captchas at different rounds.
They had 3 tries and one minute to submit for each try.
Were they to fail the 3 tries, the task ended and they would not receive any bonus.
\begin{figure}[H]
    \centering
    \includegraphics[width=\textwidth]{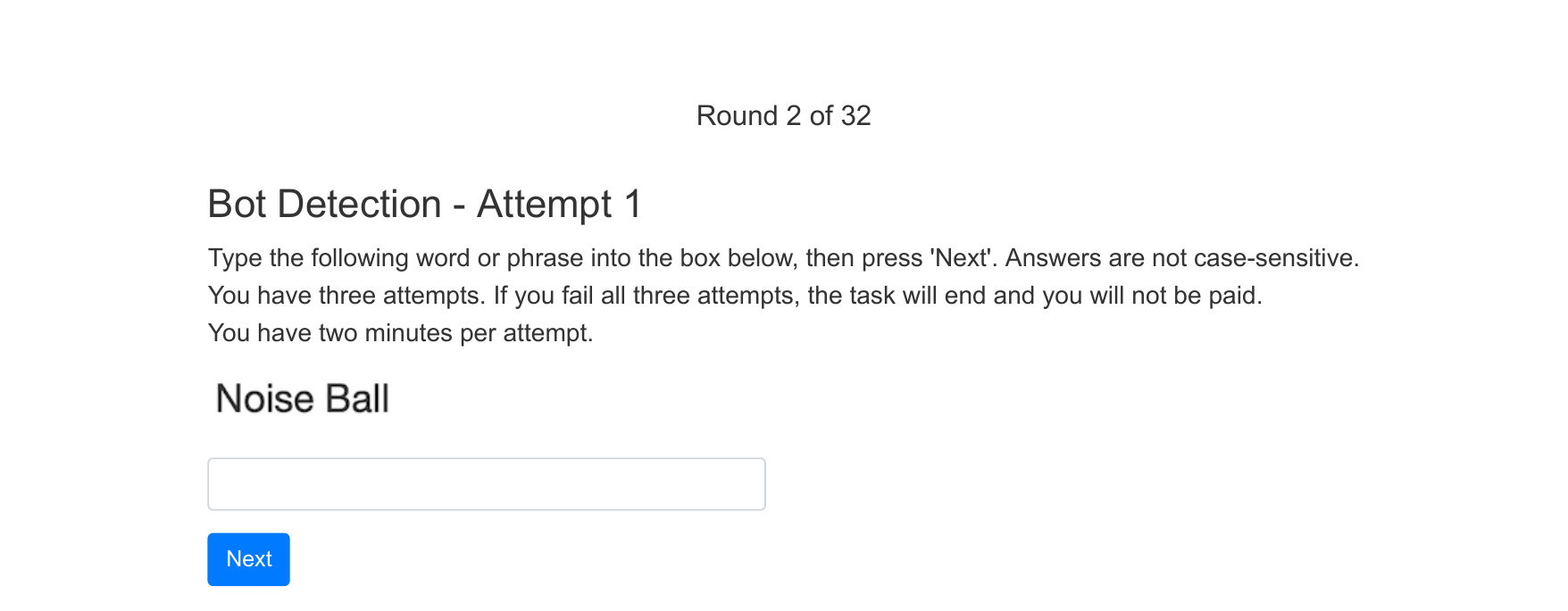}
\end{figure}

\subsubsection{Rounds}
The rounds were described in \hyref{section:framework:design}[Section].
\begin{figure}[H]
    \centering
    \includegraphics[width=\textwidth]{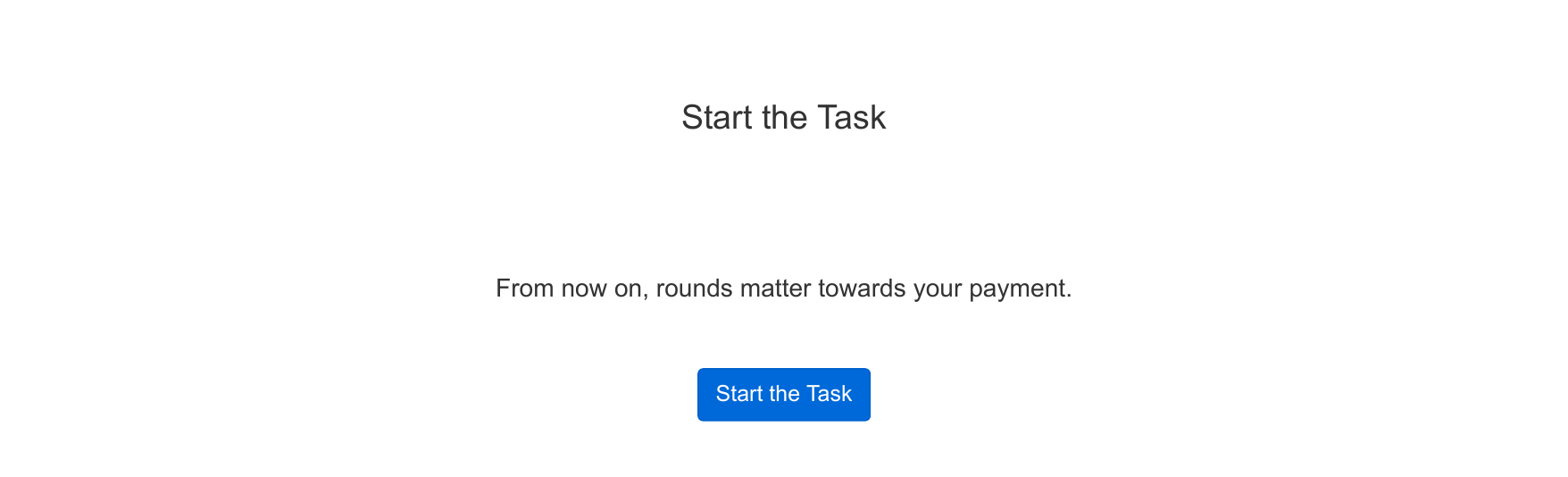}
    \includegraphics[width=\textwidth]{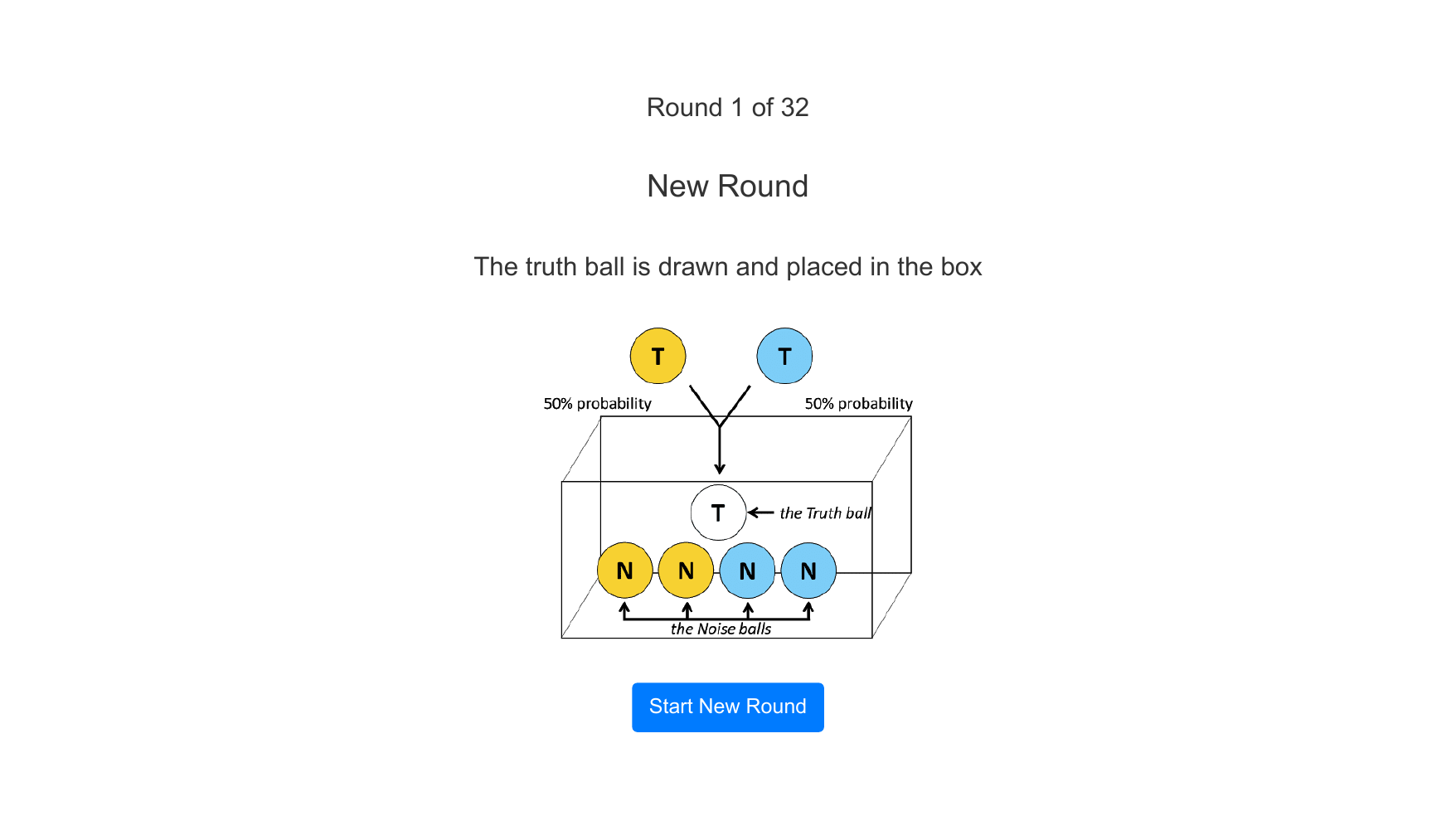}
\end{figure}
\begin{figure}[H]
    \centering
    \includegraphics[width=\textwidth]{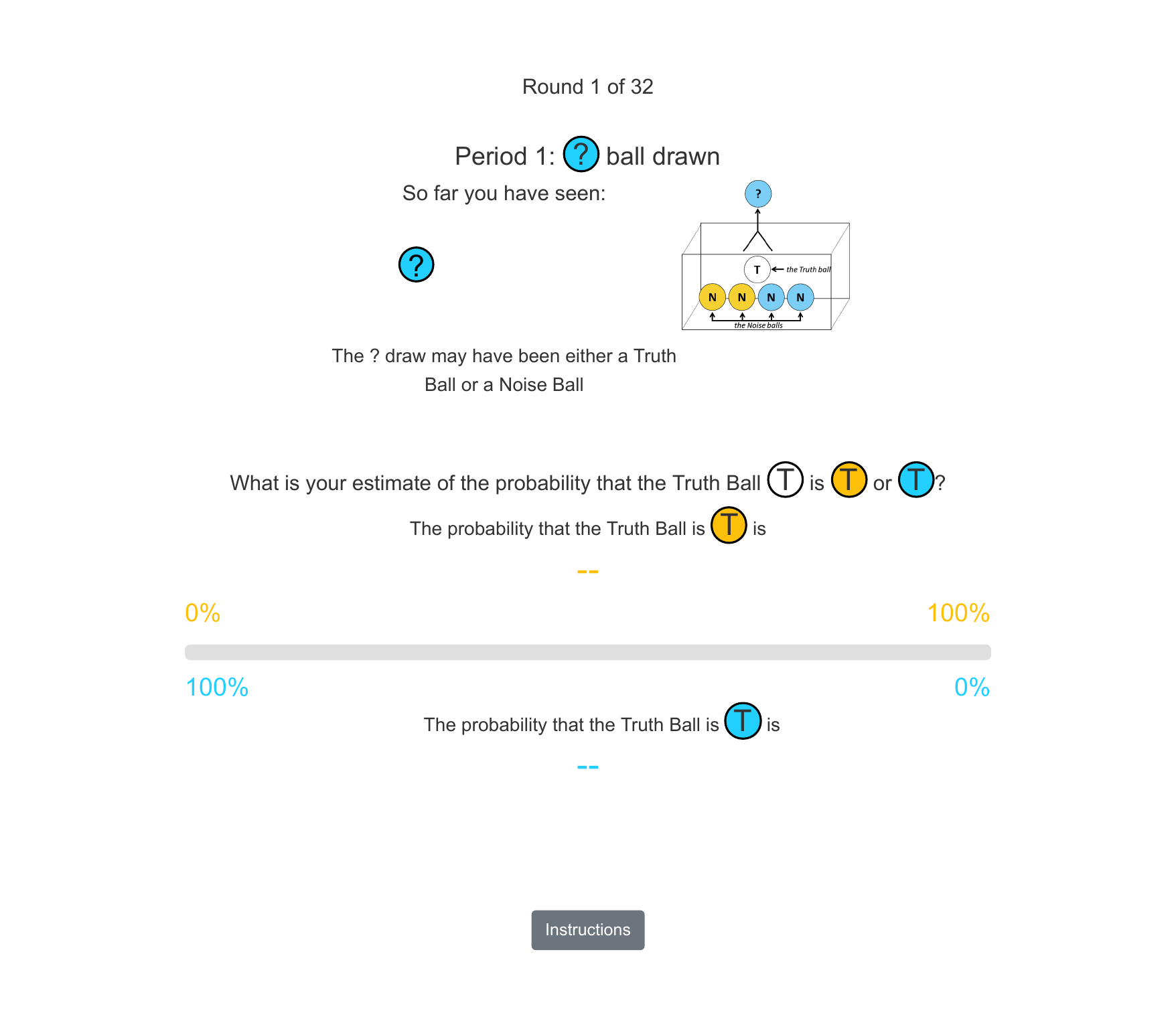}
\end{figure}

\subsubsection{Final Period Elicitation Only}
Were the participants to be in the treatment arm in which beliefs were elicited only at the last period of each round, the last period would be just as before.
In periods in which there was no belief elicitation, they would observe just the ball draw:
\begin{figure}[H]
    \centering
    \includegraphics[width=\textwidth]{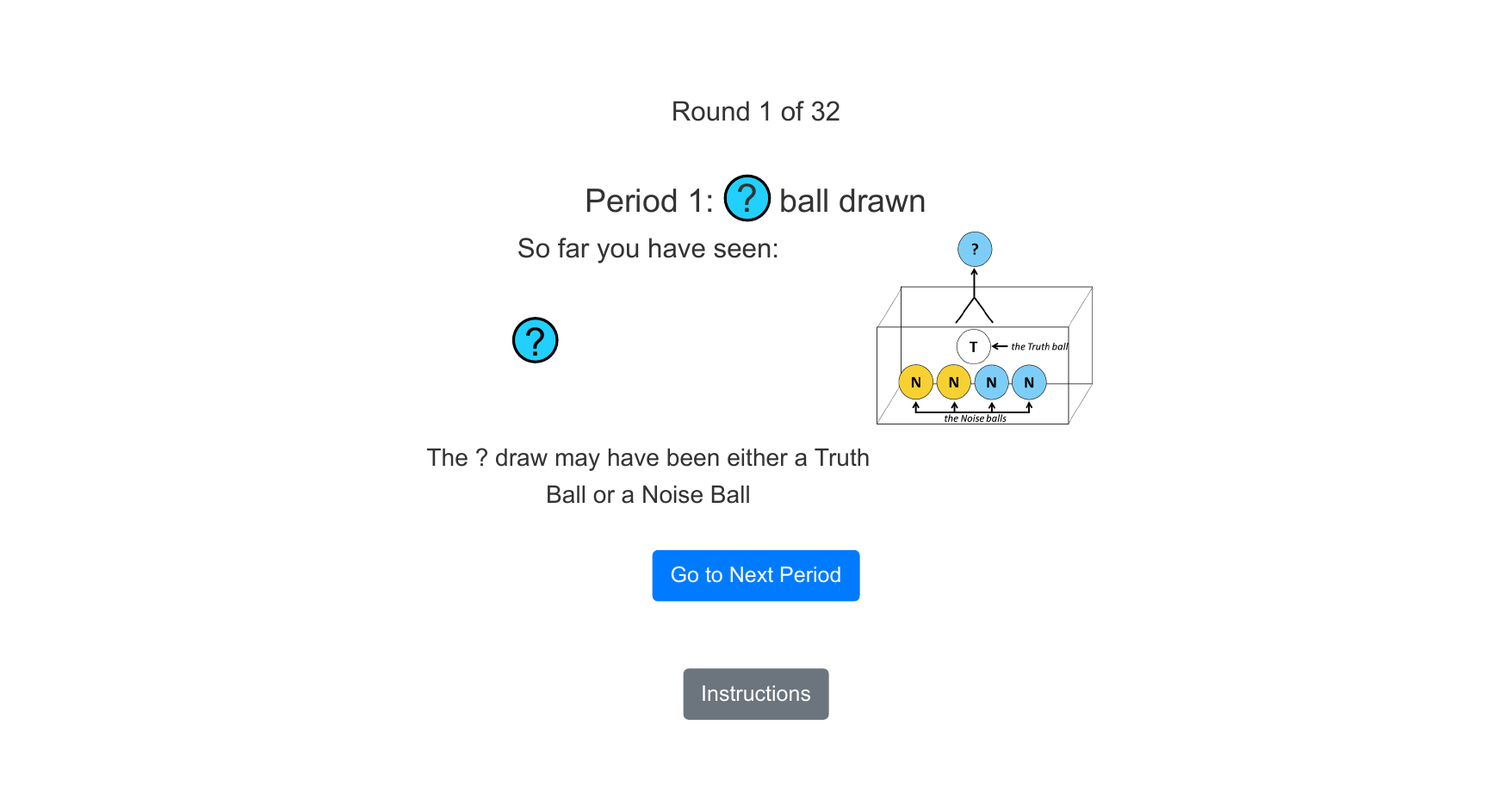}
\end{figure}

\subsubsection{Quantitative Questions}
After the main task, the participants had to answer three questions meant to assess their quantitative ability; these were incentivized.
\begin{figure}[H]
    \centering
    \includegraphics[width=\textwidth]{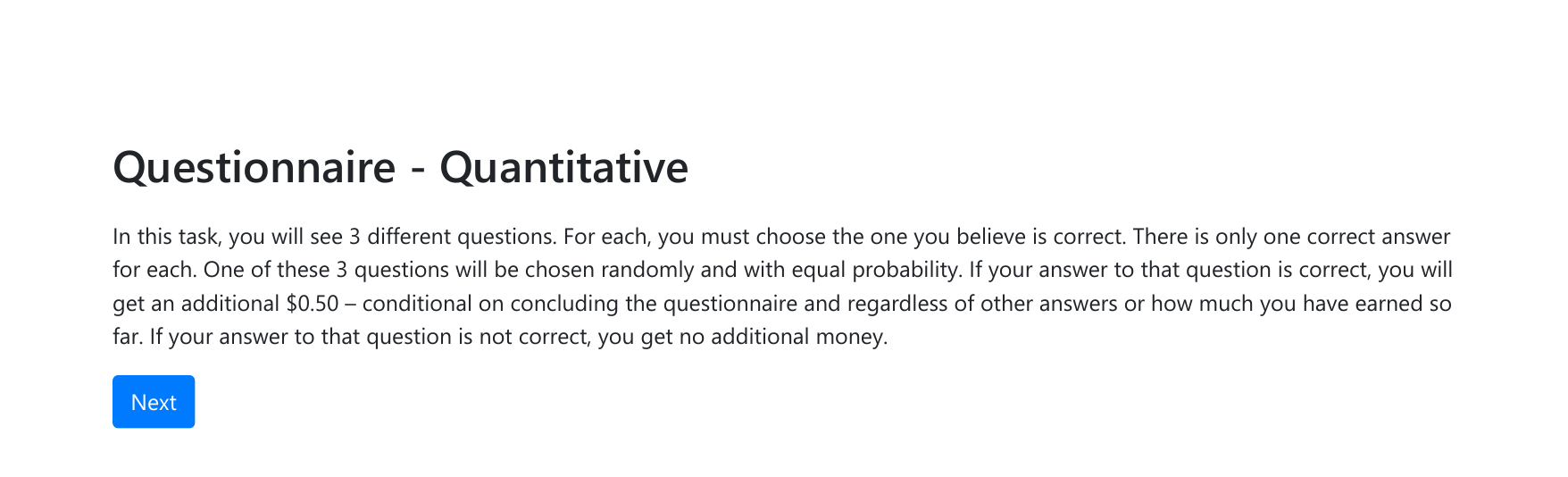}
\end{figure}
\begin{figure}[H]
    \centering
    \includegraphics[width=\textwidth]{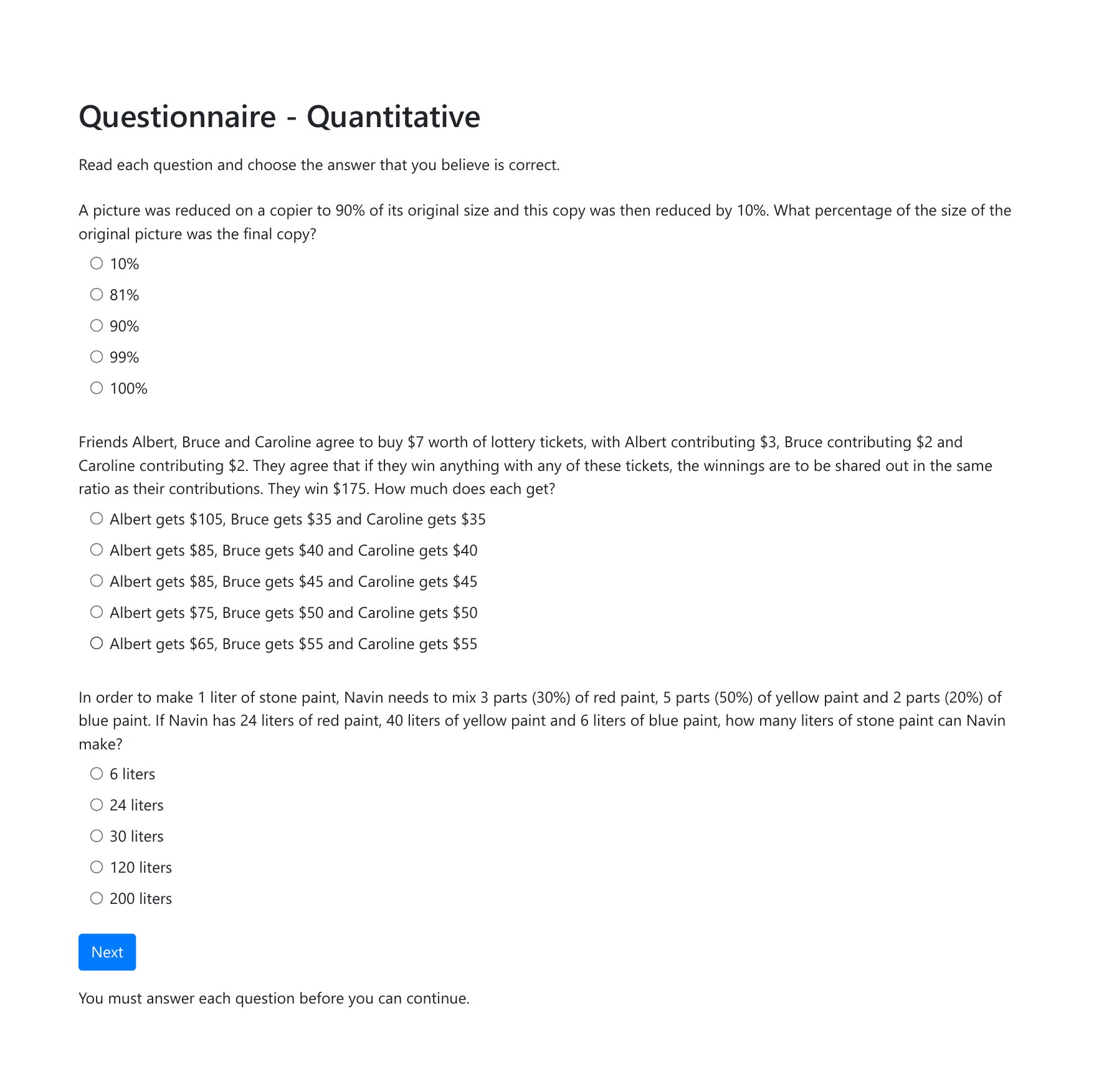}
\end{figure}

\subsubsection{Debrief and Payments}
Following the task, we gathered participants comments, socio-demographic information, and informed them of the payment they would receive.
\begin{figure}[H]
    \centering
    \includegraphics[width=\textwidth]{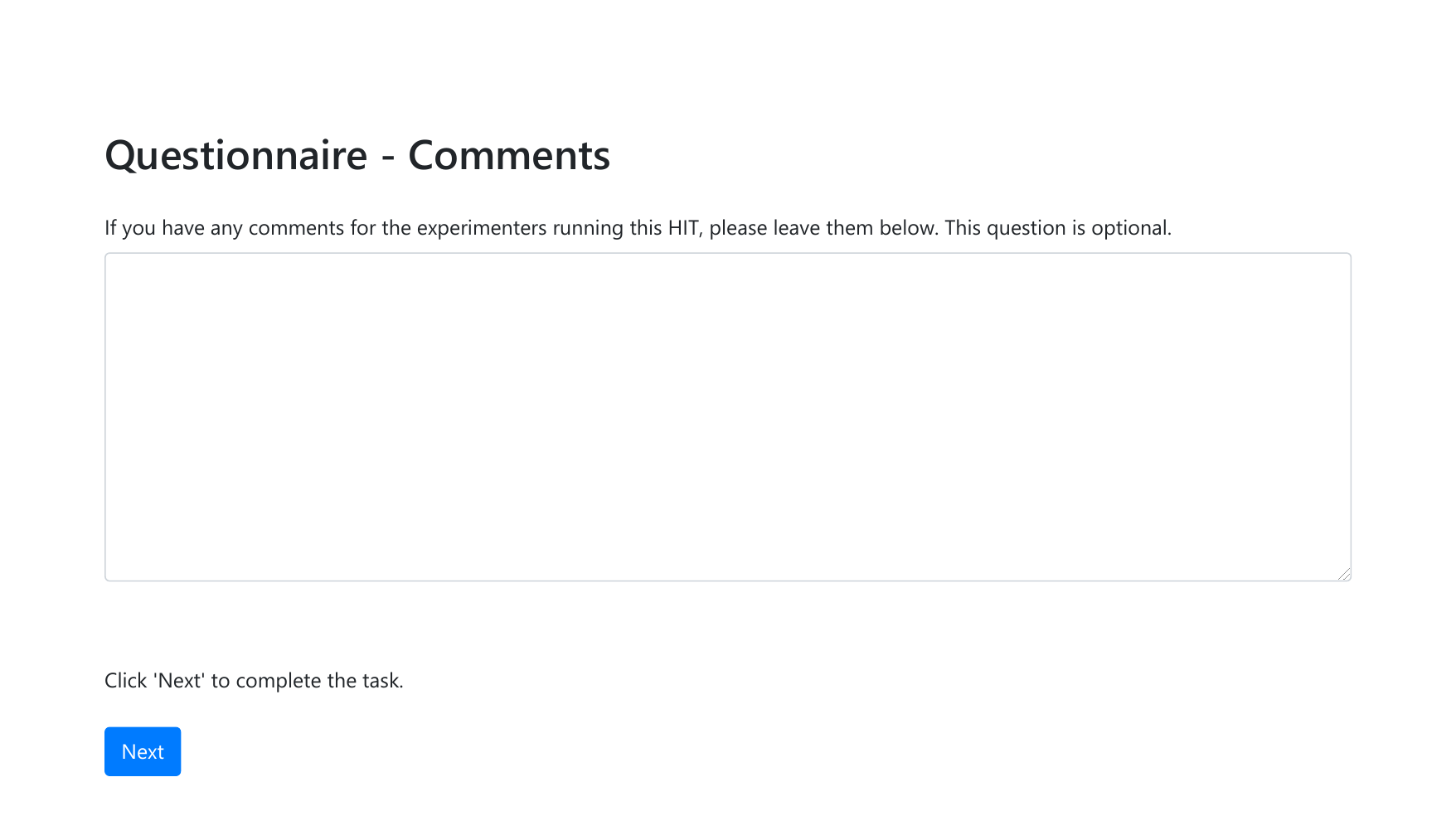}
\end{figure}
\begin{figure}[H]
    \centering
    \includegraphics[width=\textwidth]{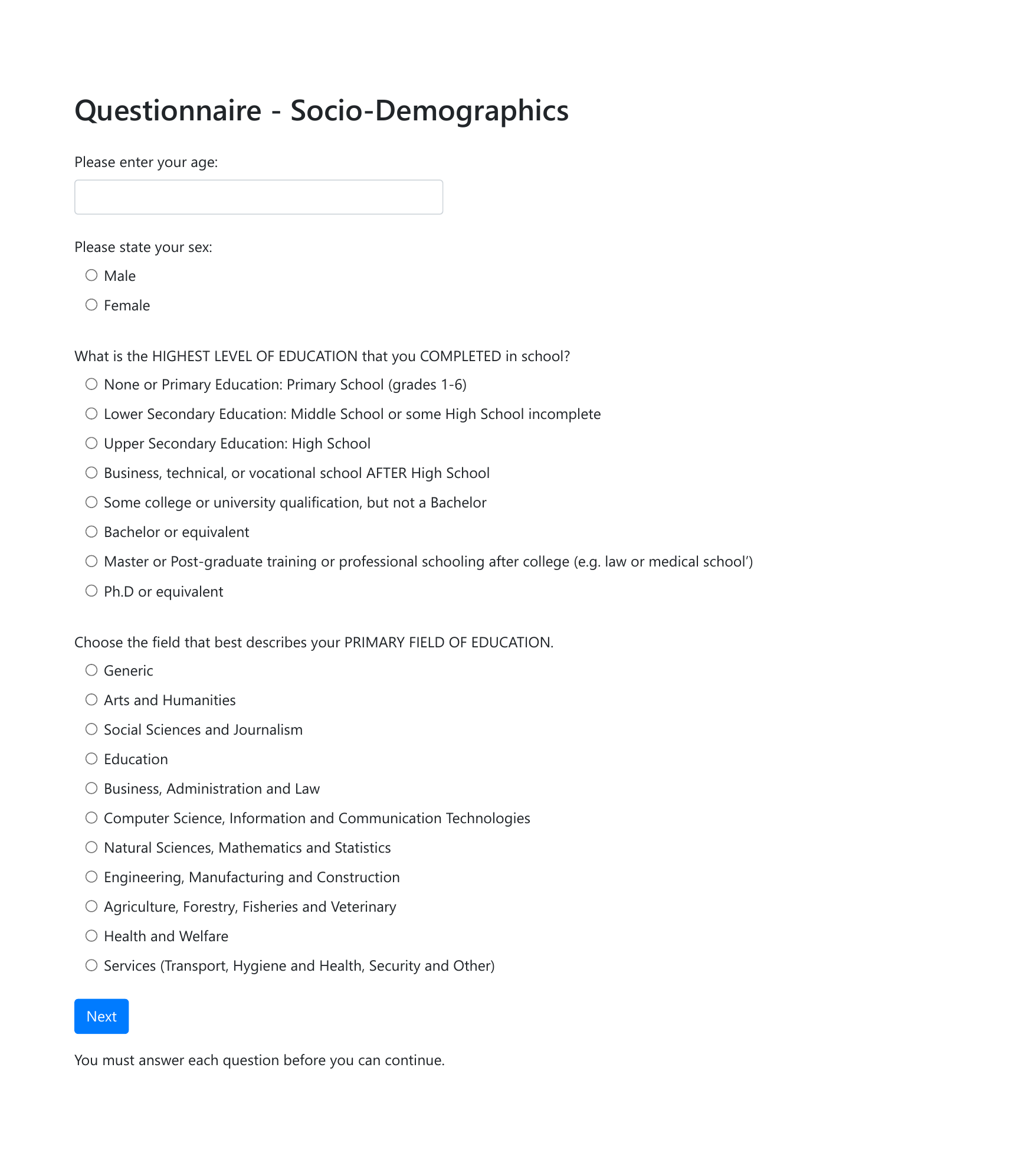}
\end{figure}
\begin{figure}[H]
    \centering
    \includegraphics[width=\textwidth]{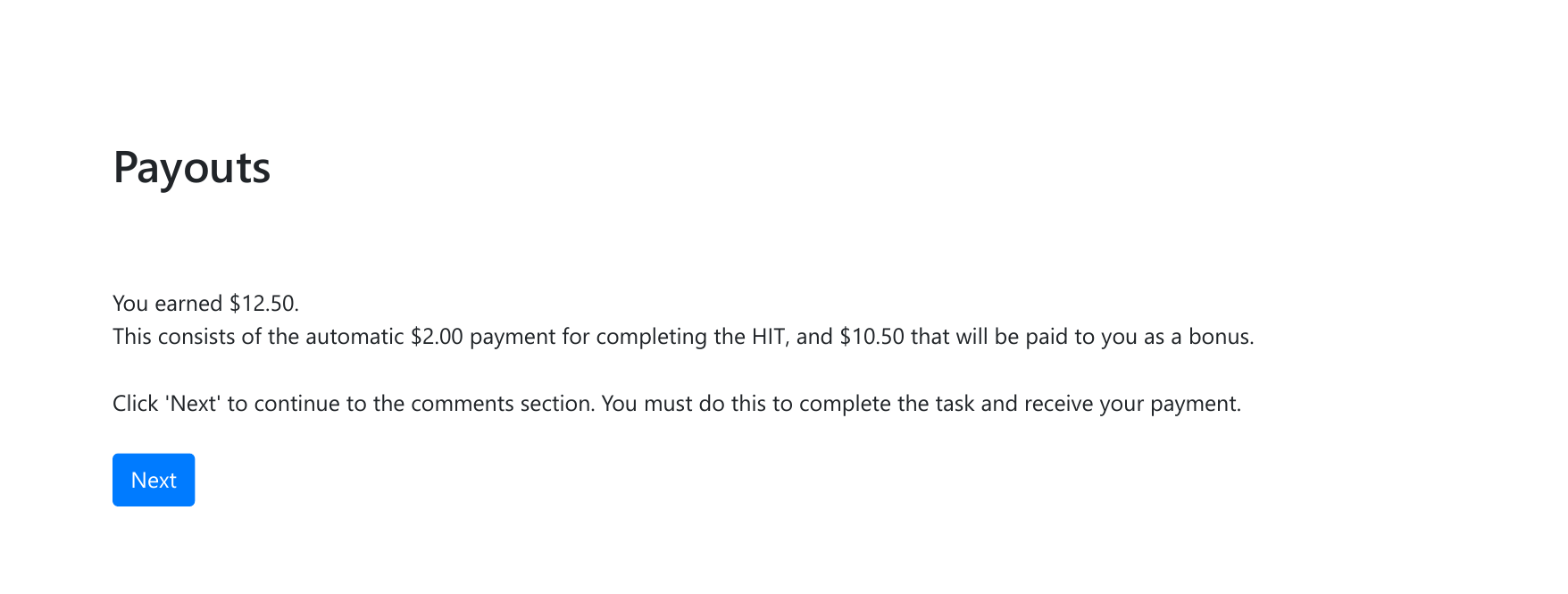}
\end{figure}

\end{document}